\documentclass[aps, prd, a4paper, 10pt, showpacs, superscriptaddress, floatfix, nofootinbib, reprint]{revtex4-1}
\tighten
\usepackage{color}
\usepackage{xcolor}
\usepackage{amsmath,amssymb,graphicx}
\usepackage{gensymb}
\usepackage{wasysym}
\usepackage{bm,lipsum}
\usepackage{times}
\usepackage{multirow}
\usepackage{microtype}
\usepackage{comment}
\usepackage[utf8]{inputenc}
\usepackage{booktabs}
\usepackage{subfigure}
\usepackage[normalem]{ulem}
\usepackage[varg]{txfonts}
\usepackage{bm,graphics,graphicx,epsfig,color,ulem}
\usepackage{array}
\usepackage{tabularx}
\usepackage[colorlinks, pdfborder={0 0 0}]{hyperref}
\definecolor{LinkColor}{rgb}{0.75, 0, 0}
\definecolor{CiteColor}{rgb}{0, 0.5, 0.5}
\definecolor{UrlColor}{rgb}{0, 0, 0.75}
\hypersetup{linkcolor=LinkColor}
\hypersetup{citecolor=CiteColor}
\hypersetup{urlcolor=UrlColor}
\allowdisplaybreaks
\newcommand{\hubble} {Hubble-Lema\^{i}tre\,}

\newcommand{\IGC}{Institute for Gravitation and the Cosmos, Department of Physics, Pennsylvania State University, University Park, Pennsylvania 16802, USA}
\newcommand{\milan}{Dipartimento di Fisica ``G. Occhialini'', Universit\'a degli Studi di Milano-Bicocca, Piazza della Scienza 3, 20126 Milano, Italy}
\newcommand{\infn}{INFN, Sezione di Milano-Bicocca, Piazza della Scienza 3, 20126 Milano, Italy}

\newcolumntype{C}{>{\arraybackslash}m{6cm}}

\newcommand{\aplus} {\texttt{HLVKI+}}
\newcommand{\voy} {\texttt{VK+HLIv}}
\newcommand{\et} {\texttt{HLKI+E}}
\newcommand{\ce} {\texttt{VKI+C}}
\newcommand{\EC} {\texttt{KI+EC}}
\newcommand{\ECS} {\texttt{ECS}}
\newcommand{\cetw} {\texttt{VKI+C..20}}
\newcommand{\cefo} {\texttt{VKI+C..40}}
\newcommand{\cetwfo} {\texttt{VKI+C..20-40}}
\newcommand{\cefofo} {\texttt{VKI+C..40-40}}

\newcolumntype{?}{!{\vrule width 1pt}}

\newcommand{\red}{\color{black}}

\begin{document}

%-----------------------------------------TITLE----------------------------------------------%

\title{Listening to the Universe with next generation ground-based gravitational-wave detectors}
\author{Ssohrab Borhanian} \email{ssohrab.borhanian@unimib.it}
\affiliation{\IGC}
\affiliation{Theoretisch-Physikalisches Institut, Friedrich-Schiller-Universit\"at Jena, Fr\"obelstieg 1, 07743 Jena, Germany}
\affiliation{\milan}
\affiliation{\infn}

\author{B. S. Sathyaprakash}
\affiliation{\IGC}
\affiliation{Department of Astronomy and Astrophysics, Pennsylvania State University, University Park, Pennsylvania 16802, USA}
\affiliation{School of Physics and Astronomy, Cardiff University, Cardiff, CF24 3AA, United Kingdom}
\date{\today}

\begin{abstract}
In this study, we use simple performance metrics to assess the science capabilities of future ground-based gravitational-wave detector networks---composed of A+ or Voyager upgrades to the LIGO, Virgo, and KAGRA observatories and proposed next generation observatories such as Cosmic Explorer and Einstein Telescope. These metrics refer to coalescences of binary neutron stars (BNSs) and binary black holes (BBHs) and include: (i) network detection efficiency and detection rate of cosmological sources as a function of redshift, (ii) signal-to-noise ratios and the accuracy with which intrinsic and extrinsic parameters would be measured, and (iii) enabling multimessenger astronomy with gravitational waves by accurate 3D localization and early warning alerts. We further discuss the science enabled by the small population of rare and extremely loud events. While imminent upgrades will provide impressive advances in all these metrics, next generation observatories will deliver an improvement of an order-of-magnitude or more in most metrics. In fact, a network containing two or three such facilities will detect half of all the BNS and BBH mergers up to a redshift of $z=1$ and $z=20$, respectively, give access to hundreds of BNSs and ten thousand BBHs with signal-to-noise ratios exceeding 100, readily localize hundreds to thousands of mergers to within $1\,{\rm deg^2}$ on the sky and better than 10\% in luminosity distance, respectively, and consequently, enable mutlimessenger astronomy through follow-up surveys in the electromagnetic spectrum several times a week. Such networks will further shed light on potential cosmological merger populations and detect an abundance of high-fidelity BNS and BBH signals which will allow investigations of the high-density regime of matter at an unprecedented level and enable precision tests of general relativity in the strong-field regime, respectively.
\end{abstract}

\maketitle

\section{Introduction:}

\subsection{Dawn of gravitational-wave astronomy}

Over the past five years and three observing runs, the Advanced Laser Interferometer Gravitational-Wave Observatory (LIGO) consisting of the LIGO-Hanford and LIGO-Livingston detectors \cite{TheLIGOScientific:2014jea} and the Virgo detector \cite{TheVirgo:2014hva} have discovered gravitational waves (GWs) from the merger of dozens of binary black holes (BBHs) \cite{Abbott:2016blz,Abbott:2016nmj,Abbott:2017vtc, LIGOScientific:2018mvr, Abbott:2020niy, LIGOScientific:2020ibl}, two binary neutron stars (BNSs) \cite{TheLIGOScientific:2017qsa, Abbott:2020uma} and two binary neutron star-black holes (NSBHs) \cite{LIGOScientific:2021qlt}. In addition to the pioneering first direct observation of GWs from a pair of merging BBHs (GW150914) \cite{Abbott:2016blz}, LIGO and Virgo have made many spectacular and surprising new discoveries. These include, amongst others, a BNS merger that was observed in the entire electromagnetic window from gamma-rays to radio waves (GW170817) \cite{TheLIGOScientific:2017qsa, GBM:2017lvd, Monitor:2017mdv}, systems in which black hole (BH) companions have masses that could not have resulted from the evolution of massive stars  (thus raising questions as to their origin, GW170729 \cite{Chatziioannou:2019dsz} and GW190521 \cite{Abbott:2020tfl, Abbott:2020mjq}), binaries that show a clear signature of sub-dominant octupole radiation in addition to the dominant quadrupole (confirming once again predictions of general relativity, GW190412 \cite{LIGOScientific:2020stg} and GW190814 \cite{Abbott:2020khf}), and a binary with a mass-ratio of almost 1:10 that challenges astrophysical BBH formation models while its secondary companion could well be the heaviest neutron star (NS) or the lightest BH observed so far (GW190814 \cite{Abbott:2020khf}). GW catalogs compiled by other groups \cite{Nitz:2018imz, Nitz:2020oeq, Roulet:2020wyq, Nitz:2021uxj, Nitz:2021zwj} have largely confirmed the population of coalescing compact binaries found by LIGO and Virgo but have reported additional events of interest (see, e.g., \cite{Zackay:2019btq, Zackay:2019tzo, Venumadhav:2019lyq, Chia:2021mxq, Olsen:2022pin})

These discoveries have already made a massive impact on our understanding of different tenets of astrophysics, fundamental physics, and cosmology. They have allowed a first glimpse into the dynamics of strongly curved spacetimes and the validity of general relativity (GR) in unexplored regimes of the theory \cite{TheLIGOScientific:2016src, Yunes:2016jcc, Abbott:2018lct, LIGOScientific:2019fpa, LIGOScientific:2020tif,  LIGOScientific:2021sio}, raised deeper questions on the formation mechanisms and evolutionary scenarios of compact objects \cite{Belczynski:2017gds, Belczynski:2018ptv, Gupta:2019nwj, Fishbach:2019bbm, Abbott:2020mjq, Abbott:2020khf, Fishbach:2020qag, Gerosa:2020bjb, Zevin:2020gbd, Zevin:2020gma, Olejak:2021iux, Baibhav:2021qzw, Mahapatra:2021hme, Fishbach:2021yvy}, provided a new tool for measuring cosmic distances that will help in precision cosmology \cite{Schutz:1986gp, Abbott:2017xzu} and in mapping the large scale structure of the Universe \cite{Mukherjee:2019wcg, Shao:2021zox, Libanore:2020fim, Vijaykumar:2020pzn}, and brought to bear a novel approach to determine the structure and properties of NSs to help in the exploration of the dense matter equation of state which governs the dynamics of NS cores \cite{Flanagan:2007ix, De:2018uhw, Abbott:2018exr, Nicholl:2021rcr, Tews:2019ioa, Raithel:2019uzi, Lim:2019som, Malik:2018zcf, Tews:2018iwm, Radice:2017lry, Kumar:2021aog}. The second part of the third observing run saw the discovery of two neutron star-black hole binaries, GW200105 (a marginal event) and GW20015 \cite{LIGOScientific:2021qlt, LIGOScientific:2021djp} thereby completing the quest for the three types of ultra-compact-object binaries, two of which have only been observed in the gravitational window. 

At the same time, multimessenger observations of GW170817---the inspiral and coalescence of a pair of NSs---have at once begun to impact on several enigmatic questions in nuclear astrophysics and fundamental physics. We now know that BNS mergers are progenitors of short, hard gamma-ray bursts \cite{Monitor:2017mdv} and sites where r-process heavy elements are produced from neutron-rich ejecta \cite{Drout:2017ijr, Coulter:2017wya, Cowperthwaite:2017dyu, Kasen:2017sxr, Soares-Santos:2017lru, Valenti:2017ngx, Arcavi:2017xiz, Tanvir:2017pws, Lipunov:2017dwd, Evans:2017mmy}, GWs travel at the speed of light to within one part in $10^{15}$ \cite{Monitor:2017mdv} which has helped rule out certain alternative theories of gravity that were invoked to explain the origin of dark energy \cite{Creminelli:2017sry, Ezquiaga:2017ekz}, and that the equation of state of dense matter cannot be too stiff, thus constraining the radius of NSs of 1.4 $M_\odot$ to be below about 14 km \cite{De:2018uhw, Abbott:2018exr}. Moreover, the luminosity distance determined by gravitational-wave observations and the redshift from the identification of the host galaxy allowed the first measurement of the Hubble constant with standard sirens \cite{Hotokezaka:2018dfi, Mukherjee:2020kki} thus beginning a new era in observational cosmology \cite{Chen:2017rfc, Borhanian:2020vyr, Mukherjee:2020hyn} envisaged more than three decades ago \cite{Schutz:1986gp}.

LIGO-Virgo discoveries have informed numerous investigations about the properties of NSs and BHs including the maximum mass of NSs \cite{Margalit:2017dij, Ruiz:2017due, Most:2018hfd, Shibata:2019ctb, Godzieba:2020tjn}, potential primordial origin of the observed BHs \cite{Clesse:2016vqa, Sasaki:2016jop, Bird:2016dcv, Carr:2016drx,  Kashlinsky:2016sdv, Saito:2009jt, Franciolini:2021nvv}, stochastic backgrounds that might be produced by the astrophysical population of BBH and BNS mergers \cite{TheLIGOScientific:2016wyq,Regimbau:2016ike,Abbott:2017xzg}, to name a few. It has already become clear that the GW window has the potential to transform our knowledge of physics and astronomy in the coming decades---some of which we hope to explore in this study.

\begin{figure}[b]
    \centering
    \includegraphics[width=\linewidth]{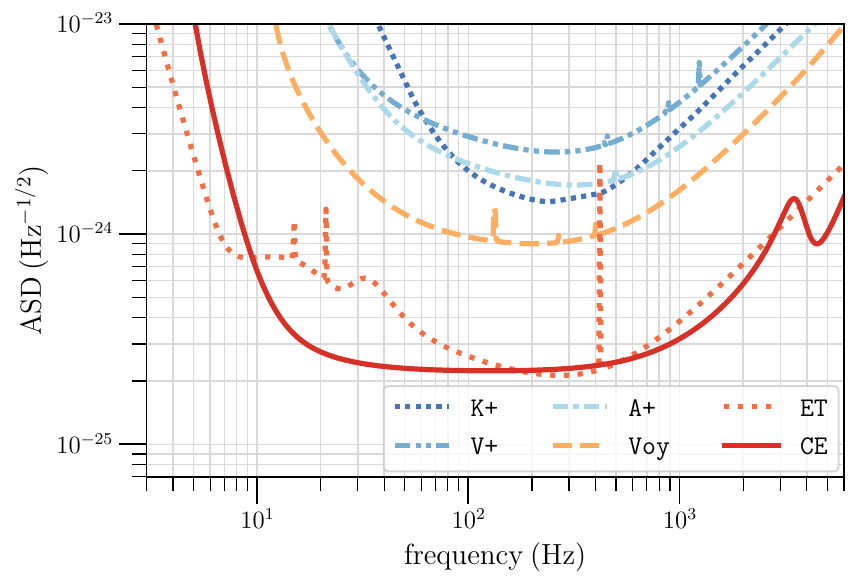}
    \caption{Amplitude spectral densities (ASDs) illustrating the sensitivities of three generations of detectors: (i) Advanced-plus upgrade of LIGO (A+: Hanford, Livingston, and India), Virgo (V+: Italy), and KAGRA (K+: Japan), (ii) Voyager upgrade of LIGO (Voy), and (iii) the next generation of detectors consisting of Cosmic Explorer (CE) and Einstein Telescope (ET). These ASDs could be realized in the next 5, 10, and 15-20 years, respectively. Refer to the text for different detector networks, composed of combinations of these detectors, considered in this paper.
    The ASDs are taken from the \texttt{kagra\_plus}, \texttt{advirgo\_plus}, \texttt{aplus}, \texttt{voyager\_cb}, \texttt{et}, and \texttt{ce2\_40km\_cb} \texttt{.txt}-files inside \texttt{ce\_curves.zip} file at \href{https://dcc.cosmicexplorer.org/public/0163/T2000007/005/ce\_curves.zip}{https://dcc.cosmicexplorer.org/public/0163/T2000007/005/ ce\_curves.zip}. The shown ET ASD is scaled by a factor of 2/3 to represent the effective sensitivity of ET's triangular design.}
    \label{fig:sensitivity curves}
\end{figure}

\subsection{Imminent upgrade and new detectors}

With the planned sensitivity improvements of Advanced LIGO and Advanced Virgo \cite{Aasi:2013wya, Reitze:2019iox}) and the addition of two new detectors, KAGRA \cite{Akutsu:2018axf, Akutsu:2020his} and LIGO-India \cite{Unnikrishnan:2013qwa, Saleem:2021iwi}), the next five years of gravitational-wave astronomy will enable significant advances in resolving some of the outstanding problems in physics and astronomy. We will explore in detail the potential of this imminent \emph{A+ detector generation} (i.e. upgrades to Advanced LIGO, Advanced Virgo, and KAGRA with sensitivity curves shown in Fig.~\ref{fig:sensitivity curves}). The \hubble tension---the discrepancy between the early and late Universe measurements of the Hubble constant \cite{Riess:2019cxk}---could be resolved by the A+ network with multimessenger observations of $\sim 50$ BNS mergers \cite{Chen:2017rfc} or with the identification of a handful of BBHs within 500 Mpc \cite{Borhanian:2020vyr}. The BNS mergers could also help gain first insights into the cold equation of state of dense cores of neutron stars. The A+ generation will observe several BBH mergers each day and a loud BBH event (with signal-to-noise ratio in excess of 100) each month (cf.\, Fig.\,\ref{fig:detection_efficiency_rate}, bottom right panel)---events that would help verify the BH no-hair theorem and further constrain GR or see hints of its violation.

\subsection{Future of gravitational-wave astronomy}

It is possible to further enhance the sensitivity of current facilities with new lasers, cryo-cooled mirror substrates, active noise cancellation techniques for gravity gradient, and the like. The concept of Voyager technology \cite{Adhikari:2020gft}, which can be installed in current facilities, aims to improve the sensitivity beyond A+ detectors by a factor of $\sim 2$ to 4 depending on the frequency.  New facilities---such as interferometers with longer arms---and instrumentation would be needed to make sensitivity improvements beyond that facilitated by the Voyager concept. Einstein Telescope (ET) \cite{Punturo:2010zz, Hild:2010id}, Cosmic Explorer (CE) \cite{Evans:2016mbw}, and Neutron Star Extreme Matter Observatory (NEMO) \cite{Ackley:2020atn} are three such concepts (jointly referred to as next generation or, simply, NG) that are currently being pursued with the hope of first facilities being built within the next 10 to 15 years.

There have been extensive studies on the capability of next generation detectors and the scientific discoveries they enable \cite{Sathyaprakash:2009xt, Punturo:2010zza, Zhao:2010sz, Sathyaprakash:2012jk,  Vitale:2016icu, Vitale:2018nif, Adhikari:2019zpy, Hall:2019xmm, Sathyaprakash:2019nnu, Sathyaprakash:2019rom, Sathyaprakash:2019yqt, Kalogera:2019sui, Maggiore:2019uih, Evans:2021gyd, Saleem:2021iwi, Bailes:2021tot, Kalogera:2021bya, Yang:2021qge, Jin:2022qnj, Jin:2020hmc, Jin:2021pcv, Zhang:2019loq}. In this study we will focus in detail on the potential of the Voyager, ET, and CE detector generations using the Voy, ET, and CE noise curves shown in Fig.~\ref{fig:sensitivity curves}.  While out of the scope of our investigation, the triangular geometry proposed for ET facilitates a particularly powerful probe of the detector's sensitivity and data quality. In a triangular configuration, ET consists of three V-shaped detectors, each with an opening angle of 60 degrees, rotated relative to each other by 120 degrees. The sum of the responses of the three detectors, called the {\em null stream} \cite{Branchesi:2023mws}, is, by construction, devoid of any GW signals \cite{Sathyaprakash:2009xt}. This null stream enables an unbiased estimation of the detector's sensitivity, free from the confusion noise from the background of unresolved GW sources, and also provides a way of measuring non-stationary, incoherent noise artifacts \cite{Goncharov:2022dgl}.

We find that a Voyager network would localize {\red $\sim 16$} BNS mergers within a redshift of 0.5 to within {\red 1 deg$^2$,} providing ample opportunity for EM follow-up, and detect thousands of BNS mergers within its per-mille redshift of {\red $z=0.9$} (see Sec.\,\ref{sec:rate} for definitions of the per-mille redshift and reach used in this paper). In contrast, the best NG network would detect almost every BNS merger, some 50,000 of them, up to a redshift of $z=0.9,$ localizing {\red a few thousand} of them to better than {\red 1 deg$^2$}, measuring distances to {\red more than ten thousand} of the mergers to an accuracy better than {\red $10\%$}. Together with EM follow-up observations, such high-fidelity measurements would provide unprecedented access to fundamental questions in physics and astronomy including the nature of NSs, central engines of gamma-ray jets, production of heavy elements in the Universe, and the expansion rate as a function of redshift. NG observatories would have a reach of $z\sim 1\textendash2$ to BNS mergers and a per-mille redshift of $z\sim 3\textendash10$. 

The Voyager network will have access to BBHs all the way to {\red $z\sim 10,$} detecting {\red 50\%} of all mergers within a redshift of {\red 1.4}. It would achieve a signal-to-noise ratio of 100 for more than {\red 50 mergers} each year. The best NG network, on the other hand, will observe {\red almost all} BBH mergers within a redshift of {\red 10}, and {\red at least every other} BBH merger up to $z\gtrsim30.$ In fact, BBH mergers with signal-to-noise ratios of 1000 or larger will be observed {\red more than $10$ times each year}. Such exquisite observations will allow extremely accurate measurements of BH masses, their spins and distances, allowing us to address a variety of questions in astrophysics and dynamical spacetimes. These include, but are not limited to, the astrophysical models of the formation and evolution of BBHs, tests of the BH no-hair theorem, exploration of the epoch of the formation of first stars at redshifts of 10$\textendash$30, etc.

\subsection{Organization of the paper}
In Sec.~\ref{sec:methods} we will introduce the detector networks we are considering in this study, the properties of the source populations we are examining, and the analysis methodology we are using to assess the measurement capabilities of detector networks. In the rest of the paper we will introduce and discuss the metrics used to assess the relative performance of different networks.
In Sec.~\ref{sec:rate} we will discuss the efficiency of detector networks in detecting signals as a function of redshift (or luminosity distance). Detection efficiency will inform us of the completeness of the observed population with respect to the full underlying population of mergers. Sec.~\ref{sec:rate} will also discuss the cosmic merger rate of compact binary coalescences inferred using current LIGO-Virgo observations, a model for the redshift evolution of the star formation rate and metallicity, and the resulting detection rates of the networks as functions of the redshift.
In Sec.~\ref{sec:visibility} we will present the evaluation of the chosen detector networks and a forecast of the science questions that could be addressed by different generations of networks.
Enabling multimessenger astronomy will require accurate 3D sky localization of compact binary coalescences, which will be the focus of Sec.~\ref{sec:mma}. This Section will also include forecasts for early warning alerts that would be possible with future detectors, with alerts sent to astronomers 10 minutes, 5 minutes, and 2 minutes before the merger occurs and the corresponding 3D sky localization of the event. Sec.~\ref{sec:rare_events} is devoted to discussing the science enabled by rare events with SNRs larger than 300 or even 1000, e.g. observing higher multipole modes, testing the black hole no-hair theorem to exquisite precession, etc. Sec.~\ref{sec:conclusion} provides a brief summary, limitations of the study, and how they might be improved and other studies that would be necessary to firm up the science potential of future detectors. 

\begin{table}[ht]
    \caption{The networks of GW detectors whose performance is evaluated using quantitative metrics discussed in Sec.\,\ref{sec:rate}--\ref{sec:rare_events}. Note that the time-scale of operation of the various networks is our best guess estimate of when a given network is likely to operate; they do not correspond to any official projections.}
    \centering
    \begin{tabular}{l|c|l}
\hline
\hline
\bf Detectors & \bf Network Name & \bf Time-scale \\
\hline
LIGO (HLI+), Virgo+, KAGRA+ & \aplus\  & 2025-2030  \\
LIGO (HLI-Voy), Virgo+, KAGRA+ & \voy\ & 2030-2035  \\
ET, LIGO (HLI+), KAGRA+ &  \et\  & 2035-2040  \\
CE, Virgo+, KAGRA+ LIGO-I+ & \ce\   & 2040-2045  \\
ET, CE, KAGRA+, LIGO-I+ & \EC\   & 2040-2045  \\
ET, CE, CE-South & \ECS\     & 2040+      \\
\hline
    \end{tabular}
    \label{tab:networks}
\end{table}

The Appendix contains further information about this study and the science capabilities of additional networks not reported in the main body of the paper. Appendix \ref{app:settings} reports the settings used in the Fisher-information software package \textsc{gwbench} \citep{Borhanian:2020ypi}; this would help other interested users to more readily reproduce the results of this paper. Appendix \ref{app:data} details how to access the raw data used in this study. Appendix \ref{app:sigmoid} provides the best fit parameters of a (modified) sigmoid function that is used to fit the efficiency of detection of the various networks. In the main body of the paper we provide only the cumulative histograms of signal-to-noise ratios and parameter error estimates for the studied BNS and BBH populations in the different networks. Thus, in Appendix \ref{app:pdf} we also present the histograms of the respective quantities. Finally, Appendix \ref{app:ce_configs} compares the results for detector networks containing one or two CEs and with arm lengths of either $20\,{\rm km}$ or $40\,{\rm km}$. Comparison of the results in this Appendix highlights the importance of having at least one $40\,{\rm km}$ Cosmic Explorer detector for the completeness of the detected catalogs and the need for the Einstein Telescope to accurately localize sources.

Unless specified otherwise, we use the geometric system of units in which Newton's constant and the speed of light are both set to unity: $c=G=1.$

\section{Networks, populations, and methodology} \label{sec:methods}

\subsection{Detector networks}
\label{sec:detector networks}

In this paper we will consider six global networks composed of at least three detectors in some combination of the aforementioned detector generations as described below and summarized in Tab.~\ref{tab:networks}:
\begin{enumerate}
    \item An A+ network \cite{Barsotti:2018} consisting of the LIGO-Hanford (H), LIGO-Livingston (L), Virgo (V), KAGRA (K), and LIGO-India (I) detectors and denoted by \aplus\ that could be operational in the next {\bf 5 to 10 years},
    \item A Voyager network \cite{Adhikari:2020gft} consisting of three LIGO detectors (HLI) operating with Voyager technology in addition to A+ versions of Virgo and KAGRA and denoted by \voy\ that could be operating in the next {\bf 10 to 15 years}, and
    \item Four networks consisting of at least one NG detector (Einstein Telescope (E) in Italy \cite{Punturo:2010zza}, Cosmic Explorer (C) in the US \cite{Reitze:2019iox}, and Cosmic Explorer South (S) in Australia) together with a combination of A+ sensitivity detectors, denoted as \et, \ce, \EC, and \ECS. The fiducial locations of the ET and CE detectors can be found in Tab.~III of \citep{Borhanian:2020ypi}. Such networks are expected to be operational in roughly {\bf 15 to 20 years}.
\end{enumerate}

\noindent Although the choice of when a specific A+ configuration is included in a NG network seems arbitrary, this resulted from our judgement that when a NG detector comes online the corresponding region's 2G+ detector(s) may no longer be in operation. For example, Virgo is not likely to be operational when ET is built, LIGO-Hanford and LIGO-Livingston may not be operating at the same time as CE. Should it prove necessary to assess the performance of a network comprised of a different combination of detectors, e.g. \texttt{HLI+C}, it would be straightforward to do so with \textsc{gwbench}, see Sec.~\ref{sec:methodology} for details.

\subsection{Source populations}

Throughout this paper, we examine two source populations comprising either BNSs or BBHs. We assume the local (i.e. $z=0$) merger rate for BNS and BBH to be $R_{\rm BNS}=320^{+490}_{-240}\,{\rm Gpc^{-3}\,yr^{-1}}$ and $R_{\rm BBH}=24^{+14}_{-7}\,{\rm Gpc^{-3}\,yr^{-1}},$ respectively \cite{Abbott:2020khf}, as determined by the first and second gravitational-wave transient catalogs \cite{LIGOScientific:2018mvr, Abbott:2020niy}. Redshift evolution of the rates is discussed in Sec.~\ref{sec:rates}.

Each population was uniformly distributed in six redshift bins, $z\in[0.02,0.5],\,[0.5,1],\,[1,2],\,[2,4],\,[4,10],\,[10,50]$, with 250,000 injections per bin. The corresponding luminosity distances $D_L$ were calculated with the \textsc{Planck18} cosmology of \textsc{astropy}. Uniform sampling over several redshift bins provides us with good parameter sampling. However, the probability density functions of the populations are scaled appropriately to capture the redshift-dependent merger rates (see Sec.~\ref{sec:resampling} for details).

The injections were further uniformly sampled over sky positions ($\alpha$ and $\cos(\delta)$) and binary orientation angles ($\cos(\iota)$ and $\psi$), with right ascension $\alpha$, declination $\delta$, inclination $\iota$, and polarization angle $\psi$. The injected spins were chosen to be aligned with the orbital angular momentum ($\chi_{1x}=\chi_{1y}=\chi_{2x}=\chi_{2y}=0$) while the $z$-components were uniformly sampled, for BNS $\chi_{1z},\,\chi_{2z} \in [-0.05,0,05]$ and for BBH $\chi_{1z},\,\chi_{2z} \in [-0.75,0,75]$. The BNS masses were chosen to be normally distributed in $[1,2]\,M_\odot$ with mean $\mu=1.35\,M_\odot$ and standard deviation $\sigma=0.15\,M_\odot$. The BBH masses were chosen to follow the \texttt{POWER+PEAK} distribution described in the Second Gravitational-Wave Transient Catalog population paper \cite{Abbott:2020gyp}, but with the secondary mass sampled uniformly in $[5\,M_\odot,m_1]$. These parameters are summarized in Tab.~\ref{tab:populations}. Note that differences in the simulated populations and network configurations require cautious comparison with other literature.

\begin{table}[hb]
\caption{Sampling parameters for the BNS and BBH populations.}
\begin{tabular}{l|l|l}
\hline
\hline
Population               & \multicolumn{1}{c|}{BNS}              & \multicolumn{1}{c}{BBH}                                 \\
\hline
$m_1$                    & \begin{tabular}[c]{@{}l@{}}Gaussian \\ ($\mu=1.35\,M_\odot$, $\sigma=0.15\,M_\odot$)\end{tabular}            & \texttt{POWER+PEAK} \cite{Abbott:2020gyp}   \\
\hline
$m_2$                    & \begin{tabular}[c]{@{}l@{}}Gaussian \\ ($\mu=1.35\,M_\odot$, $\sigma=0.15\,M_\odot$)\end{tabular}            & uniform in $[5\,M_\odot,m_1]$          \\
\hline
$\chi_{1x}$, $\chi_{2x}$ & \multicolumn{2}{l}{$0$}     \\
\hline
$\chi_{1y}$, $\chi_{2y}$ & \multicolumn{2}{l}{$0$}    \\
\hline
$\chi_{1z}$, $\chi_{2z}$ & uniform in $[-0.05,0.05]$ & uniform in $[-0.75,0.75]$           \\
\hline
$z$                      & \multicolumn{2}{l}{\begin{tabular}[c]{@{}l@{}}uniform in six bins:\\ $[0.02,0.5]$, $[0.5,1],$ $[1,2]$, $[2,4]$, $[4,10]$, $[10,50]$\end{tabular}} \\
\hline
$D_L$                    & \multicolumn{2}{l}{convert $z$ via \textsc{astropy.Planck18}} \\
\hline
$\cos(\iota)$            & \multicolumn{2}{l}{uniform in $[-1,1]$} \\
\hline
$\alpha$                 & \multicolumn{2}{l}{uniform in $[0,2\pi]$} \\
\hline
$\cos(\delta)$           & \multicolumn{2}{l}{uniform in $[-1,1]$} \\
\hline
$\psi$                   & \multicolumn{2}{l}{uniform in $[0,2\pi]$} \\
\hline
$t_c$, $\varphi_c$       & \multicolumn{2}{l}{$0$} \\
\hline
\hline
\end{tabular}
\label{tab:populations}
\end{table}

\subsection{Analysis methodology}
\label{sec:methodology}

We assess the performance of the chosen GW networks using a fast, Fisher-information \cite{Cutler:1994ys,Poisson:1995ef,Balasubramanian:1995bm,Arun:2004hn} based Python package, \textsc{gwbench}, that we developed and made publicly available \cite{Borhanian:2020ypi}; Appendix~\ref{app:settings} reports the settings used in \textsc{gwbench} and Sec.~\ref{sec:limitations} summarizes the limitations of the Fisher-information formalism.
The Fisher information formalism approximates the parameter posteriors to be Gaussian (assuming Gaussian noise), provides an analytic recipe to calculate the covariance matrix of the posteriors, and thus allows the computation of measurement error estimates for the gravitational waveform parameters.
The implementation in \textsc{gwbench} incorporates the effect of Earth's rotation, which is important for signals lasting for longer than about an hour, and numerical differentiation schemes, enabling access to a host of gravitational waveform models in \textsc{LAL}, the LSC Algorithm Library \cite{lalsuite}---ultimately enabling the use of more sophisticated waveform models and, therefore, more accurate estimates.
\textsc{gwbench} packs this in an easy-to-use fashion and further gives access to signal-to-noise ratio (SNR) calculations, various detector locations and noise curves, and basic sampling methods. 
Alternatively, the gravitational-wave community has developed other Fisher-information based Python packages such as \textsc{gwfast} \cite{Iacovelli:2022mbg} and \textsc{gwfish} \cite{Dupletsa:2022scg} that can be used to perform similar analyses.

Thus, the package allows us to make a comprehensive evaluation of the six chosen detector networks using well-defined performance metrics: SNR $\rho$, 90\%-credible sky area $\Omega_{90}$, and measurement error estimates of gravitational waveform parameters. The latter depend on the waveform models employed: For the BNS population we use the \texttt{IMRPhenomD\_NRTidalv2} model \cite{Dietrich:2019kaq} to accurately capture tidal effects. For the BBH population we use the \texttt{IMRPhenomXHM} \cite{Garcia-Quiros:2020qpx, Garcia-Quiros:2020qlt} model to accurately capture the dynamics of sub-dominant modes beyond the quadrupole that are important for highly mass-asymmetric systems which our BBH population includes. We perform the Fisher analysis over 9 parameters for both populations: chirp mass $\mathcal{M}$, symmetric mass ratio $\eta$, luminosity distance $D_L$, time and phase of coalescence $t_c,\, \phi_c$, and the aforementioned binary orientation angles $\iota,\, \psi$ and sky positions $\alpha,\, \delta$. The Fisher analysis is performed in the frequency domain over the range $f_L=5\,{\rm Hz}$ to $f_U=N\,f_{\rm isco}$ in $\mathrm{d}f=1/16\,{\rm Hz}$ steps. The frequency at the innermost, stable, circular orbit for a binary of total mass $M$ is defined as $f_{\rm isco}=({6^{3/2}\,\pi M})^{-1}.$ The factor $N$ is chosen to be 4 for \texttt{IMRPhenomD\_NRTidalv2} and 8 for \texttt{IMRPhenomXHM} to include the full ringdown dynamics. In either case, we truncated $f_U$ above 1024 Hz to improve performance since the detector noise curves make waveform contributions above these frequencies negligible. Since the used Virgo+ sensitivity curve starts from $10\,{\rm Hz}$ we discarded injections with $f_U<11\,{\rm Hz}$ which only affects the most massive BBHs at very large redshifts.

The performance metrics are chosen in order to assess the scientific capabilities of the Cosmic Explorer design at accomplishing specific science goals outlined in \citep{Evans:2021gyd}. The raw data for a number of networks including the reference ones studied in this paper are available online (see Appendix \ref{app:data}).

The accuracy of the 90\%-credible sky area estimates from \textsc{gwbench} has been tested against the well-established \textsc{bayestar} code \cite{Singer:2015ema} for BNS signals in Ref.\,\cite{Magee:2022kkc}. Their findings suggest that the sky-localization estimates using Fisher analysis are trustworthy for a large population of events such as in this study. Indeed, they find that the agreement between Fisher and \textsc{bayestar} for the 90\%-credible sky area estimates improves strongly for larger SNR signals, see Fig.~2 in \cite{Magee:2022kkc}: The \textsc{gwbench} and \textsc{bayestar} estimates agree at 90th percentiles to within a factor $\sim6$ and $<6$ for signals with SNR of at least 15 and 25, respectively. Given that all well-localized events ($\Omega_{90}\leq1\,{\rm deg^2}$) in our population, see Fig.~\ref{fig:visibility_scatter}, have SNRs $\rho\geq30$ for BNSs ($\geq20$ for BBHs), we believe our conclusions for the sky localization capabilities of the investigated networks to be well-justified for such events. Furthermore, we note that their test was performed for a three-detector network of lower sensitivity than the weakest network in our study (\aplus), while all but one (\ce) of our networks contain five detectors.

The \textsc{gwbench} package employs numerical inversion to calculate covariance matrices from Fisher information matrices. In particular, the code requires the condition number of a Fisher matrix---the fraction of its largest and smallest eigenvalue---to be less than $10^{15}$, otherwise it discards the event from further analysis \cite{Rodriguez:2011aa}. Such large condition numbers can occur for loud signals when two waveform parameters have vastly different error scales, making such events likely to be classified as ill-conditioned and hence discarded in our study. Consequently, the expected event numbers for signals in the high-SNR/well-measured tail of the population quoted throughout this paper should be regarded as conservative lower bounds. The exceptions are SNR statements and network efficiencies since a signal's SNR is calculated independently of its Fisher matrix.

\section{Merger and detection rates: compact binaries throughout the cosmos}
\label{sec:rate}
This section summarizes the assumptions and the essential elements of the simulations carried out in this study. We begin with a discussion of the response of a detector to an incident signal and the need to include the motion of the detector when computing the response. We then examine the efficiency of a network of detectors, which determines a network's detection rate, followed by a definition of the average and maximum distance up to which a network can observe compact binary coalescences. We conclude with the computation of detection rate as a function of redshift for different detector networks.
\subsection{Detector response including its motion relative to a source}
\paragraph{Antenna pattern}
Gravitational-wave detectors are quadrupole antennas and their sensitivity to sources has the same anisotropic response as that of a quadrupole (see, e.g., Ref.\,\cite{Sathyaprakash:2009xs}) with the additional complication that GWs are metric perturbations with two independent polarizations $h_+$ and $h_\times.$ The response of a GW detector to signals coming from a direction $(\alpha,\delta)$ is \cite{Dhurandhar:1988abc, Tinto:1987abc, Finn:2008np}
\begin{eqnarray}
    h^{(A)}(t, \boldsymbol{\mu}) & = & F_+^{(A)}(\alpha,\delta, \psi; R_A, \alpha_A, \beta_A, \gamma_A)\, h_+(t, \boldsymbol{\lambda}) \nonumber \\
    & + & F_\times^{(A)}(\alpha,\delta,\psi; R_A, \alpha_A, \beta_A, \gamma_A)\,h_\times(t,\boldsymbol{\lambda}),
    \label{eq:response}
\end{eqnarray}
where $F_+^{(A)}$ and $F_\times^{(A)}$ are the quadrupole antenna pattern functions of a detector indexed by $A$ in the long-wavelength approximation (which is sufficiently accurate even in the case of NG detectors that will be tens of kilometers long, except for small corrections that might be needed at kilohertz frequencies), $(\alpha,\, \delta)$ are the source's right ascension and declination in the geocentric coordinate system, $\psi$ is the polarization angle, $(R_A, \alpha_A, \beta_A, \gamma_A)$  are the altitude, latitude, longitude, and the angle from local north to the $x$-arm of detector $A$, $\boldsymbol\lambda=\{\lambda_k\},\, k=1,\dots,n_\lambda,$ is the parameter vector describing the strain at the location of the detector and $\boldsymbol{\mu} = \{\mu_K\}=\{\alpha, \delta, \psi, \boldsymbol{\lambda}\},$ $K=1,\ldots,n_\mu,$ is the full parameter vector including the source's position in the sky and the polarization angle.

\paragraph{Strain parameters}
For a compact binary system on a quasi-circular inspiralling orbit, $\boldsymbol{\lambda}$ consists of the luminosity distance $D_L$ to the source, the orientation of the orbit relative to the line-of-sight $\iota,$ the redshifted masses $(m_1,\, m_2)$ and spins $(\boldsymbol{S}_1, \, \boldsymbol{S}_2)$ of the companions, the epoch of coalescence $t_c$ taken to be the time when the amplitude of the response, which grows until the compact binary coalesces and then decays down rapidly, reaches its maximum value, and $\varphi_c$ the phase of the signal at that epoch.

It is often convenient to use dimensionless spins $(\boldsymbol{\chi}_1,\, \boldsymbol{\chi}_2),$ which, in geometric units, are given by  $(\boldsymbol{\chi}_1,\, \boldsymbol{\chi}_2)=  (\boldsymbol{S}_1/m_1^2,\, \boldsymbol{S}_2/m_2^2).$ The magnitudes of the dimensionless spin vectors lies between $[0,\,1].$ Likewise, in the post-Newtonian expansion of the waveform phase the chirpmass ${\cal M}=(m_1\,m_2)^{3/5}/(m_1+m_2)^{1/5}$ and symmetric mass ratio $\eta=m_1m_2/(m_1+m_2)^2$ appear more naturally than the component masses \cite{Blanchet:2013haa}.

\paragraph{Time-dependent antenna pattern}
The position of the source in the sky is given in the geocentric coordinate system with respect to which the detector's position, described by angles $(\alpha_A(t),\,\beta_A(t))$, varies due to Earth's rotation. The relative motion of a detector with respect to a source induces amplitude modulation in the observed signal due to the changing antenna pattern functions in the direction of a source \cite{Chan:2018csa}. The value of the antenna pattern changes at most by a magnitude of order $\Delta h \sim |\sin(2\Delta\alpha)|,$ where $\Delta\alpha$ is the change in the right ascension of the source relative to the detector; $\Delta h \sim 1$ over a six hour period. Since we can determine the amplitude of a signal to an accuracy of order $1/\rho,$ where $\rho$ is the SNR, the amplitude modulation could be important for signals that last for more than $\sim 30$ minutes at an SNR of $\rho =10,$ and for even shorter periods for louder events. For signals that last for periods greater than 30 minutes, time-dependence of the response could further improve the localization of the source and is most relevant for BNS sources in ET.

\paragraph{Detector location-dependent phase factor}
In addition, a detector location-dependent phase factor should be included to account for the changing arrival times of the signal at the detectors relative to the geocentric coordinate system. This is accomplished by multiplying the Fourier transform of the response function $\tilde h(f)$ by the appropriate phase factor \cite{Chan:2018csa}:
$$\tilde h(f) \rightarrow \tilde h(f) \exp\left[ \mathrm{i}2\pi f \frac{\boldsymbol{r}(t) \cdot \boldsymbol{ n}(\alpha,\delta)}{c}\right ],$$
where $\boldsymbol{r}(t)$ is the position vector of the detector on Earth, $\boldsymbol{n}$ is a unit vector in the direction of the source, and $c$ is the speed of light. This phase factor is precisely what allows the triangulation of the source's sky position and is most effective when three or more non-collocated detectors are available.

\paragraph{Frequency modulation}
The frequency modulation due to Earth's rotation, however, is not important as the fractional change in frequency is expected to be negligible over the observation period of a signal: $\Delta f/f =v_{\rm rot}/c \ll 1/(f P_{\rm max}),$ where $v_{\rm rot}/c=1.5 \times 10^{-6}$ is Earth's rotational velocity at the equator, and $P_{\rm max}$ is the maximum duration for which a signal lasts in the detector's sensitivity bandwidth. $P_{\rm max}$ is the largest for BNS signals and varies from a few minutes to several hours depending on the lower-frequency cutoff $f\gtrsim1\,{\rm Hz}$ of a detector below which contribution to the SNR of a signal is negligible. Thus, the change in frequency is not discernible for compact binary coalescence sources and frequency modulation can be neglected.

\subsection{Network efficiency}
Having defined the response of a detector to a GW signal we next turn to defining the efficiency of a network of detectors, which would be required in computing the detection rate of a network. The \emph{efficiency} $\epsilon(z)$ of a detector, or a network of detectors, at a given redshift or luminosity distance is defined as the fraction of all sources within that redshift for which the matched filter SNR $\rho$ of the network is larger than a preset threshold $\rho_*.$ To this end, the matched filter SNR $\rho$ of a network of $n_D$ detectors to an incident signal is defined as:
\begin{equation}
    \rho^2 = \sum_{A=1}^{n_D} \rho_A^2,\quad \rho_A^2 = 4 \int_{f_L}^{f_U} \frac{\left |\tilde h^{(A)}(f)\right |^2}{S_h(f)}\,{\rm d}f, \label{eq:network_snr}
\end{equation}
where $h^{(A)}$ is the response function of detectors $A=1,\ldots, n_D,$ given in Eq.\,(\ref{eq:response}), $f_L$ and $f_U$ are the detector- and signal-dependent lower and upper frequency cutoffs chosen so that there is negligible SNR outside that interval. The SNR depends on both the intrinsic and extrinsic parameters of the source: the source's masses and spins, distance, sky position, and orientation.

\begin{figure*}
\includegraphics[width=0.49\textwidth]{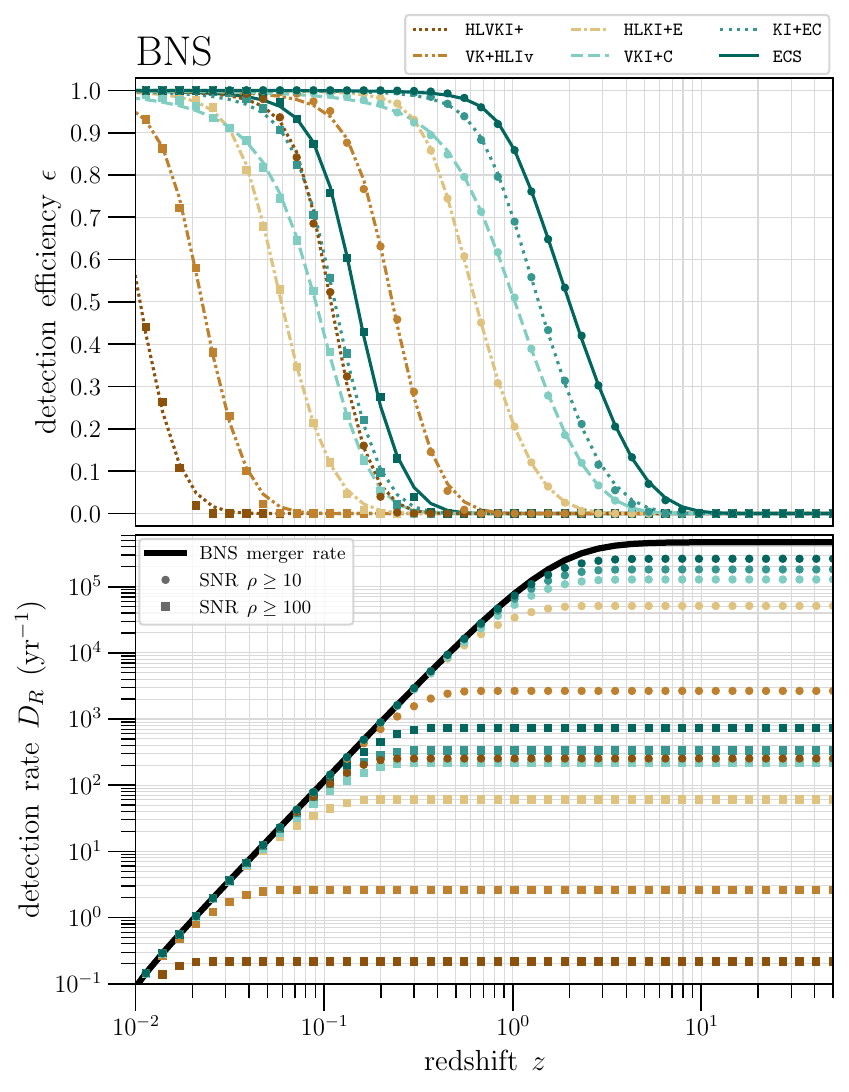}
\includegraphics[width=0.49\textwidth]{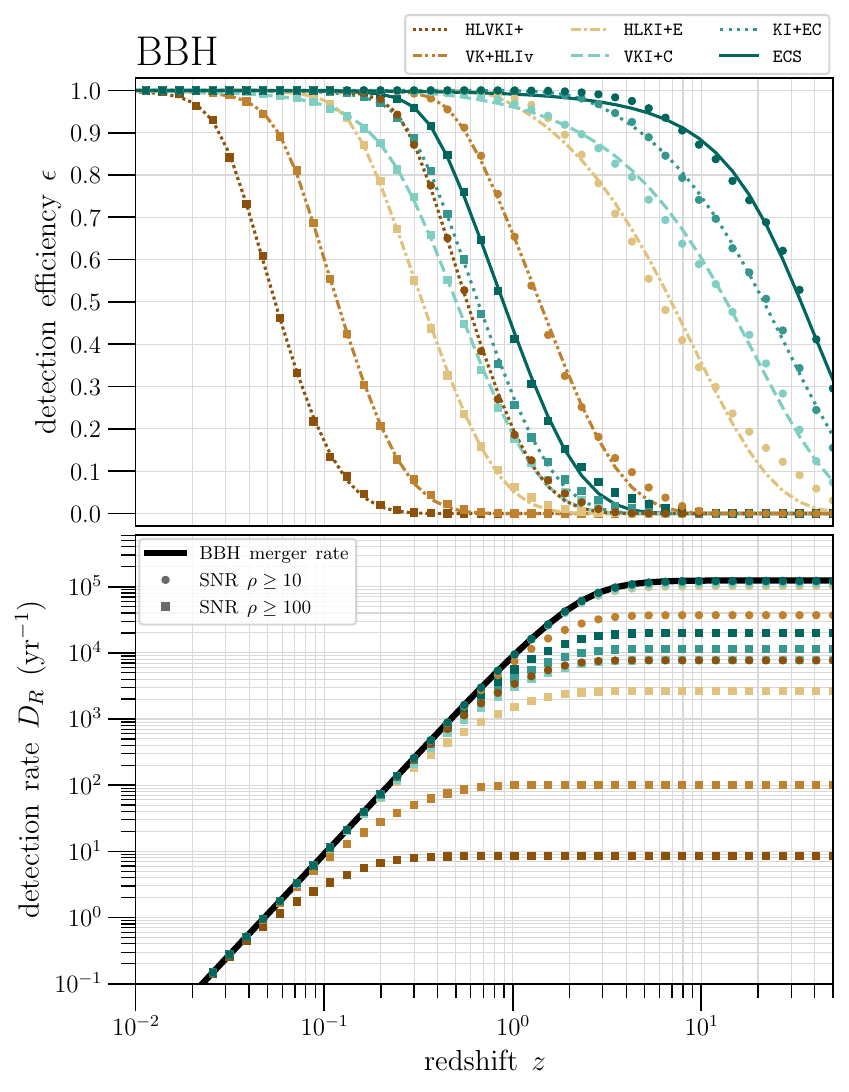}
\caption{Detection efficiencies $\epsilon$ and detection rates $D_R$, see Eqs.  \eqref{eq:efficiency} and \eqref{eq:det_rate}, respectively, of the six studied A+, Voyager, and NG networks are plotted as functions of redshift $z$. The efficiency specifies the value at the indicated redshift $z$ while the detection rate corresponds to the integrated rate up to that redshift. The circles (squares) denote the values for events with SNR $\rho\geq10$ ($\rho\geq100$). The thick, black lines in the rate panels are the cosmic BNS and BBH merger rates, see Sec.~\ref{sec:rate}. The fit lines in the efficiency panels are sigmoid fits with $f_{\rm sigmoid}=\Big(\frac{1+b}{1+b\,\mathrm{e}^{ax}}\Big)^c$. The fit parameters for the curves are given in Appendix \ref{app:sigmoid}.}
\label{fig:detection_efficiency_rate}
\end{figure*}

In order to compute the efficiency of a network at a given redshift $z$ we inject signals in the interval $z$ and $z+{\rm d}z$ with their intrinsic and extrinsic parameters distributed as in Tab.\,\ref{tab:populations} and count the number of sources for which the SNR was larger than $\rho_*$:
\begin{equation}
     \epsilon(z,\,\rho_*) = \frac{1}{N_z} \sum_{k=1}^{N_z} \Pi\left(\rho_k-\rho_*\right), \quad \Pi(x) =
     \left \{ \begin{array}{l}
          0,\, \mbox{ if } x\le 0\\
          1,\, \mbox{ if } x>0,
     \end{array} \right .
     \label{eq:efficiency}
\end{equation}
where $N_z$ is the number of sources in the simulation in the redshift interval $[z,\, z+dz],$ $\rho_k$ is the SNR of the $k^{\rm th}$ event in the injection list, and $\Pi(x)$ is the Heaviside step function. We repeat this process from $z=0.01$ to $z=50,$ thereby obtaining the efficiency curves  shown in Fig.\,\ref{fig:detection_efficiency_rate}, top two panels, for different detector networks considered in this study. The well-known \emph{sigmoid} functions with three parameters are a good fit to the efficiency curves
\begin{equation}
    f_{\rm sigmoid}=\left(\frac{1+b}{1+b\,\mathrm{e}^{ax}}\right)^c.
\end{equation}
Appendix \ref{app:sigmoid} lists the best-fit parameters $a,\,b,$ and $c$ for the various networks. Note that for a given source population the network efficiency depends only on the SNR threshold $\rho_*$ and not how sources are distributed as a function of redshift. The detection rate of networks for a different source distribution than the one considered in this study can be computed using the efficiency curves provided here.

The network efficiency is plotted for two choices of the SNR threshold, $\rho_*=10$ and $\rho_*=100.$ The lower value corresponds to the smallest SNR at which we assume a confident detection can be made by a detector network while the larger value is included to show redshift/distance from within which one can observe signals with high-fidelity. In Fig.\,\ref{fig:detection_efficiency_rate}, there are a pair of fitting lines with the same color for a given network but distinguished by circles for $\rho_*=10$ and squares for $\rho_*=100$, the one with greater efficiency corresponds to the lower $\rho_*.$

The network efficiency can be used to characterize the \emph{completeness} of the survey of a detector network. For example, while the \voy\ network is {\red 50\%} complete for BNS sources up to a redshift of $z\simeq 0.23$, the \ECS\ network achieves {\red 50\%} completeness up to a redshift of $z\simeq2$, as can be seen from the top panels of Fig.~\ref{fig:detection_efficiency_rate} and Tab.~\ref{tab:reach}. Most remarkably, the \ECS\ network will be {\red almost 90\%} complete to BNS mergers within $z=1$---the redshift limit up to which many current and future EM telescopes will have the ability to carry out follow-up observations of the mergers. Access to such a complete sample would help in understanding the physics and astrophysics of BNS mergers with little observational bias.

\subsection{The \emph{reach} and \emph{per-mille redshift} of a network}
In the following, we refer to the \emph{reach} $z_r$ of a network as the redshift at which the detection efficiency is $\epsilon=0.5$, which means the network is capable of detecting at least half of the source population at any redshift up to its reach at a threshold SNR $\rho_*$. Further, we present the \emph{per-mille redshift} $z_{pm}$ as the redshift for which the detection efficiency at the SNR threshold $\rho_*$ drops to $\epsilon=0.001$. This may be seen as a proxy for the networks' horizon redshift.

In Tab.~\ref{tab:reach} we provide the reach and per-mille redshift of all the networks for BNS and BBH mergers for two threshold SNR values: {\em detections} with $\rho\geq10$ and exceptionally loud signals with $\rho\geq100$. Note that the GW literature has many different definitions of the reach (see, e.g., Ref.\,\cite{Chen:2017wpg}) which may not all agree with the one defined here.

\subsection{Merger and detection rates}
\label{sec:rates}
The merger rate of compact binaries in the local Universe comes from the  first and second Gravitational-Wave Transient Catalogs GWTC-1 and GWTC-2 \cite{LIGOScientific:2018mvr, Abbott:2020niy}. They contain 2 BNS and 44 stellar-mass BBH mergers as enumerated in Tab.~1 of Ref.\,\cite{Abbott:2020gyp}. The inferred local merger rates (i.e. at redshift $z=0$) were found to be $R_{\rm BNS}=320^{+490}_{-240}\,{\rm Gpc^{-3}\,yr^{-1}}$ for BNS mergers and  $R_{\rm BBH}=24^{+14}_{-7}\,{\rm Gpc^{-3}\,yr^{-1}}$ for BBH mergers \cite{Abbott:2020khf}.

LIGO and Virgo rarely observe coalescences at cosmological distances (i.e. $z\gtrsim 0.5$) but future detectors will. Several factors affect the rate as a function of redshift. Following Ref.\,\cite{Belczynski:2016obo} we take into account three principal ones: (1) The rate at which stars form, (2) the delay between the formation of a compact binary and its merger, and, in the case of BBHs, (3) the variation of metallicity with redshift.

The star formation rate (SFR) $\psi(z)$ as a function of redshift is not known precisely but several phenomenological models are available capturing the observational uncertainty \cite{Madau:1996yh, Hopkins:2006bw, Fardal:2006sd, Wilkins:2008be} (for a recent review see, e.g., Ref.\,\cite{Madau:2014bja}). The SFR initially increases as a function of redshift, peaking around $z\sim 2,$ after which it falls off. We will take the compact binary formation rate, and hence the merger rate, to essentially follow the SFR except that there will be a delay $t_d$ between the epoch $t_f$ when a compact binary forms and the epoch when it merges. Furthermore, the merger rate is calibrated by demanding its local value at $z=0$ to be consistent with the merger rate determined by LIGO and Virgo observations given above. Finally, metallicity plays a crucial role in the formation of BHs due to its effect on stellar winds, which must be folded into the calculation of rates.  In addition, we must be mindful of the fact that the cosmological volume element $({\rm d}V/{\rm d}z)$ corresponding to a redshift interval ${\rm d}z$ is itself a function of redshift due to the Hubble expansion and the Universe was smaller at earlier times.

\begin{table}[hb]
\begin{tabular}{l|c|c|c|c|c|c|c|c}
\hline
\hline
           & \multicolumn{4}{c|}{BNS}                                 & \multicolumn{4}{c}{BBH}                                 \\
\hline
SNR $\rho\quad \;$ & \multicolumn{2}{c|}{$\geq10$} & \multicolumn{2}{c|}{$\geq100$} & \multicolumn{2}{c|}{$\geq10$} & \multicolumn{2}{c}{$\geq100$} \\
\hline
\hline
   & $\;\;\: z_r\;\;\:$        & $\;\; z_{pm}\;\; $       & $\;\; z_r\;\; $         & $\;\; z_{pm}\;\; $       & $\;\;\: z_r\;\;\: $        & $\;\; z_{pm}\;\; $       & $\;\;\: z_r\;\;\: $        & $\;\; z_{pm}\;\; $        \\
\hline
\hline
\aplus       &  0.11  & 0.39  & 0.01  & 0.04      &  0.57  & 3.5   & 0.05  & 0.31      \\ \hline
\voy         &  0.23  & 0.92  & 0.02  & 0.09      &  1.4   & 10    & 0.12  & 0.7       \\ \hline
\et          &  0.63  & 3.3   & 0.06  & 0.27      &  7     & $>50$ & 0.33  & 2.2       \\ \hline
\ce          &  1     & 6.6   & 0.09  & 0.39      &  14    & $>50$ & 0.5   & 3.6       \\ \hline
\EC          &  1.4   & 8     & 0.11  & 0.44      &  21    & $>50$ & 0.65  & 4.3       \\ \hline
\ECS         &  2     & 13    & 0.15  & 0.59      &  34    & $>50$ & 0.89  & 6.1       \\ \hline
\hline
\end{tabular}
\caption{The reach $z_r$ and per-mille redshift $z_{pm}$ of the considered networks for BNS and BBH signals with SNRs $\rho\geq10$ or $\rho\geq100$. Here we define the reach (per-mille redshift) as the redshift at which a given network detects 50\% {\red (0.1\%)} of the injections with the specified SNR or louder. Please refer to the detection efficiency panels of Fig.~\ref{fig:detection_efficiency_rate} for a visual representation.}
\label{tab:reach}
\end{table}

Compact binaries that form at redshift $z_f$ merge at redshift $z$ after a delay time $t_d$ (see, e.g., Ref.\,\cite{deFreitasPacheco:2005ub, Regimbau:2012ir}). For a given redshift $z$ and delay $t_d$ the redshift when the binary forms can be found by solving
\begin{equation}
    t_d - \frac{1}{H_0} \int_z^{z_f} \frac{{\rm d}z'}{(1+z')E(z')}=0,\quad E(z') = \left [ \Omega_\Lambda + \Omega_M(1+z')^3 \right ],
\end{equation}
where $\Omega_M$ and $\Omega_\Lambda$ are the dark matter and dark energy densities and we have assumed a flat Universe in which dark energy is interpreted as a cosmological constant \cite{Sahni:1999gb}. The delay time is not the same for all binaries and depends on the configuration of the compact binary when it forms and is modelled by a probability density $P(t_d).$  The merger rate density $\dot n(z)$ (normally computed in units of Gpc$^{-3}$ yr$^{-1}$) in the source's frame is obtained by integrating the SFR over all possible time delays,
\begin{equation}
    \dot n(z) = A\,\int_{t_d^{\rm min}}^{\,t_d^{\rm max}} \psi(z_f(z,t_d))\, P(t_d)\, {\rm d}t_d,
\end{equation}
where $t_d^{\rm min}$ and $t_d^{\rm max}$ denote the smallest and largest possible time delays and the normalization constant $A$ is chosen so that local rate $\dot n(0)$ is consistent with the rate inferred using the GWTC-2 catalog \cite{Abbott:2020gyp}: $\dot n_{\rm BNS}(0) = 320 \,\rm Gpc^{-3}yr^{-1}$ and $\dot n_{\rm BBH}(0) = 23\,\rm Gpc^{-3}yr^{-1},$ where we have taken the central value of the rate posteriors ignoring the measurement uncertainties.

The delay-time distribution is not very well known but could be inferred accurately from future observations \cite{Safarzadeh:2019znp, Safarzadeh:2019pis, Safarzadeh:2019zif}. As is customary, we take $P(t_d)$ to be Jeffry's prior $P(t_d)\propto 1/t_d$ \cite{deFreitasPacheco:2005ub}. With normalization it becomes
\begin{equation}
P(t_d)=\frac{1}{\ln\left ({t_d^{\rm max}}/{t_d^{\rm min}} \right )t_d}.
\end{equation}
For this prior, most of the contribution to the density comes from near $t_d^{\rm min}$ which we set to be $t_d^{\rm min}=20\,\rm Myr$ for BNS and $t_d^{\rm min}=10\,\rm Myr$ for BBH mergers. We take $t_d^{\rm max}=10\,\rm Gyr$ in both cases.

From the merger rate density $\dot n(z)$ one computes the merger rate per redshift bin using
\begin{equation}
R(z) = \dot n(z) \,\frac{{\rm d}V}{{\rm d} z}
\end{equation}
where ${\rm d}V(z)/{\rm d}z$ is the comoving volume element. An observer in the local Universe would measure the rate of mergers to be a factor $(1+z)$ smaller than in the source's frame, i.e. $R_{\rm obs}(z)=R(z)/(1+z).$ 

A network of detectors, however, would not observe all the mergers that would occur at a given redshift but only a fraction $\epsilon(z,\,\rho_*)$ determined by the network's efficiency given in Eq.\,(\ref{eq:efficiency}), which in turn depends on the SNR threshold $\rho_*.$ Thus, the detection rate, i.e. the number of detections per year, $D_R(z,\,\rho_*)$ observed up to redshift $z$ is given by:
\begin{equation}
    D_R(z,\,\rho_*)
    = \int_0^z\epsilon(z',\,\rho_*)\,\frac{\dot n(z')}{(1+z')} \,\frac{{\rm d}V(z')}{{\rm d}z'}\,dz'.
    \label{eq:det_rate}
\end{equation}

The BBH merger rate was chosen to follow the `Madau-Dickinson-Belczynski-Ng' rate for field BHs described in \cite{Ng:2020qpk}. For the BNS population we take the SFR to be that of Madau-Dickinson but neglect the effect of metallicity evolution as this is not as important for the formation of NSs.

The bottom panels of Fig.~\ref{fig:detection_efficiency_rate} plot the detection rate $D_R$ as a function of redshift for two choices of the threshold: $\rho_*=10$ and $\rho_*=100.$ We will discuss the distributions of SNR and measurement error estimates in the next section.

\begin{figure*}
\includegraphics[width=\textwidth]{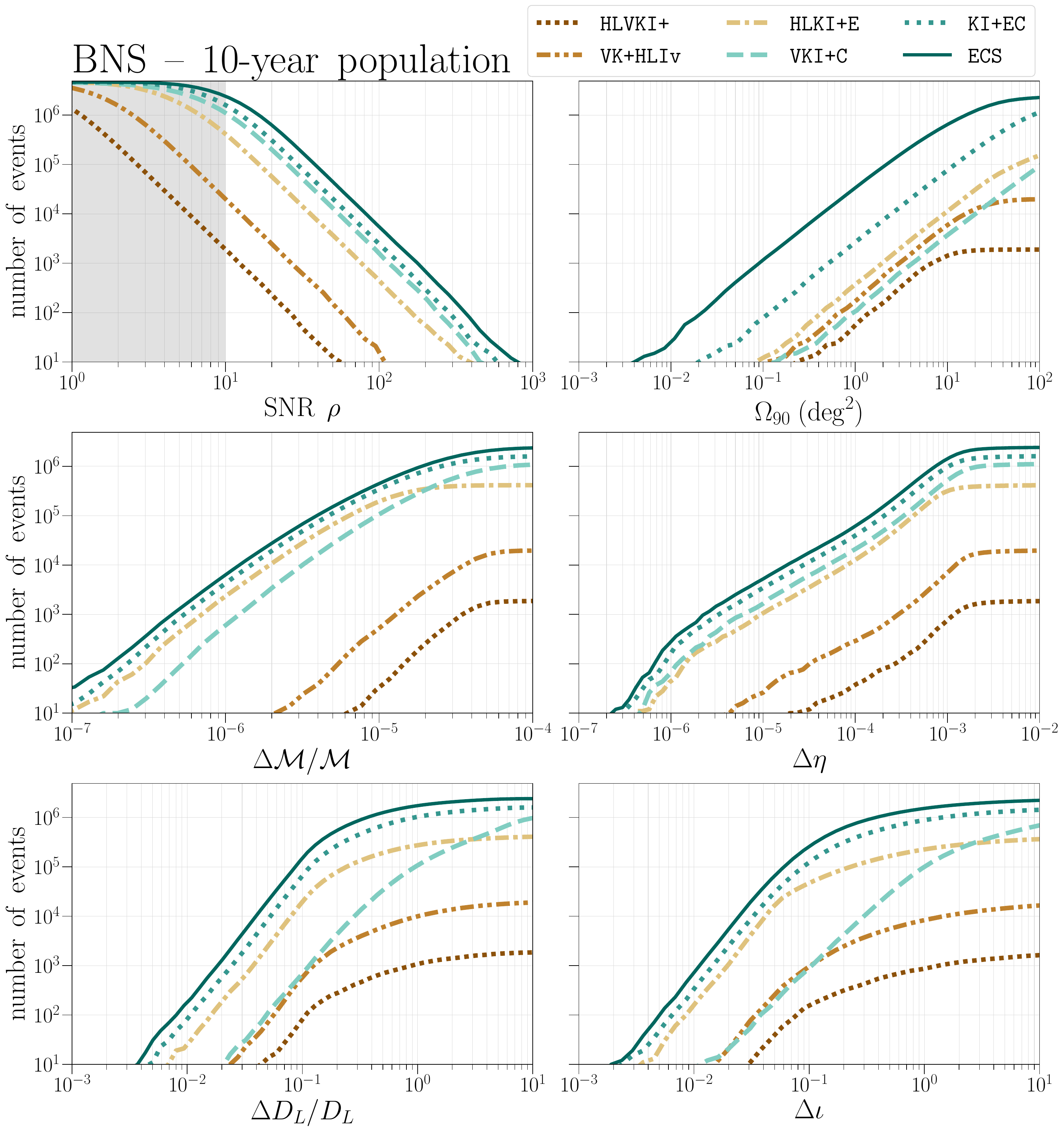}
\caption{Cumulative histograms for the SNR $\rho$, 90\%-credible sky area $\Omega_{90}$, fractional errors on chripmass $\Delta \mathcal{M}/\mathcal{M}$ and luminosity distance $\Delta D_L/D_L$, and absolute errors on symmetric mass ratio $\Delta \eta$ and inclination angle $\Delta\iota$ for BNS mergers observed in the six studied A+, Voyager, and NG networks. The histograms were generated from $\sim 4.7\times10^{6}$ injections sampled according to Sec. \ref{sec:resampling}. The non-SNR panels are obtained for events with SNR $\rho\geq10$, indicated by the non-shaded region in the SNR panel. The SNR panel is flipped to highlight the behavior for large values.}
\label{fig:BNS_visibility}
\end{figure*}

\begin{figure*}
\includegraphics[width=\textwidth]{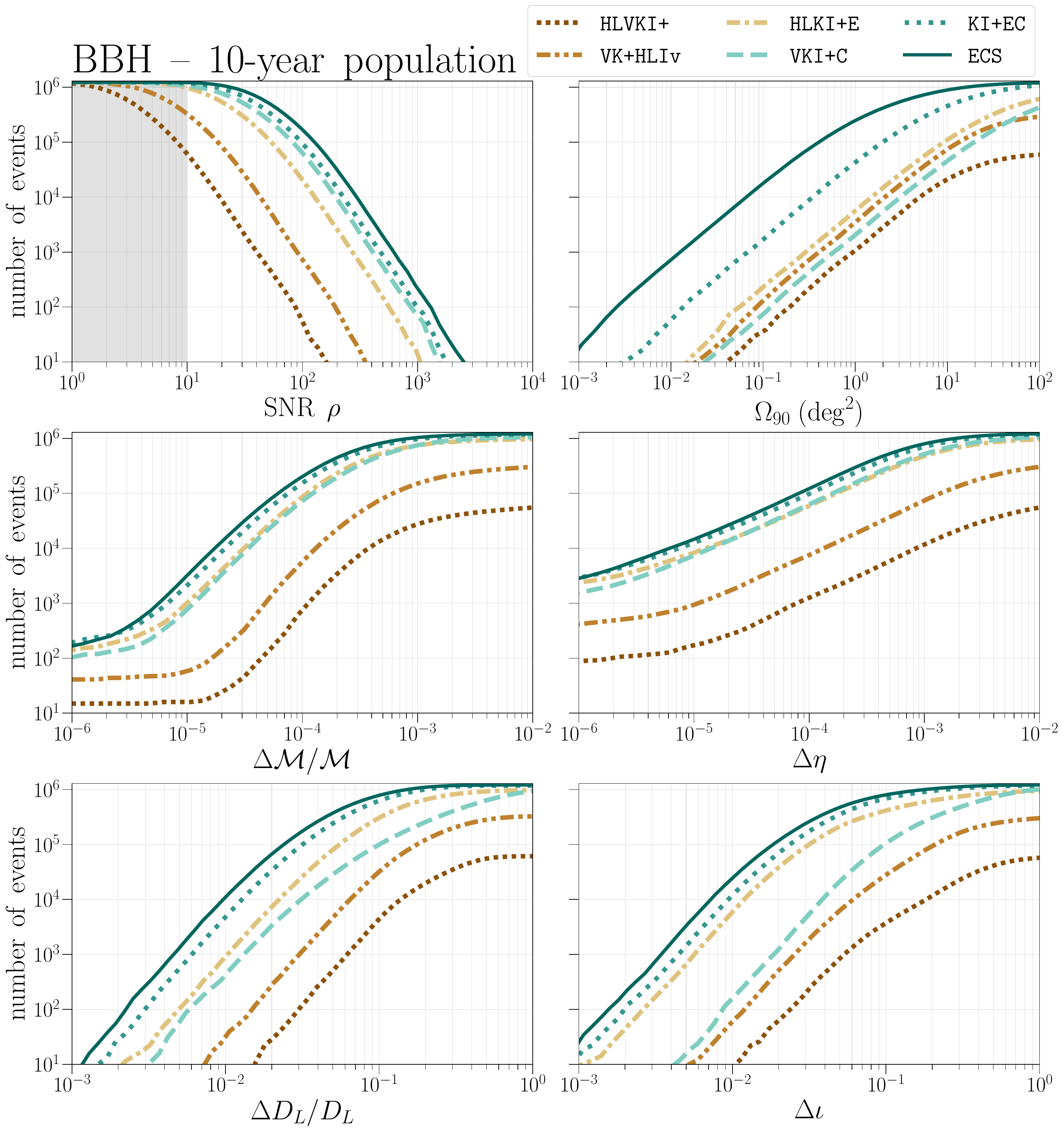}
\caption{Cumulative histograms for the SNR $\rho$, 90\%-credible sky area $\Omega_{90}$, fractional errors on chripmass $\Delta \mathcal{M}/\mathcal{M}$ and luminosity distance $\Delta D_L/D_L$, and absolute errors on symmetric mass ratio $\Delta \eta$ and inclination angle $\Delta\iota$ for BBH mergers observed in the six studied A+, Voyager, and NG networks. The histograms were generated from $\sim 1.2\times10^{6}$ injections sampled according to Sec. \ref{sec:resampling}. The non-SNR panels are obtained for events with SNR $\rho\geq10$, indicated by the non-shaded region in the SNR panel. The SNR panel is flipped to highlight the behavior for large values.}
\label{fig:BBH_visibility}
\end{figure*}

\section{Visibility and measurement quality of compact binary mergers in future detectors}
\label{sec:visibility}

In the following we summarize the overall capabilities of the six detector networks for the BNS and BBH populations throughout the chosen redshift range $z\in[0.02,50]$.
As such we will make `full-population' statements for the expected detection rates that each network should achieve given a set of targets for the performance metrics specified below.

Figs.~\ref{fig:BNS_visibility} and \ref{fig:BBH_visibility} present cumulative histograms for six performance metrics: SNR $\rho$, 90\%-credible sky area $\Omega_{90}$, fractional errors on chirpmass $\Delta\mathcal{M}/\mathcal{M}$ and luminosity distance $\Delta D_L/D_L$, and absolute errors on symmetric mass ratio $\Delta \eta$ and inclination angle $\Delta \iota$.
Fig.~\ref{fig:BNS_visibility} summarizes the BNS injections, while Fig.~\ref{fig:BBH_visibility} shows the results for the BBHs.
In both cases, the cumulative histograms present the \emph{relevant} data from the 1.5 million injections that were simulated in the redshift range and are shown to the 0.01\%-level.
Here, \emph{relevant} data refers to (i) the entire injection set in the case of the SNR sub panel, with ($\rho<1$)-events truncated, and (ii) the detectable events with $\rho\geq10$ for the remaining metrics.
We show the respective histograms in Appendix \ref{app:pdf}.

\subsection{Incorporation of redshift-dependent merger rates}
\label{sec:resampling}

While the uniform sampling per redshift bin provides us with good parameter sampling, it also means that we need to scale the probability density functions of the populations to capture the redshift-dependent merger rates as discussed in the previous section.
For this purpose we divide the redshift range in {\red $N=150$} sub-bins, containing each {\red $n_i\approx10,000$} injections.
Each sub-bin has an associated merger rate $R_i,\,i=1,\dots,N$ computed following Sec.~\ref{sec:rate}, allowing us to define sub-bin probabilities as $p_i=R_i/R$ with $R=\sum_{i=1}^N R_i$. We then sample the $N$ sub-bin indices $i$ with probabilities $p_i$ up to the desired number of BNS and BBH mergers{\red, e.g. corresponding to 10 years of mergers}, thus providing us with a number of samples $n_i$ for each index $i$.
Finally, we uniformly sample $n_i$ injections from the simulated BNS and BBH mergers in the $i$-th sub-bin, thus resulting in a random sample of injections in each sub-bin with the desired total number of mergers in the total redshift range.

In order to mitigate sampling effects in the distribution tails we perform all analyses using a \textbf{10-year random sample} of BNS and BBH mergers drawn from the simulated populations at the corresponding cosmic merger rates for the redshift range of interest. We show these 10-year samples in all histograms and scatter plots throughout this paper to better represent what each network is statistically capable to achieve and we mention this again in the captions of the respective figures. Similarly, all presented cumulative histograms are calculated from these larger samples to improve the statistics. In contrast, \textbf{we cite yearly detections rates} for the studied networks both in the text and tables, in order to represent rates for a time-scale more akin to the observing runs of the LIGO and Virgo detectors.

\subsection{Signal visibility}

The cosmic merger rate of BNS is of the order of $4.7\times10^5$ per year, but depending on the network we can only expect completeness as low as {\red $\sim 0.04\%$ in \aplus\ and $\sim 0.4\%$ in \voy, but up to $\sim50\%$} with three NG detectors (\ECS), translating to {\red $\mathcal{O}(10^2)$, $\mathcal{O}(10^3)$, and up to $\mathcal{O}(10^5)$} BNS detections per year, respectively, summarized in Tab.~\ref{tab:rates_snr}. These raw numbers illustrate the stark difference in volume and corresponding merger rates that each network's reach encompasses: {\red $z_r\sim 0.11$ for \aplus, $z_r\sim 0.23$ for \voy, and $z_r\sim 2$} for \ECS\ (see Fig.~\ref{fig:detection_efficiency_rate} and Tab.~\ref{tab:reach}). Further, Tab.~\ref{tab:rates_snr} indicates that we can expect a host of BNS events in the `GW170817-class' with SNRs $\rho\geq30$, while SNR values above 100 are only regularly observable in NG networks. In fact, the NG networks containing a CE detector will observe such loud events at rates of the order 1/10 to 1/4 of {\em all} yearly BNS detections in the Voyager network.

\begin{table}[ht]
    \centering
        \caption{Cosmic merger rates (per year) of BNS and BBH mergers in the Universe and the number that would be observed by different detector networks each year with $\rho \geq 10, 30, 100,$ where $\rho$ is the SNR ratio. Due to uncertainty in the various quantities that go into the calculation these numbers are no more accurate than one or two significant figures.}
    \begin{tabular*}{\columnwidth}{@{\extracolsep{\fill}}l?r|r|r?r|r|r}
\hline
\hline
                    & \multicolumn{3}{c?}{BNS} & \multicolumn{3}{c}{BBH} \\
\hline
Cosmic rate         & \multicolumn{3}{c?}{$4.7 \times 10^5$} & \multicolumn{3}{c}{ $1.2 \times 10^5$} \\
\hline
\hline
SNR $\rho$          & \multicolumn{1}{c|}{$ \geq10$} &  \multicolumn{1}{c|}{$ \geq30$} & \multicolumn{1}{c?}{$\geq100$} &  \multicolumn{1}{c|}{$\geq10$} &  \multicolumn{1}{c|}{$ \geq30$} & \multicolumn{1}{c}{$\geq100$}\\
\hline
\hline
\aplus    & 190      & 6        & 1        & 6,100    & 240      & 6       \\
\voy      & 2,000    & 71       & 2        & 33,000   & 2,900    & 74      \\
\et       & 41,000   & 1,700    & 45       & 97,000   & 31,000   & 2,100   \\
\ce       & 110,000  & 6,000    & 160      & 110,000  & 53,000   & 6,400   \\
\EC       & 160,000  & 9,400    & 250      & 120,000  & 69,000   & 9,500   \\
\ECS      & 240,000  & 20,000   & 550      & 120,000  & 87,000   & 17,000  \\
\hline
\hline
    \end{tabular*}
    \label{tab:rates_snr}
\end{table}

The picture for BBH mergers, see the SNR panel of Fig.~\ref{fig:BBH_visibility} and Tab.~\ref{tab:rates_snr}, is less skewed in favor of the NG networks in comparison to the Voyager network since its redshift reach {\red $z_r\sim1.4$} extends into the peak of the Madau-Dickinson-Belczynski-Ng merger rate density; see \cite{Ng:2020qpk}: Both \aplus\ and \voy\ will detect {\red $\mathcal{O}(10^3)$ and $\mathcal{O}(10^4)$} BBH mergers per year with completeness of {\red 5\% and 28\%}, respectively, while the NG networks will observe {\red $\mathcal{O}(10^5)$} BBH mergers with completeness between {\red 81\% (\et) and 99\%} (\ECS). In fact, all considered networks will detect BBH coalescences with SNRs above 100 albeit at per-mille redshifts of {\red $z_{pm}\sim 0.31, 0.7$, and up to $6.1$} for \aplus, \voy, and \ECS, respectively, and corresponding reaches of {\red $z_r\sim 0.05, 0.12$, and $0.89$}. Hence, the difference between these networks comes down to the NG networks' reach for BBHs extending beyond redshift {\red 7 (or even 14} with at least one CE detector) and the resulting completeness of the observable population at high redshifts: \EC\ and \ECS\ will detect essentially all BBH signals at the cosmic merger rate of {\red $\sim1.2\times10^5\,{\rm yr^{-1}}$}.

\begin{figure*}
\includegraphics[width=0.9\textwidth]{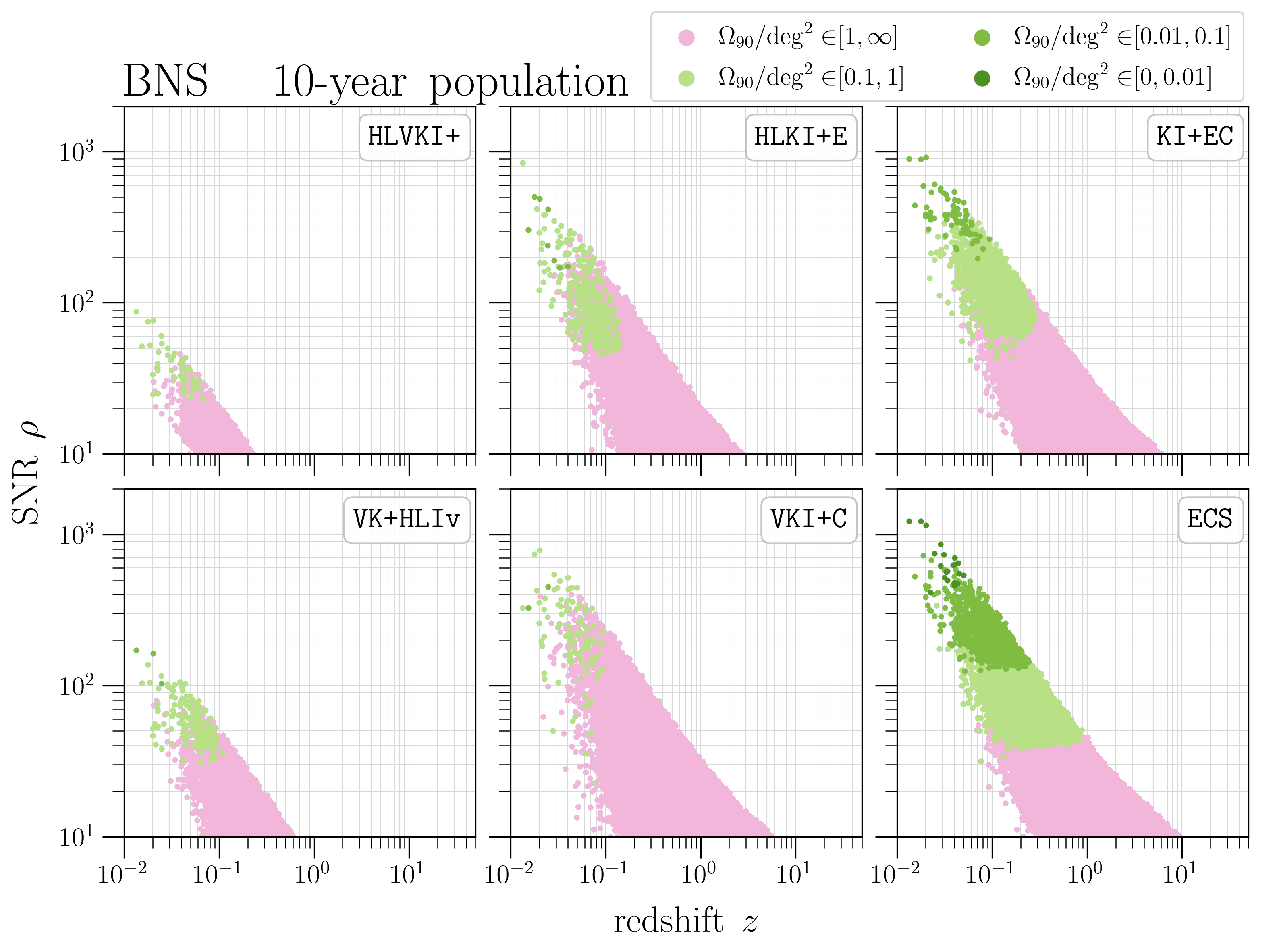}
\includegraphics[width=0.9\textwidth]{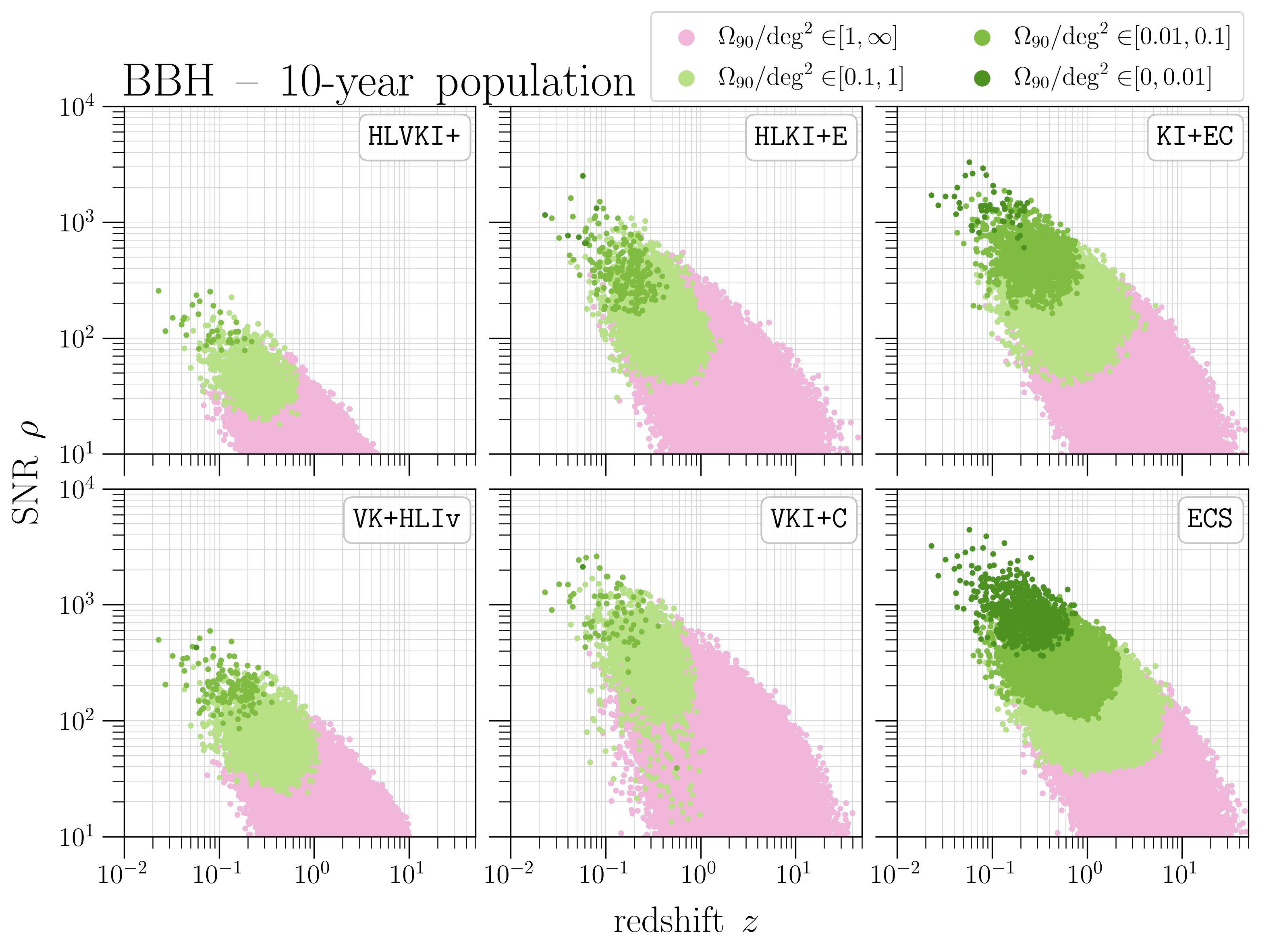}
\caption{The scatter plots illustrate the correlations between redshift $z$, SNR $\rho$, and 90\%-credible sky area $\Omega_{90}$ for BNS (\emph{top}) and BBH (\emph{bottom}) mergers in the six studied A+, Voyager, and NG networks.}
\label{fig:visibility_scatter}
\end{figure*}

While the yearly rates for all six networks far exceed the number of events observed with the current generation of detectors \cite{LIGOScientific:2018mvr, Abbott:2020niy}, the differences between these future networks are still crucial in their impact for astrophysics, cosmology, tests of GR, and dense matter physics. Fig. \ref{fig:visibility_scatter} captures these differences and presents what signal loudness and sky localization distributions to expect for GW observations from the studied networks throughout the redshift range over 10 years of observations.

Only NG networks will deliver an abundance of exceptionally loud BNS signals with SNRs above 100; even to cosmological distances of $z\lesssim0.4$. Such events will allow us to probe the nuclear physics with high fidelity, constrain the dense-matter equation of state, and explore the BNS post-inspiral signal.

For both the BNS and BBH coalescences the differences in reach and per-mille redshifts, see Tab.~\ref{tab:reach}, imply that not only a larger, but also older part of the BNS and BBH populations can be studied with NG networks. Especially, \EC\ and \ECS\ will observe almost all BNS and BBH mergers up to luminosity distances of {\red $\lesssim2 \,{\rm Gpc}$ and $\lesssim25 \,{\rm Gpc}$}, respectively. Their per-mille redshifts lie at {\red $z_{pm}\gtrsim 8$ for BNSs and beyond 50} for BBHs, in contrast to the Voyager network with {\red $z_{pm}\sim 0.9$ and $z_{pm}\sim 9.7$}, respectively. This means that a NG network could observe BNS coalescences from roughly 500 Myr and BBH mergers from less than 50 Myr after the Big Bang, thus expanding their observational potential deep into the realm of the dark ages! While the expectations for mergers in this regime are very low, population III stars and primordial BHs could pose potential progenitor and source systems, to which all networks other than NG would be blind.

\begin{table}[ht]
    \centering
      \caption{Detection rates of BNS and BBH mergers from the full redshift range $z\in[0.02,50]$ to be observed by different detector networks each year with $\Omega_{90}/{\rm deg^2}\leq 1,\, 0.1,\, 0.01$ as well as $\Delta D_L/D_L \leq 0.1,\, 0.1$, where $\Omega_{90}$ is the 90\%-credible sky area and $D_L$ the luminosity distance. These detection rates are calculated for events with SNR $\rho\geq10$. Due to uncertainty in the various quantities that go into the calculation these numbers are no more accurate than one or two significant figures.}
    \begin{tabular*}{\columnwidth}{@{\extracolsep{\fill}}l?r|r|r?r|r}
\hline
\hline
Metric \phantom{AAA} & \multicolumn{3}{c?}{$\Omega_{90}$  $({\rm deg^2})$ \phantom{A}} & \multicolumn{2}{c}{$\Delta D_L/D_L$}\phantom{A} \\
\hline
Quality &  $\leq1$\phantom{A} &  $\leq 0.1$\phantom{A} &  $\leq 0.01$\phantom{A} &  $\leq 0.1$\phantom{A} &  $\leq0.01$\phantom{A}\\
\hline
\hline
\multicolumn{6}{c}{\emph{BNS}} \\
\hline
\hline
\aplus    & 5        & 1        & 0        & 8        & 0       \\
\voy      & 17       & 1        & 0        & 58       & 0       \\
\et       & 37       & 1        & 0        & 1,900    & 3       \\
\ce       & 11       & 1        & 0        & 67       & 0       \\
\EC       & 270      & 8        & 0        & 6,600    & 8       \\
\ECS      & 3,400    & 120      & 3        & 15,000   & 18      \\
\hline
\hline
\multicolumn{6}{c}{\emph{BBH}} \\
\hline
\hline
\aplus    & 110      & 4        & 0        & 440      & 0       \\
\voy      & 350      & 13       & 0        & 3,200    & 3       \\
\et       & 560      & 23       & 1        & 31,000   & 92      \\
\ce       & 210      & 8        & 0        & 10,000   & 45      \\
\EC       & 4,400    & 170      & 6        & 64,000   & 490     \\
\ECS      & 24,000   & 1,800    & 72       & 78,000   & 1,100   \\
\hline
\hline
    \end{tabular*}
    \label{tab:rates_sky_area_DL}
\end{table}

Lastly, the shear abundance of loud events up to far redshifts will further enable astronomers and cosmologists to better understand source population demographics as well as trace and correlate the large-scale structure of the Universe with these mergers. Louder and more abundant signals will be a treasure trove for tests of GR which benefit from the outright signal strengths but also the potential of signal binning.
We examine rare, extremely loud signals with SNRs {\red $\rho\gtrsim300$} in Sec.~\ref{sec:rare_events}.

\subsection{Measurement quality}

The three-dimensional localization of a source, on the sky and in luminosity distance, is crucial in enabling a multitude of science, especially for BNS mergers which can exhibit observable electromagnetic (EM) counterparts. Hence, we summarize the expected detection rates of events with 90\%-credible sky area $\Omega_{90}/{\rm deg^2}\leq 0.01,\,0.1,\,1$ and fractional luminosity distance errors $\Delta D_L/D_L \leq 0.01,\,0.1$ in Tab.~\ref{tab:rates_sky_area_DL}. The visual representations in form of cumulative histograms can be found in the $\Omega_{90}$- and $\Delta D_L/D_L$-panels of Fig.~\ref{fig:BNS_visibility}. We will further expand our examination of the potential to enable multi-messenger astronomy at redshifts {\red $z\leq0.5$} in Sec.~\ref{sec:mma}.

GW190814 \cite{Abbott:2020khf} is the best-localized, observed GW event so far, with a 90\%-credible sky area of {\red $19\,{\rm deg^2}$} and a luminosity distance error of about 19\%.
In comparison, both \aplus\ and \voy\ will detect {\red $\mathcal{O}(1)$ to $\mathcal{O}(10)$} BNS sources per year with sky areas below $1\,{\rm deg^2}$ and 10\% luminosity distance errors, with BBH-numbers being $\sim20$ (sky area)  or $\sim50$ (distance errors) times larger.
\aplus\ and \voy\ will further be capable to observe annually {\red $\mathcal{O}(1)$ BNS and $\mathcal{O}(1)$ to $\mathcal{O}(10)$ BBH} mergers, respectively, with sky areas below $0.1\,{\rm deg^2}$.

The picture for the NG networks is distinctly different than it was for the signal visibility. Since these networks differ in the number of NG detectors per network with one in \et\ and \ce, two in \EC, and three in \ECS\ and since sky localization and luminosity distance measurements improve dramatically with more detectors in a network, the number of more sensitive NG detectors has a strong effect on the measurement quality. This is illustrated by the increase of green-colored points in the \EC\ and \ECS\ panels of Fig.~\ref{fig:visibility_scatter}, indicating their improved sky localization capabilities.

Consequently, these networks will detect BNS coalescences localized to within a 90\%-credible sky area smaller than $(1,\,0.1)\,{\rm deg^2}$ on the order of {\red $(\mathcal{O}(10),\,\mathcal{O}(1))$} per year in \et\ and \ce, {\red $(\mathcal{O}(10^2),\,\mathcal{O}(1))$} per year in \EC, and {\red $(\mathcal{O}(10^3),\,\mathcal{O}(10^2))$} per year in \ECS, with the latter being the only network to observe a handful of BNS per year with $\Omega_{90}\leq0.01\,{\rm deg}^2$. Furthermore, fractional luminosity distance errors smaller than $(0.1,\,0.01)$ will be observed at rates of the order {\red $(\mathcal{O}(10),\,\mathcal{O}(0))$} per year in \ce, {\red $(\mathcal{O}(10^3),\,\mathcal{O}(1))$} per year in \et\ and \EC, and {\red $(\mathcal{O}(10^4),\,\mathcal{O}(10))$} per year in \ECS. 
The rates for BBHs are approximately one order of magnitude larger per year {\red($\lesssim 20\times$)}, if permitted by the cosmic merger rates. The notable exceptions are that \ce\ will measure the luminosity distance of {\red $\mathcal{O}(10^4)$} {\red ($\sim150\times$)} while all NG networks observe {\red $\sim30$--60 times} more BBH than BNS mergers per year down to sub-1\% accuracies in luminosity distance. 
Finally, while \EC\ will be able to pin-point some BBH mergers to within $\Omega_{90}\leq0.01\,{\rm deg}^2$, only \ECS\ will be capable to do so for BNS mergers as well.

The sky localization capabilities of all networks, but \ce, scale with the networks' sensitivities (Tab.~\ref{tab:rates_snr}). In contrast, while \ce\ observes significantly more events than the Voyager network \voy, if not limited by the cosmic merger rate (e.g. {\red 80} times the number of BNS and BBH mergers with $\rho\geq 100$), both networks perform roughly equally in terms of sky localization. This is a clear indication of the importance of both the number of detectors in a network and their sensitivity. With only four detector sites and only one detector beyond the A+ generation, \ce's sky localization capabilities cannot surpass a network of five detectors, three of which operating at Voyager sensitivities. Ultimately, it performs equally due to its farther reach and thus larger base detection rate.
In Appendix \ref{app:ce_configs} we examine the differences between networks with four different CE configurations: i) a single 20 km arm length, ii) the same single 40 km arm length as in \ce, iii) two CEs with 20 and 40 km arm lengths, and iv) two CEs with 40 km arm lengths.

As was mentioned above, the three-dimensional localization of a GW source is extremely important for BNS coalescences which exhibit EM counterparts, such as the gamma-ray burst GRB 170817A and kilonova AT2017gfo associated with the GW event GW170817 \cite{GBM:2017lvd}. Unfortunately, such coincident detections with bright, EM transients like GRBs, require the emission of the burst in our direction while being in the field-of-view (FOV) of operating telescopes. Thus, if this initial pointer is missed or poorly localized, the EM follow-up will be hampered. As such the second BNS event GW190425, observed by the LIGO and Virgo detectors, did not appear to have a coincident gamma-ray burst, its location was not determined, and the potential counterpart was not studied in the EM spectrum.

If the GW signal itself already were to point to the source, by pin-pointing the sky location to within a 90\%-credible sky area that telescopes can quickly survey, a strong, coincident EM would not be required to find the fainter counterpart. Further, the identification of the signal's host galaxy allows the astronomy community to improve their surveys of the Universe's large-scale structure both locally but also in the distant Universe with BBHs. Besides, measurements of the host's redshift in conjunction with an accurate estimate of the luminosity distance from the GW signal could enable high-fidelity measurements of the Hubble constant in the local Universe to the level needed to resolve the \hubble tension with a single compact binary coalescence \cite{Borhanian:2020vyr}. We examine the potential of the chosen networks in enabling multi-messenger astronomy in the next section.

\section{Enabling multi-messenger astronomy}\label{sec:mma}

The EM follow-up campaign of the BNS event GW170817 \cite{GBM:2017lvd} resulted in the identification of the counterpart and observation of the afterglow in the entire EM spectrum, providing a treasure trove of data that has impacted several areas in fundamental physics \cite{Creminelli:2017sry, Monitor:2017mdv, Ezquiaga:2017ekz, De:2018uhw, Abbott:2018exr}, astrophysics \cite{Monitor:2017mdv, Drout:2017ijr, Coulter:2017wya, Cowperthwaite:2017dyu, Kasen:2017sxr, Soares-Santos:2017lru, Valenti:2017ngx, Arcavi:2017xiz, Tanvir:2017pws, Lipunov:2017dwd, Evans:2017mmy} and cosmology \cite{Abbott:2017xzu, Hotokezaka:2018dfi}. As such the synergy of GW and EM observations of compact binary coalescences in the readily EM-observable Universe, $z\leq0.5$, will be of paramount importance in the coming decades. In this section we examine how well each network will localize BNS and BBH mergers at redshifts $z\leq0.5$ in the GW window enabling the potential for EM follow-up irrespective of a loud EM transient such as a gamma-ray burst. Further, we illustrate each network's capabilities for early warning of BNS mergers, i.e. both the detection and sky localization of a compact binary during the inspiral ahead of the merger.

Tab.~\ref{tab:em_fov} presents the FOVs for 13 current or planned EM telescopes that have the capability to slew and follow-up GW detections. When considering these FOVs it is important to remember that they represent the sky area the respective telescope can observe without any tiling: It is not uncommon for EM telescope to observe larger FOVs during a follow-up campaign by tiling the search region with up to $\sim 10$ segments.

\begin{table}[hb]
\caption{Field-of-views (FOVs) for various electromagnetic telescopes.}
\begin{tabular}{lc}
\hline
\hline
EM telescope\phantom{AAA}   & \phantom{A} FOV  (${\rm deg^2}$) \phantom{A} \\
\hline
\hline
Rubin Observatory \cite{RubinObservatory:2022}                & 9.6     \\
EUCLID \cite{Euclid:2022}                                     & 0.54    \\
Athena \cite{Athena:2022}                                     & 0.35   \\
Nancy Roman Grace Space Telescope \cite{NancyRomanGrace:2022} & 0.28    \\
Chandra X-ray Observatory \cite{Chandra:2022}                 & 0.15    \\
Lynx \cite{Lynx:2022}                                         & 0.134   \\
Swift--XRT \cite{SwiftXRT:2022}                               & 0.12   \\
$30\,{\rm m}$-Telescope \cite{30mTelescope:2022}                       & 0.11    \\
Keck \cite{Keck:2022}                                   & 0.11    \\
VLT \cite{VLT:2022}                                           & 0.054   \\
ELT \cite{ELT:2022}                                           & 0.028   \\
GMT \cite{GMT:2022}                                           & 0.008   \\
HST--WFC3 \cite{HSTWFC3:2022}                                 & 0.002   \\
\hline
\hline
\end{tabular}
\label{tab:em_fov}
\end{table}
\subsection{Three-dimensional localization}

\begin{figure*}
\includegraphics[width=\textwidth]{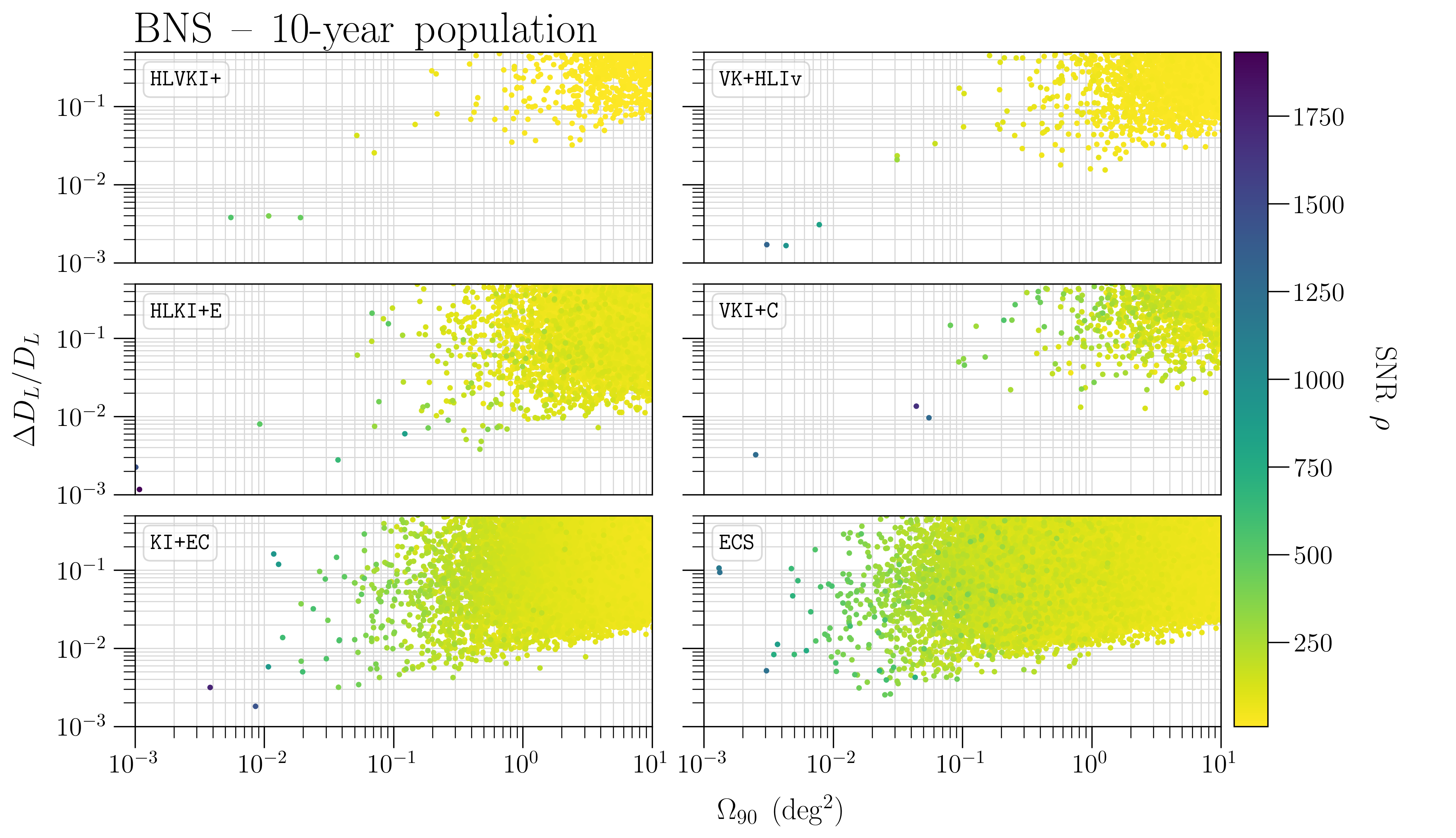}
\includegraphics[width=0.93\textwidth]{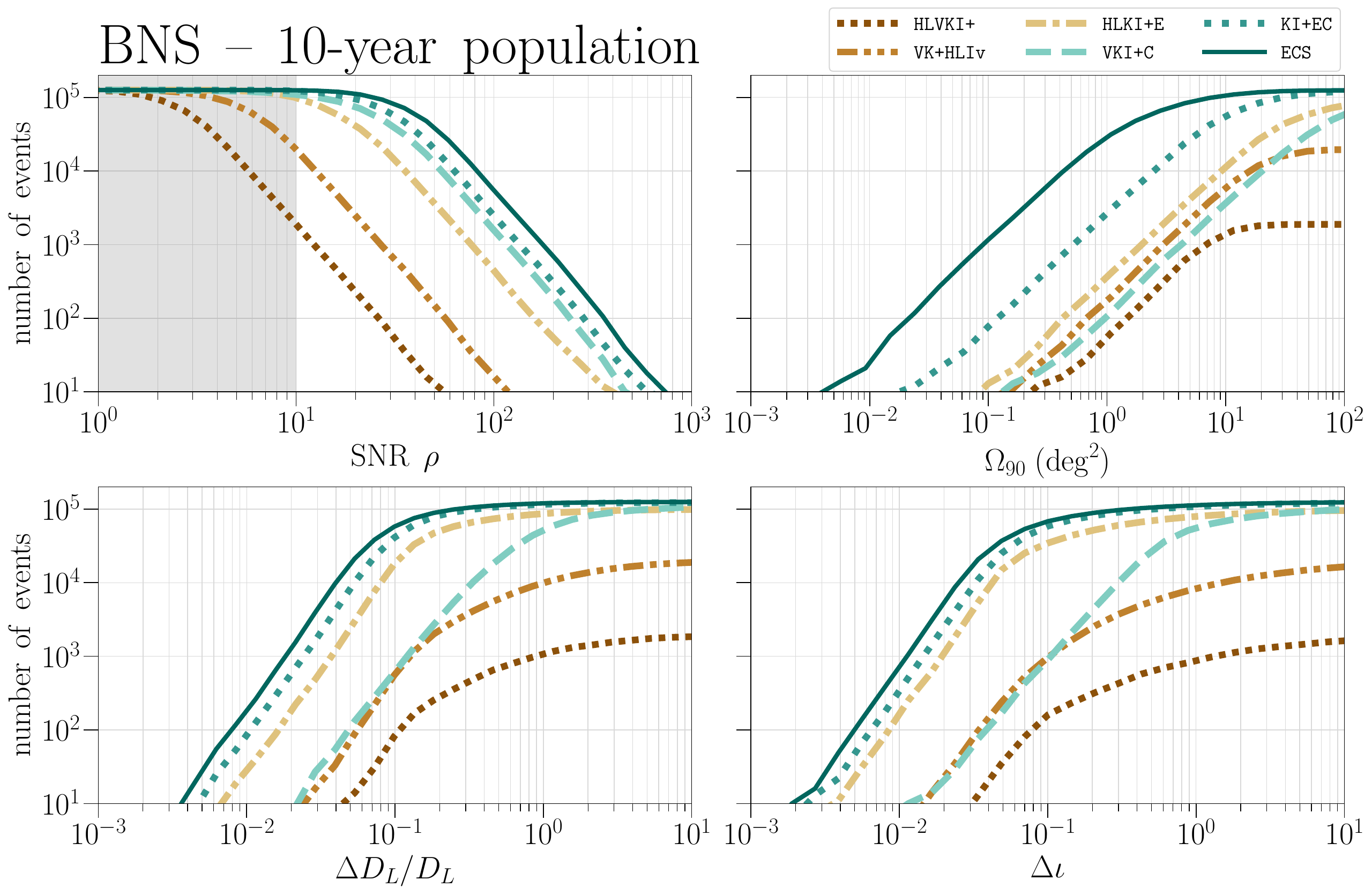}
\caption{\emph{Top}: The scatter plots illustrate the correlations between SNR $\rho$, 90\%-credible sky area $\Omega_{90}$, and fractional luminosity distance error $\Delta D_L/D_L$ for BNS mergers with SNR $\rho\geq10$ in the six studied A+, Voyager, and NG networks for redshifts $z\leq0.5$. The color bar indicates the SNR of the events.\\
\emph{Bottom:} Cumulative histograms for the SNR $\rho$, 90\%-credible sky area $\Omega_{90}$, fractional luminosity distance errors $\Delta D_L/D_L$, and absolute errors the inclination angle $\Delta\iota$ for BNS mergers observed in the six studied A+, Voyager, and NG networks for redshifts $z\leq0.5$. The histograms were generated from $\sim 1.2\times10^{5}$ injections sampled according to Sec. \ref{sec:resampling}. The non-SNR panels are obtained for events with SNR $\rho\geq10$, indicated by the non-shaded region in the SNR panel. The SNR panel is flipped to highlight the behavior for large values.}
\label{fig:mma_BNS}
\end{figure*}

\begin{figure*}
\includegraphics[width=\textwidth]{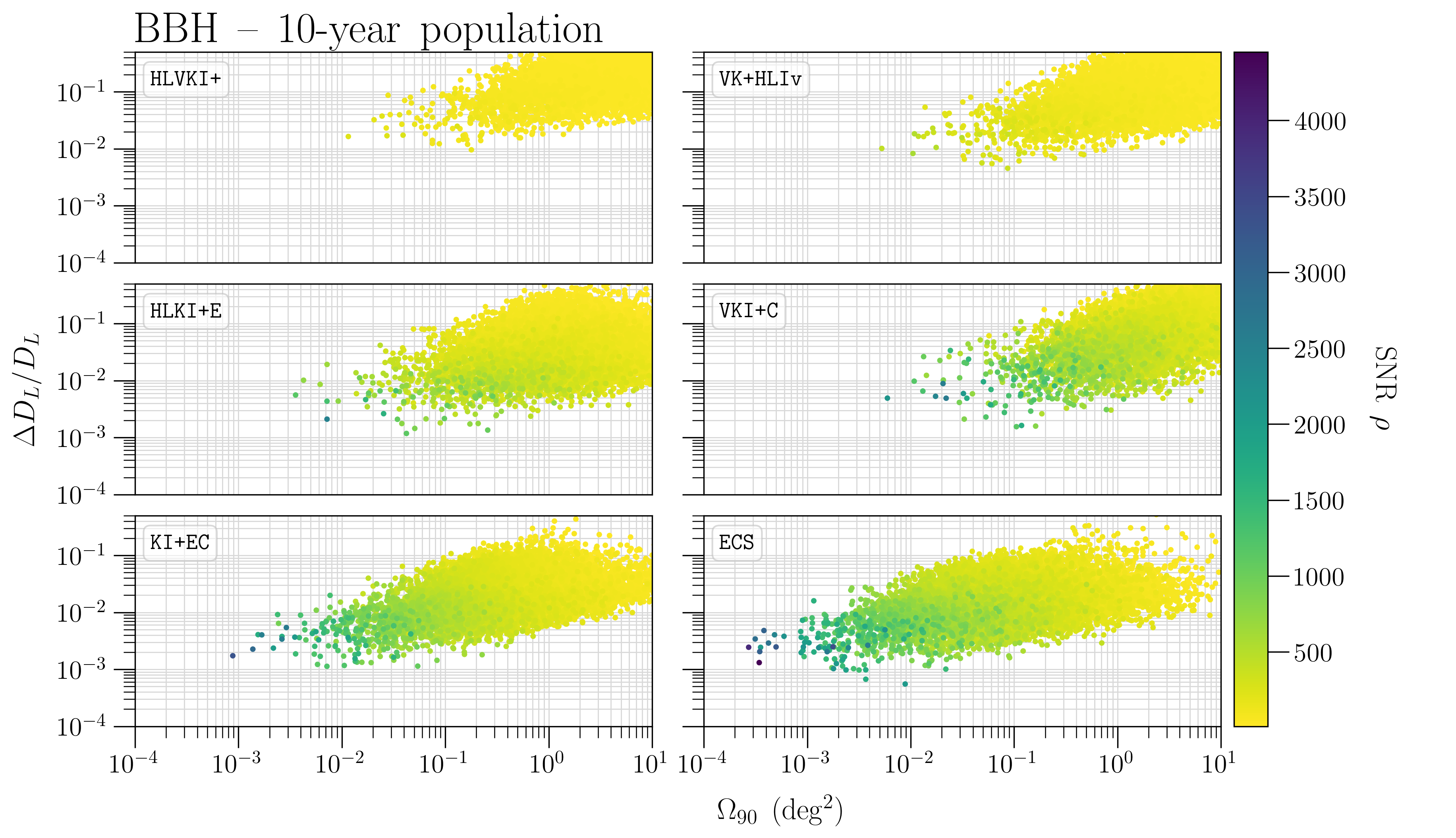}
\includegraphics[width=0.93\textwidth]{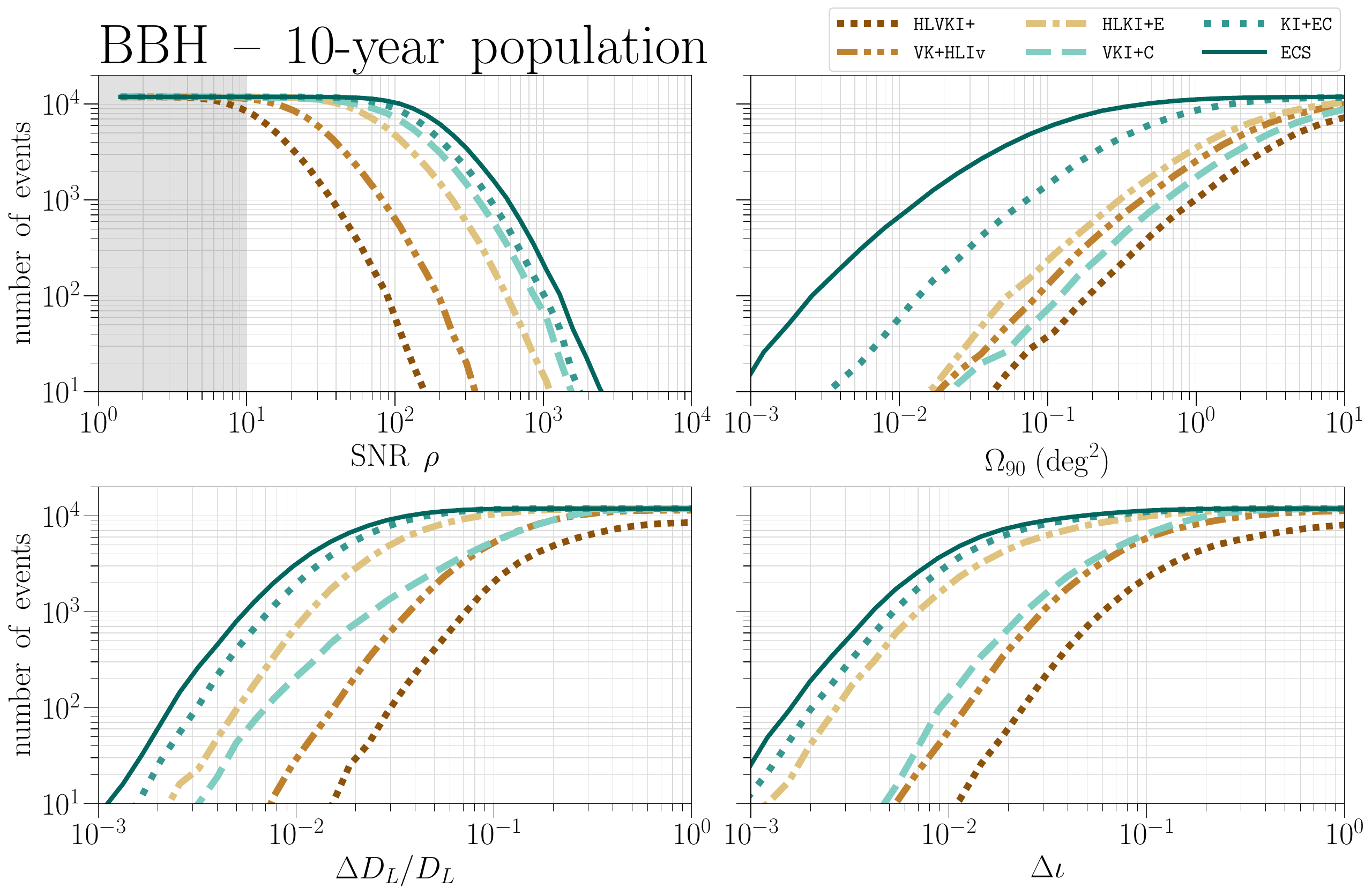}
\caption{\emph{Top}: The scatter plots illustrate the correlations between SNR $\rho$, 90\%-credible sky area $\Omega_{90}$, and fractional luminosity distance error $\Delta D_L/D_L$ for BBH mergers with SNR $\rho\geq10$ in the six studied A+, Voyager, and NG networks for redshifts $z\leq0.5$. The color bar indicates the SNR of the events.\\
\emph{Bottom:} Cumulative histograms for the SNR $\rho$, 90\%-credible sky area $\Omega_{90}$, fractional luminosity distance errors $\Delta D_L/D_L$, and absolute errors on the inclination angle $\Delta\iota$ for BBH mergers observed in the six studied A+, Voyager, and NG networks for redshifts $z\leq0.5$. The histograms were generated from $\sim 1.2\times10^{4}$ injections sampled according to Sec. \ref{sec:resampling}. The non-SNR panels are obtained for events with SNR $\rho\geq10$, indicated by the non-shaded region in the SNR panel. The SNR panel is flipped to highlight the behavior for large values.}
\label{fig:mma_BBH}
\end{figure*}

The localization of the GW signal's sky position is the key metric governing the feasibility of a follow-up campaign with EM telescopes. Additionally, GW observations can provide accurate distance measurements that are independent of the cosmic distance-ladder calibration issues and can help distinguish between a number of potential hosts \cite{Singer:2016eax}. Finally, the visibility of certain EM phenomena such as gamma-ray bursts and jets depends on the binary's orientation with respect to the observer's line of sight. Thus, in this section we present the potential of the chosen networks to measure these metrics in form of the SNR, 90\%-credible sky area, as well as luminosity distance and inclination angle errors for GW signals at redshifts $z\leq0.5$.

Fig.~\ref{fig:mma_BNS} presents the distribution and correlation of SNR $\rho$, sky area $\Omega_{90}$, and fractional luminosity distance error $\Delta D_L/D_L$ for BNS signals as well as the cumulative histograms for all four metrics, including absolute errors on the inclination angle $\iota$. We also summarize the expected detection rates of events with 90\%-credible sky area $\Omega_{90}/{\rm deg^2}\leq 0.01,\,0.1,\,1$ and fractional luminosity distance errors $\Delta D_L/D_L \leq 0.01,\,0.1$ in Tab.~\ref{tab:mma_rates}.

The cosmic merger rate for BNSs up to redshift $z=0.5$ is approximately 12,000 per year. The top part of Fig.~\ref{fig:mma_BNS} and Tab.~\ref{tab:mma_rates} indicate that networks containing less than {\red two} \emph{distinct} NG detector sites (\aplus, \voy, \et, and \ce) will localize, if at all, rare, {\em golden}\footnote{We refer to uncommonly rare events with high measurement fidelity at rates below 3 per year as {\em golden}.}
BNS signals ({\red $\lesssim 1$} per year) to within $\Omega_{90}\leq0.1\,{\rm deg^2}$. Luminosity distances should be measured to better than 10\% accuracy {\red $\mathcal{O}(1)$ (\aplus), $\mathcal{O}(10)$ (\voy, \ce), and $\mathcal{O}(10^3)$ (\et)} per year; and only \et\ will be capable to consistently observe {\red a few} golden events per year to sub-1\% accuracy.
The discrepancy between the NG networks \et\ and \ce\ stems from the sub-10 Hz sensitivity of ET and its ability to measure both polarizations by itself.
\EC\ and \ECS\ with two and three NG sites will determine the sky positions of {\red a few or $\sim$ 100} BNS mergers per year to $\leq0.1\,{\rm deg^2}$, respectively, and measure the luminosity distance of {\red $\mathcal{O}(10)$} events per year to sub-1\% accuracies.
Ultimately, \ECS\ is the only network to consistently localize golden BNS mergers to within $\Omega_{90}\leq0.01\,{\rm deg^2}$. In fact, the scatter plot in Fig.~\ref{fig:mma_BNS} indicates most of these observations will come with a sub-10\% accuracy in the luminosity distance while a few golden events will push it down below 1\%.

\begin{table}[ht]
    \centering
     \caption{Detection rates of BNS and BBH mergers up to redshift $z=0.5$ to be observed by different detector networks each year with $\Omega_{90}/{\rm deg^2}\leq 1,\, 0.1,\, 0.01$ as well as $\Delta D_L/D_L \leq 0.1,\, 0.1$, where $\Omega_{90}$ is the 90\%-credible sky area and $D_L$ the luminosity distance. These detection rates are calculated for events with SNR $\rho\geq10$. Due to uncertainty in the various quantities that go into the calculation these numbers are no more accurate than one or two significant figures. The bare merger rates for BNSs and BBHs up to redshift $z=0.5$ are $\sim12,000\,{\rm yr^{-1}}$ and $\sim1,200\,{\rm yr^{-1}}$, respectively.}
    \begin{tabular*}{\columnwidth}{@{\extracolsep{\fill}}l?r|r|r?r|r}
\hline
\hline
Metric        & \multicolumn{3}{c?}{$\Omega_{90}$  $({\rm deg^2})$ \phantom{A}} & \multicolumn{2}{c}{$\Delta D_L/D_L$}\phantom{A} \\
\hline
Quality &  $\leq1$\phantom{A} &  $\leq 0.1$\phantom{A} &  $\leq 0.01$\phantom{A} &  $\leq 0.1$\phantom{A} &  $\leq0.01$\phantom{A}\\
\hline
\hline
\multicolumn{6}{c}{\emph{BNS}} \\
\hline
\hline
\aplus    & 5        & 0        & 0        & 8        & 0       \\
\voy  \phantom{AA}      & 16       & 1        & 0        & 56       & 0       \\
\et       & 36       & 1        & 0        & 1,800    & 2       \\
\ce       & 10       & 1        & 0        & 56       & 0       \\
\EC       & 260      & 8        & 0        & 4,100    & 8       \\
\ECS      & 2,800    & 110      & 2        & 5,600    & 17      \\
\hline
\hline
\multicolumn{6}{c}{\emph{BBH}} \\
\hline
\hline
\aplus    & 100      & 4        & 0        & 210      & 0       \\
\voy      & 270      & 14       & 0        & 560      & 3       \\
\et       & 350      & 23       & 1        & 1,100    & 72      \\
\ce       & 180      & 8        & 0        & 550      & 21      \\
\EC       & 870      & 150      & 6        & 1,200    & 200     \\
\ECS      & 1,100    & 580      & 69       & 1,200    & 320     \\
\hline
\hline
    \end{tabular*}
    \label{tab:mma_rates}
\end{table}

While BNS mergers are the headlight events for multi-messenger astronomy, sparked by GW170817 and expected EM counterparts, the EM follow-up of BBH coalescences is equally intriguing, especially since astrophysicists want to explore the unclear origins of the massive BBHs LIGO and Virgo have observed so far. As such, Fig.~\ref{fig:mma_BBH} and Tab.~\ref{tab:mma_rates} summarize the potential for multi-messenger astronomy with BBHs up to redshift $z=0.5$. With a cosmic merger rate of only 1,200 BBH mergers per year, all networks localize {\red 8\% to 92\%} of these signals to within $1\,{\rm deg^2}$. Moreover, BBHs localized to within $0.1\,{\rm deg^2}$ will be observed by all networks, with \et\ seeing such an event every other week, while \EC\ and \ECS\ pushing the rates to once every other day and almost twice a day, respectively. In fact, \EC\ will pinpoint BBHs to within $0.01\,{\rm deg^2}$ a handful of times per year and \ECS\ at least once a week! Similarly, all studied networks are capable of determining the luminosity distance of {\red 18\% to all} of the BBHs to 10\% or less, while better than 1\%-level accuracies are much rarer: {none to a few} golden events in \aplus\ or \voy, respectively, {\red $\mathcal{O}(10)$} in both \et\ and \ce, and {\red $\mathcal{O}(100)$} in \EC\ and \ECS. Again, \EC\ and \ECS\ will actually detect such an event about once a day. The strong difference between the BNS and BBH rates for the given measurement accuracies in \aplus\ are a result of the network's vastly different redshift reaches for BNS ($z_r\approx 0.11$) and BBH ($z_r\approx0.6$).

In conclusion, all six networks will observe at least some well-localized BNS mergers per year, yet only those with at least two NG sites will provide EM astronomers with an abundance of events to follow up on a daily basis in addition to a few golden events with very high 3D-localization accuracy. While even the A+ and Voyager networks will localize a large number of well-localized BBH events, \EC\ and \ECS\ will elevate nearly all BBHs within a redshift of $z=0.5$ to this level and enable the follow-up of dark siren events---in the absence of an EM counterpart---on a daily basis! As such, GW170817-like follow-up campaigns could become common-place and increasingly more dependent on the EM telescopes' availability and slewing capabilities. We also want to stress that while the rates that \EC\ and \ECS\ enable could be deemed as unnecessarily high, in reality, not every GW event can be followed-up due to maintenance outages of EM telescope, conflicts with other observations, or the potential of objects covering the EM counterpart or the source galaxy amongst other things.

\subsection{Early warning alerts}

The EM follow-up campaign of the event GW170817 was successful in spite of the fact that the earliest observations took place many hours after the epoch of merger \cite{Drout:2017ijr}, thereby missing critical data from the fireball that would have been launched moments after the merger as evidenced by the detection of gamma ray bursts by the Fermi gamma ray observatory \cite{LIGOScientific:2017zic, Ajello:2018mgd} and the INTEGRAL satellite \cite{Savchenko:2017ffs} a mere 1.7 s after merger. The alert from LIGO and Virgo with the full 3D localization of the event was delayed by a little over 4.5 hrs \cite{GW170817:GCN} (see, in particular, GCN Circular
Number 21513). Six groups reported optical observations carried out between 10.89 hrs and 11.57 hrs after the epoch of merger \cite{GBM:2017lvd}.  During the third observing run, GW alerts have been sent out with a average latency of about {\red  5 to 7 minutes} (see Appendix A of Ref.\,\cite{LIGOScientific:2021djp}) and there is effort to reduce the latency to less than a minute.  So far, GW170817 remains the only GW event with an EM counterpart.

\paragraph{Motivation for observing events at the onset of merger}
From an astrophysical point of view, there are compelling reasons to begin EM observation right at the onset of merger (see, e.g., Ref.\,\cite{Sachdev:2020lfd} for a summary) but that would require sending alerts before the epoch of merger \cite{Cannon:2011vi} to allow EM telescopes to slew to the right part of the sky. Early X-ray observations could resolve the initial state of the merger remnant, namely if a hypermassive NS forms first before collapsing to a BH or if the collapse is prompt. Immediate optical and infrared observations could inform the nature of the dynamical ejecta and outflow, formation of the accretion disc, and the onset of r-process nucleosynthesis. Radio observations could shed light on the magnetosphere interactions between the two NSs before merger and test the hypothesis that some fast radio bursts result in the aftermath of a BNS merger.  As we shall argue below, it should be possible to send out alerts before the epoch of coalescence \cite{Cannon:2011vi, Chan:2018csa, Tsutsui:2020bem, Kapadia:2020kss, Singh:2020lwx, Tsutsui:2021izf, Li:2021mbo, Nitz:2021pbr, Singh:2022tlh, Akcay:2018aqh, Baltus:2021nme} and efforts are underway to accomplish this during the fourth observing run of the LIGO and Virgo detectors \cite{Sachdev:2020lfd, Magee:2021xdx, Yu:2021vvm}.

\paragraph{Coalescence time scale}

Gravitational waves from the inspiral phase of BNSs last for tens of minutes to hours in ground-based detectors depending on the lower-frequency cutoff. The time left until coalescence, often referred to as \emph{coalescence time}, starting from a frequency $f_L,$ is given by \cite{Sathyaprakash:1991mt}
\begin{eqnarray}
    \tau
    & \simeq & \left (\frac{0.25}{\eta} \right ) \left ( \frac{2.8\,M_\odot}{M} \right )^{5/3}
    \left (\frac{5\,\rm Hz}{f_L} \right )^{8/3}\,6.4 \times 10^3 \mbox{ s},
    \label{eq:coalescence time}
\end{eqnarray}
where, as before, $\eta$ is the symmetric mass ratio and $M$ is the observed total mass of the system related to its intrinsic mass via $M\equiv (1+z)M_{\rm int}$. Thus, asymmetric binaries last longer compared to symmetric ones but the time-scale is a sharp function of both the total mass and the starting frequency.

\begin{figure}[t]
    \centering
    \includegraphics[width=0.49\textwidth]{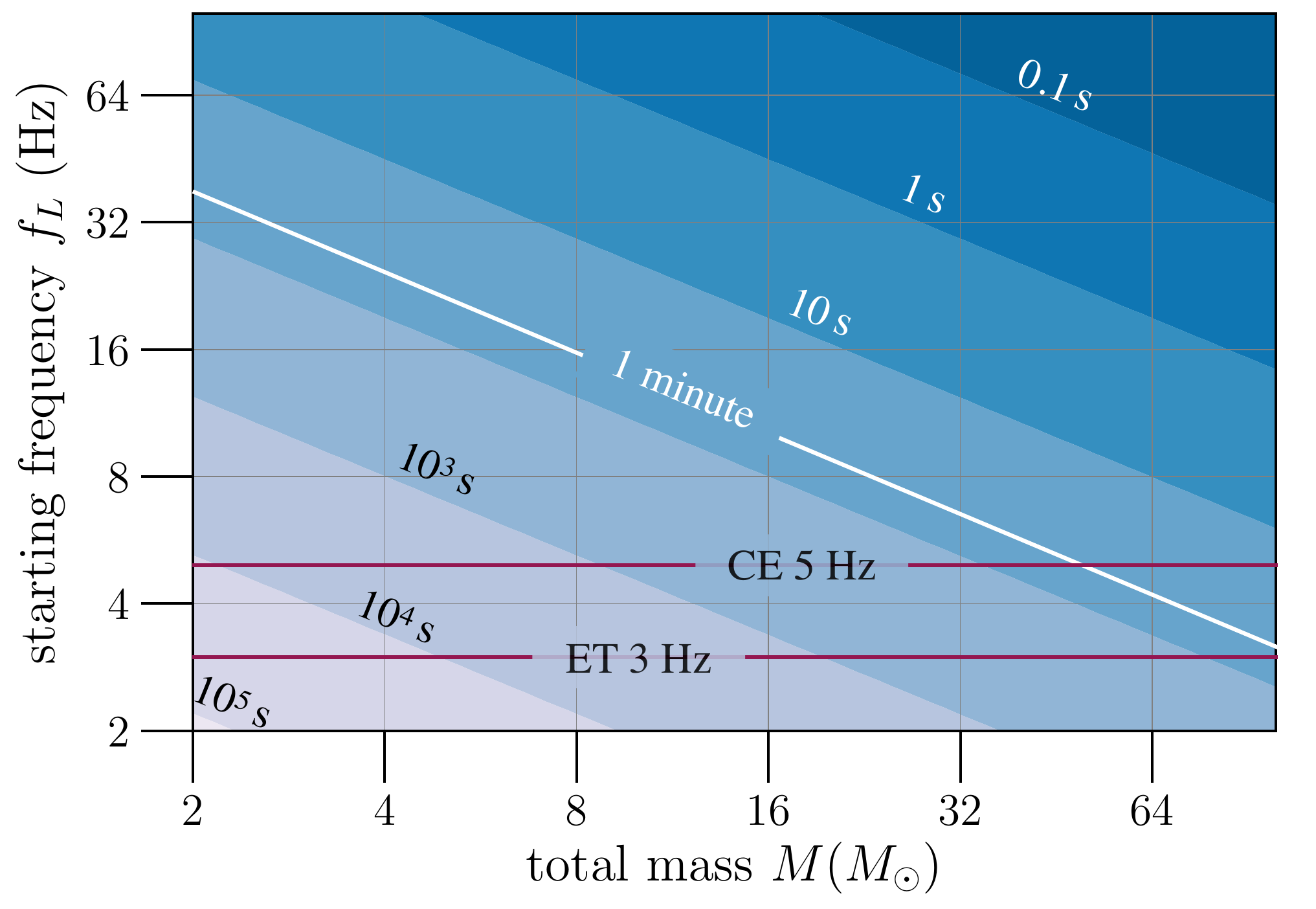}
    \caption{The plot shows the duration (in seconds) for which a GW signal from an equal-mass binary of total mass $M$ lasts until coalescence starting from frequency $f_L.$
    }
    \label{fig:duration}
\end{figure}

Figure \ref{fig:duration} plots the duration of a compact binary signal as a function of the total mass and the starting frequency for equal mass (i.e., $\eta=1/4$) binaries. In current detectors the lower-frequency cutoff is $f_L=20\,\rm Hz$ and sources are detected at $z\ll 1.$ Thus, a typical BNS would last for a few minutes. With A+ detectors, which are expected to have a lower-frequency cutoff of 10 Hz, this increases to about 15 minutes. For CE, however, a lower-frequency cutoff of 5 Hz is appropriate and BNSs at $z=0$ would last for slightly less than two hours, while in the case ET $f_L=3\,\rm Hz$ is more appropriate, in which case $\tau \sim 6.9\,\rm hrs.$

Thus, as a detector's low-frequency sensitivity improves signals last longer and some of these could be detected well before the epoch of coalescence, making it possible to send \emph{early warning} (EW) alerts to EM telescopes to observe the events either before or right at the onset of coalescence. Due to the sharp dependence of the time-scale on the total mass, it is far more plausible to send EW alerts for lower-mass systems than it is to do so for higher-mass binaries.

\paragraph{Early warning and localization}
Current algorithms are able to filter the data through a template bank within about $30\,{\rm s}$ after data acquisition. This includes time required for data transfer and application of denoising algorithms. With lower frequencies and longer duration templates filtering the data could take longer and we assume a latency of $60\,{\rm s}$ for data processing. To slew telescopes to the direction would also involve some latency and we assume that with automation this would be as low as $60\,{\rm s}$. In what follows we will consider three EW times:  $\tau_{\rm EW}=600 \, {\rm s}$, $\tau_{\rm EW}=300\, {\rm s}$ and $\tau_{\rm EW}=120 \, {\rm s}$ before merger. Given the EW time, Eq.\,(\ref{eq:coalescence time}) can be inverted to determine the frequency $f_{\rm EW}$ from which the system has time $\tau_{\rm EW}$  left until coalescence:
\begin{equation}
    f_{\rm EW}
    \simeq \left (\frac{0.25}{\eta} \right )^{3/8} \left ( \frac{2.8\,M_\odot}{M} \right )^{5/8}
    \left (\frac{120\,\rm s}{\tau_{\rm EW}} \right )^{3/8}\,22.2 \mbox{ Hz}
    \label{eq:coalescence frequency}
\end{equation}
In computing the Fisher matrix integrals for EW alerts, we use a lower frequency of $f_L=5\,\rm Hz$ for all detectors except Virgo+ for which it is set to be $f_L=10\,Hz$ (see Sec.\,\ref{sec:methodology}) and an upper frequency cutoff of $f_U=f_{\rm EW}.$

\begin{figure*}[b]
\includegraphics[width=0.99\textwidth]{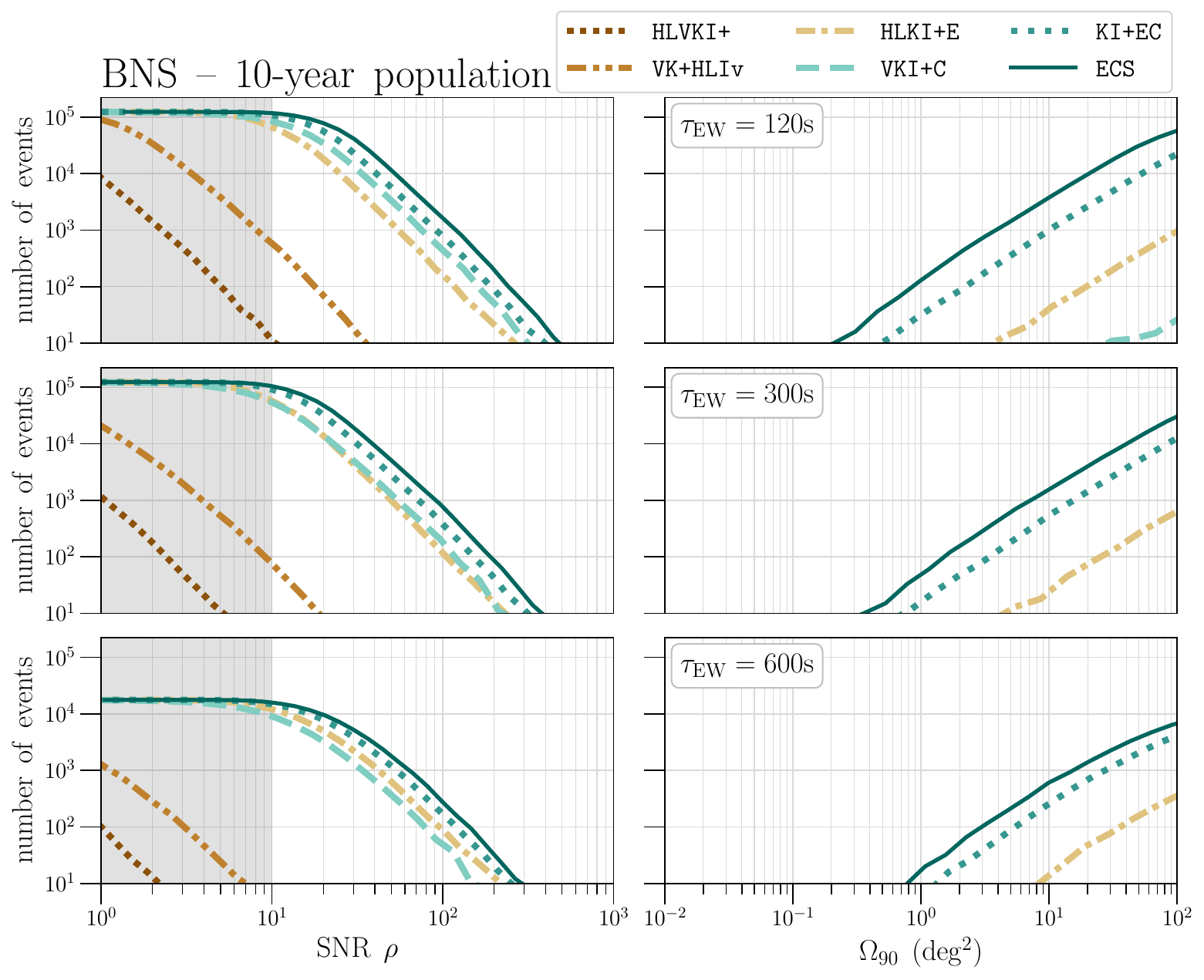}
\caption{Cumulative histograms for the SNR $\rho$ and 90\%-credible sky area $\Omega_{90}$ for BNS mergers observed in the six studied A+, Voyager, and NG networks for redshifts $z\leq0.5$ 2, 5, and 10 minutes before merger. The histograms were generated from $\sim 1.2\times10^{5}$ injections sampled according to Sec. \ref{sec:resampling}. The $\Omega_{90}$ panel is obtained for events with SNR $\rho\geq10$, indicated by the non-shaded region in the SNR panel. The SNR panel is flipped to highlight the behavior for large values. \\
The cumulative histograms of the SNR in the $\tau_{\rm EW}=600\,{\rm s}$ panel do not reach the value 1 because $\sim85\%$ of the events exhibited a cutoff-frequency $f_{\rm EW} < 11\,{\rm Hz}$. Since this is too close to the lower frequency cutoff of the used V+ sensitivity curve used in half our networks, we chose to abort these runs for all networks.}
\label{fig:ew_snr_sa90}
\end{figure*}

\begin{table*}[b]
    \centering
    \begin{tabular*}{0.75\textwidth}{l @{\extracolsep{\fill}}|  cccc|cccc|cccc}
    \hline \hline
EW time & \multicolumn{4}{c|}{$\tau_{\rm EW}=120$ s} &
  \multicolumn{4}{c|}{$\tau_{\rm EW}=300$ s} &
  \multicolumn{4}{c}{$\tau_{\rm EW}=600$ s} \\
\hline
$\Omega_{90}$ (deg$^2$)\phantom{ABC} &
$\le 100$ &
$\le 20$ &
$\le 10$ &
$\le 1$ \phantom{A} &
$\le 100$ &
$\le 20$ &
$\le 10$ &
$\le 1$ \phantom{A} &
$\le 100$ &
$\le 20$ &
$\le 10$ &
$\le 1$ \phantom{A} \\
    \hline
    \hline
\et       & 50    & 5     & 2     & 0    \phantom{ABC} & 28    & 3     & 1     & 0    \phantom{ABC} & 3     & 0     & 0     & 0    \\
\ce       & 2     & 0     & 0     & 0    \phantom{ABC} & 0     & 0     & 0     & 0    \phantom{ABC} & 0     & 0     & 0     & 0    \\
\EC       & 1,800 & 210   & 79    & 3    \phantom{ABC} & 880   & 99    & 37    & 1    \phantom{ABC} & 46    & 8     & 3     & 0    \\
\ECS      & 5,100 & 850   & 330   & 11   \phantom{ABC} & 2,400 & 320   & 120   & 4    \phantom{ABC} & 82    & 16    & 7     & 0    \\\hline \hline
    \end{tabular*}
    \caption{Number of BNS mergers events for redshifts $z\leq0.5$ observed in the six studied A+, Voyager, and NG networks that can be localized to within a small region on the sky ($\Omega_{90}/{\rm deg^2} \leq 100, 20, 10, 1$) 2, 5, and 10 minutes before merger. For a given early warning time $\tau_{\rm EW}$ we determine the corresponding starting frequency $f_{\rm EW}$ using Eq. \eqref{eq:coalescence frequency}. The \aplus\ and \voy\ networks do not meet the requirement of angular resolution for any of the early warning times considered.}
    \label{tab:ew}
\end{table*}

In order to follow-up GW events, EM telescopes would need to be given the 3D localization of the events with a fairly good accuracy. The best optical and infrared telescopes, such as the Rubin Observatory, have a FOV of $10\,{\rm deg^2}$, while others have narrower FOV of $\sim 1\,{\rm deg^2}$. X-ray and radio observatories have still narrower FOV of {\red $\sim 0.1\,\rm deg^2$}. Follow-up observations typically cover a sky area equivalent to several times their FOV and we assume the number of such follow-up grids to be between 10 to 100. We will, therefore, consider the number of events that can be localized to within $\Omega_{90} = 100\,{\rm deg^2}$, $20\,{\rm deg^2},$ $10\,{\rm deg^2}$, and $1 \,{\rm deg^2}.$

Current GW observatories can localize only few events to such narrow regions in the sky {\em even with the full-signal power} and while their planned upgrades (A+, Voyager) will indeed see many events with a $\Omega_{90}\leq100\,{\rm deg^2}$, see Fig.~\ref{fig:mma_BNS}, this unfortunately does not translate to early warning: Less than $\lesssim0.5$\% of events can be detected 2 minutes before coalescence in the case of the Voyager network, see Fig.~\ref{fig:ew_snr_sa90} which plots the cumulative density plots of the SNR (left panels) and $\Omega_{90}$ (right panels) for the detector networks considered in this study and BNS sources up to a redshift of 0.5.

The NG networks (\et, \ce, \EC, \ECS) increase the detected fraction of events 2 minutes before coalescence to {\red $\sim$(53\%, 68\%, 86\%, 94\%)}. In fact, these networks will detect a considerable fraction of events also 5 minutes {\red $\sim$(45\%, 43\%, 73\%, 83\%)} and 10 minutes {\red $\sim$(10\%, 7\%, 12\%, 13\%)} before merger. This abundance of early detections translates to the ability to generate EW alerts with good localization accuracy for a small fraction of BNS events from 2 to 10 minutes before the epoch of coalescence. Tab.~\ref{tab:ew} lists the number of these BNS events that can be localized each year to within $\Omega_{90}$ of (100,\, 20,\, 10,\, 1)\,${\rm deg^2}$. We have left out \aplus\ and \voy\ networks as they do not have any significant number of detections with the required sky localization at least 2 minutes before coalescence. The \ce\ network can only meet the requirement of good sky localization for a handful events two minutes before merger.

From Tab.~\ref{tab:ew}, it is clear that ET plays a crucial role in the localization of events. The \et\ network is able to localize a {\red few {\em golden events} per year} to within $10\,{\rm deg^2}$ 2 and 5 minutes before merger, while the \ECS\ (\EC) network will observe such events {\red almost daily (weekly) and twice a week (every other week)} 2 and 5 minutes before merger, respectively. Thus, optical and infrared telescopes such as the Rubin observatory will have plenty of opportunity to observe mergers as they happen.

These numbers decrease by up to two orders of magnitude for $\Omega_{90} \le 1\,\rm deg^2,$ yet the NG observatories \EC\ and \ECS\ meet this constraint for a few {\em golden} events each year 2 and 5 minutes before the merger, thereby providing a number of events for early observation by EUCLID, Athena, Nancy Grace Roman, Chandra, Lynx, Swift-XRT, $30\,{\rm m}$-Telescope, and Keck, and even by optical and infrared telescopes, such as the VLT, ELT, and GMT, with their smaller FOVs thanks to the potential of tiling.

\section{Rare and loud events} \label{sec:rare_events}
The quality of science delivered by a GW network is determined by a combination of a large number of events at moderate SNR and a population of loud events, even if a small number, that would be useful in obtaining answers to certain key physics questions. The Advanced LIGO and Virgo network makes most of its observations at or near the threshold SNR \cite{LIGOScientific:2018mvr, Abbott:2020niy}. The loudest event so far is the BNS merger GW170817 \cite{TheLIGOScientific:2017qsa} and it has undoubtedly delivered the best science to date, impacting many branches in physics and astronomy \cite{GBM:2017lvd, Abbott:2018lct,Cowperthwaite:2017dyu, Valenti:2017ngx,De:2018uhw, Abbott:2018exr, Creminelli:2017sry, Ezquiaga:2017ekz, Abbott:2017xzu}. The number of events observed until now is also low---about one per week \cite{Abbott:2020niy}. As shown in Sec.\,\ref{sec:visibility}, with the \aplus\ network the number of BNS (BBH) events will increase to 100s (1000s) per year but there will not be (m)any high fidelity BNS or BBH signals with SNR well beyond $100.$ This is because, the number of events at an SNR of $\rho_2$ (300 or 1000), relative to an SNR of $\rho_1$ (100), would be roughly a factor $(\rho_2/\rho_1)^3$ (respectively, 27 or 1000) smaller. The \voy\ network will observe a handful of BBH events with $\rm SNR>300$ but not any high-SNR BNS mergers.

NG observatories will usher in an era of precision measurements by observing large populations of signals that are needed to mitigate statistical uncertainties and systematic biases for some of the inferences (e.g. precision cosmology); with hundreds of high-SNR {\red ($\rho\geq300$)} BBH events that could help in detecting subtle signatures of new physics, e.g. dark matter, violation of GR, etc.  In this section we briefly discuss some of the most impactful science enabled by NG observatories.

\begin{figure}
    \centering
    \includegraphics[width=\columnwidth]{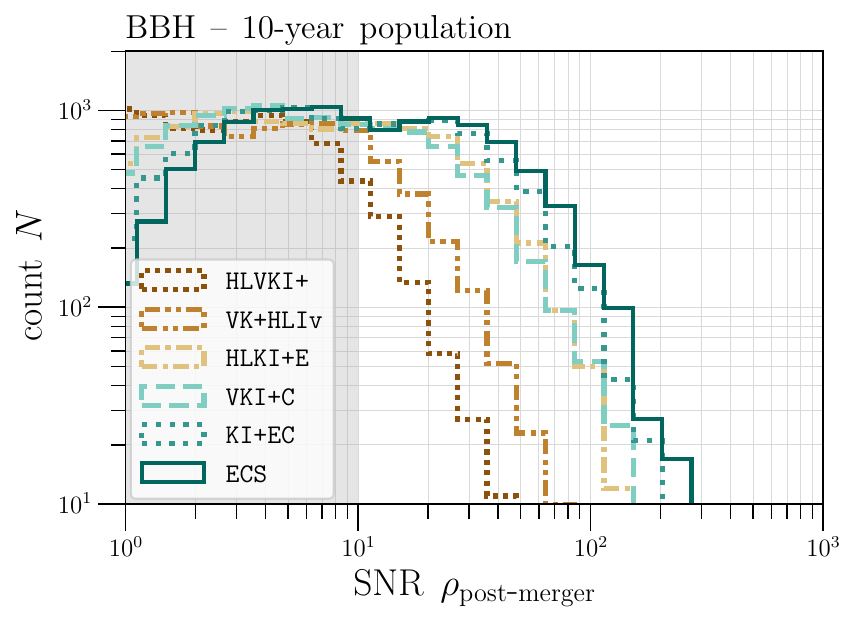}
    \caption{Histograms of the SNR $\rho_{{\rm post}\textnormal{-}{\rm merger}}$ in the post-merger signal of the BBH population in the six studied A+, Voyager, and NG networks for redshifts $z\leq0.5$ and a observation time of \textbf{10 years}. The histograms were generated from $\sim 1.2\times10^{4}$ injections sampled according to Sec. \ref{sec:resampling}. }
    \label{fig:pm}
\end{figure}

\paragraph{Understanding the nature of black holes}
BHs are unlike other macroscopic objects. Perturbed BHs return to their quiescent state by emitting GWs whose spectra is completely determined by the BH's mass and spin angular momentum via a theorem called the \emph{black hole no-hair theorem} \cite{Israel:1967wq, Carter:1971zc}. Thus, observation of how the remnant of a compact binary coalescence settles down to its final state could tell us about its nature, but to do so, it is necessary to observe not just the fundamental mode of the GW spectrum but the higher modes and overtones excited during the coalescence \cite{Dreyer:2003bv,Berti:2007zu,Gossan:2011ha}. Unfortunately, the amplitude of these sub-dominant modes and overtones is generally far lower than the dominant, fundamental quadrupole mode and detecting them would require high-SNR events \cite{Kamaretsos:2011um, Kamaretsos:2012bs, London:2014cma}. For example, the loudest BBH event so far, GW150914, had an estimated SNR of between 4 to 8.5 in the ringdown part of the signal \cite{TheLIGOScientific:2016src}---depending on when the ringdown signal is assumed to begin---compared to an SNR of 24 in the full inspiral-merger-ringdown signal \cite{Abbott:2016blz}. Thus, the ringdown signal alone was not loud enough to accurately measure the parameters of even the dominant mode. To test the black hole no-hair theorem one would need to measure the complex frequencies of at least two modes with the SNR in weaker modes in excess of 15 to 20.  This would require SNRs of several hundreds or more in the full signal (see, e.g., Fig.\,1 of Ref.\,\cite{Gossan:2011ha}, cf.\,Tab.\,\ref{tab:rates_snr}) or an SNR of 30 or more in the post-merger signal. More recently, it has been suspected that it might also be possible to detect the overtones of quasi-normal modes \cite{Giesler:2019uxc, Bhagwat:2019dtm, Ota:2019bzl, Dhani:2020nik, Dhani:2021vac, Cotesta:2022pci}, which would require even higher SNRs in the post-merger signal, which, as we shall see below, is only accessible to NG observatories (cf.\,Fig.\,\ref{fig:pm}).

The {\em post-merger} phase of a BBH waveform is conventionally defined as the signal that follows after the waveform reaches its peak amplitude; where the amplitude evolution is given by the Euclidean norm of $h_+ - \mathrm{i}\,h_\times$. In this study we compute this just for the quadrupole moment. Fig.\,\ref{fig:pm} plots the SNR histogram in the post-merger phase of the BBH population up to redshift $z=0.5$ observed over a duration of ten years. It is clear that only networks containing at least one NG observatory will have access to several high-fidelity events with SNR $\rho_{{\rm post}\textnormal{-}{\rm merger}}\geq100$ each year, while earlier generations of detectors will rarely see such events. Thus, NG observatories will be unique in their ability to map out the detailed structure of dynamical horizons via the complex quasi-normal mode spectrum expected in black hole mergers.

In addition to testing the no-hair theorem with perturbed BHs, it is possible to test GR by checking the consistency of the binary parameters estimated using different multipoles provided there is enough SNR from higher order modes \cite{Dhanpal:2018ufk, Islam:2019dmk}. For example, the octupole mode of the inspiral signal from GW190814 is estimated to have an SNR of about 6.6 compared to the total SNR of 21.4 in the full signal \cite{Abbott:2020khf}. In contrast, the ECS network would have, for the same event, an SNR of about 1500 in the full signal and 400 in higher order modes. Such events will determine the nature of the merger remnant with exquisite precision.

\paragraph{Inferring dense matter equation of state and QCD phase transition}
Understanding the equation of state of dense nuclear matter is one of the open problems in fundamental physics (see, e.g. Ref.\,\cite{Lattimer:2020tot}). GW observations can determine an effective tidal deformability of a BNS system but not the tidal deformabilities of each of the companions \cite{Hinderer:2009ca, Damour:2012yf}. Inferring the radius of each companion would require additional assumptions that may not be valid \cite{Kastaun:2019bxo, Godzieba:2020bbz}. BNS mergers with SNRs in excess of several hundreds, which will be rare but abundant in the NG era, will enable accurate inference of the tidal deformabilities of both NSs without requiring additional assumptions \cite{Smith:2021bqc}. Thus, we expect NG observatories to measure radii of NSs to within a few hundred meters and infer the equation of state of cold dense nuclear matter to a high degree of precision \cite{Maggiore:2019uih}. The post-merger oscillations could also carry the signature of the \emph{hot} dense matter equation of state that could be determined by NG observatories by accurately measuring the complex oscillation frequencies \cite{Takami:2014zpa, Bose:2017jvk, Breschi:2019srl, Easter:2020ifj}.

The remnant that forms after the coalescence of a BNS could sometimes be a hypermassive NS with core densities possibly exceeding several times the nuclear density \cite{Shibata:2005ss,Shibata:2003ga}. At such densities, matter could undergo quark-deconfinement phase transitions \cite{Prakash:1995uw, Prakash:1995uw}, from the hadronic phase to quark-gluon plasma, and this signature might be present in the post-merger GWs emitted by the hypermassive remnant \cite{Most:2018eaw, Blacker:2020nlq}. NG networks would observe post-merger signals with SNRs of 20 or more \cite{Srivastava:2022slt} depending on the equation-of-state and thereby shed light on the QCD (quantum chromodynamics) phase transition.

\paragraph{Testing general relativity}
General relativity is consistent with laboratory experiments and astronomical observations over a wide range of field strengths \cite{Will:2014kxa}. Yet, the theory raises a number of fundamental questions that have not found satisfactory answers \cite{Berti:2015itd, Yunes:2013dva}. These include BH information loss \cite{Unruh:2017uaw} and non-unitary evolution of quantum states \cite{Hawking:1976ra}, the late-time accelerated expansion of the Universe and the nature of the cosmological constant or dark energy \cite{Riess:1998cb, Perlmutter:1998np}, BH and big bang singularities that pose a major conceptual hurdle in predictability \cite{Hawking:1969sw}, to name a few. By directly probing BH horizons and the way remnant objects approach their final state it will be possible to probe predictions of GR to higher precision \cite{Berti:2007zu}. NG observatories with hundreds of high-SNR BBH events with {\red $\rm SNR>300$} will not only detect many subtle effects predicted in GR but allow precision tests of the theory \cite{Perkins:2020tra}. For example, by measuring the final state of the BH and comparing it to the properties of the progenitor binary when the companion stars are widely separated it will be possible to test strong-field predictions in the full non-linear GR \cite{Ghosh:2016qgn, Ghosh:2017gfp}. Additionally, it will be possible to test predictions of alternative gravity theories invoked to explain the Universe's recent (i.e. $z\lesssim 1$) accelerated expansion, the presence of dipole radiation, and constraints on the Brans-Dicke parameter \cite{Abernathy:2010, Zhang:2017sym}. With signals that arrive from very large redshifts ($z\gtrsim 10$) it will be possible to set tighter bounds on the graviton mass \cite{Arun:2009pq} and Lorentz violations \cite{Samajdar:2017mka}.

\paragraph{Probing population III stars and primordial black holes} 
Next generation observatories will have access to events at redshifts larger than 15---an epoch when the first stars were formed from primordial hydrogen and helium and devoid of heavier elements \cite{Schaerer:2001jc}. Detecting merging BBHs from that epoch could shed light on the properties of population III stars, especially their mass spectrum and redshift distribution. NG observatories will have the unique capability to not only detect a majority of the binaries from this epoch (with up to 70\% efficiency at a redshift of $z=20$) but decisively measure distances and infer their redshift \cite{Chen:2019irf, Ng:2021sqn}. Indeed, black holes formed in the early Universe (primordial black holes) have been proposed to be sources of gravitational waves observed by LIGO and Virgo (see, e.g., Ref.\,\cite{Clesse:2015wea, Sasaki:2016jop, Carr:2016drx, Garcia-Bellido:2017fdg} for a discussion of the origin of primordial black holes), and NG detectors would have the best sensitivity high-redshift BBH coalescences in the mass range 10-100 $M_\odot,$ at redshifts $z=20$-50, when no population III stars could have existed \cite{Kalogera:2019sui, Ng:2020qpk}. The first BHs from this era could well be the ones that seeded the supermassive BHs found in the nuclei of most galaxies today (see, e.g., Ref.\,\cite{Volonteri:2002vz} for a discussion of the origin of supermassive black holes). NG detectors could help understand the formation and growth of seed BHs through early cosmic history if their masses lie below about 100 $M_\odot$ \cite{Natarajan:2019ipd, Barack:2018yly, Volonteri:2010wz, Valiante:2020zhj, Volonteri:2021sfo}.

\section{Conclusion}\label{sec:conclusion}

\subsection{Summary of results}

\paragraph{Visibility of full cosmic populations} Due to the vastly different reaches, see Fig.~\ref{fig:detection_efficiency_rate} and Tab.~\ref{tab:reach}, the visibility of BNS mergers differs greatly between the three generations of observatories, see Tab.~\ref{tab:rates_snr}. These range from detections of $\rho\geq10$-events every other day with A+ over $\rho\geq30$-events once a week with Voyager to at least one BNS signal with SNR $\rho\geq100$ every other day in NG networks containing a CE detector. The lower sensitivity of ET results in less frequent detections, once every week, of such loud BNS events in \et.

Since all networks have farther reaches for BBHs signals, the BBH detection rates outpace the BNS by at least one order of magnitude in the networks without a CE detector---yielding two $\rho\geq30$-events every three days with A+, weekly $\rho\geq100$-events with Voyager, and almost six $\rho\geq100$-events per day with a single ET---while the rates of networks containing a CE detector are bound by the cosmic merger rate, \emph{CE networks will observe nearly all BBHs up to redshift $z=10$}. These networks, containing either one CE, one CE and ET, or two CEs and one ET, will observe about approximately 20 to 50 BBH signals with $\rho\geq100$ per day. As such the differences in other metrics are discriminating factors for BBH detections.

\paragraph{Measurement quality - three-dimensional localization} While the sky localization and distance estimation metrics generally follow the tendencies of the networks' visibility metric, see Tab.~\ref{tab:rates_sky_area_DL}, there are two stark exceptions. The NG networks containing either only one CE or only one ET exchange their roles: while ET has a lower sensitivity for BNS and BBH mergers than of CE, its low-frequency sensitivity and geometry are very advantageous for sky localization and also distance estimation.

Hence we can expect the following sky localization rates for BNS mergers from the examined networks: The A+ detector network will only observe consistently events to within $\Omega_{90} \le 1\,\rm deg^2$ a few times per year, while \ce, \voy, and \et\ should achieve this for 10--30 BNS signals per year. Consistent detection of BNSs localized to within $\Omega_{90} \le 0.1\,\rm deg^2$ requires at least two NG detector sites and even then \EC\ will only observe a handful of such events per year, while \ECS\ will push this number to more than twice a week, and only \ECS\ will achieve sky localizations of BNS signals to better than $\Omega_{90} \le 0.01\,\rm deg^2$, but only for rare, {\em golden} events.

While coalescences of BNSs are expected to be more abundant compared to BBHs, the intrinsic loudness of BBH mergers increases the detection rates of well-localized events: A+ and single CE networks will localize a few events per year to within $\Omega_{90} \le 0.1\,\rm deg^2$, while we should observe such signals once every four weeks in Voyager networks and once every other week in \et. Finally, BBHs localized to within $\Omega_{90} \le 0.01\,\rm deg^2$ are consistently detected only in networks containing two or three NG detectors, with \EC\ and \ECS\ observing such events a few times per year to more than once a week, respectively! Thus, while the visibility did not allow for conclusive discrimination of the networks containing a CE detector, the sky resolution clearly favors those with at least two, preferably three NG detectors.

Finally, luminosity distance estimation accuracies better than 10\% show a wide range for BNS mergers: once every five week with A+, once a week with Voyager or one CE, and 5, 18, and 41 times per day in \et, \EC, and \ECS, respectively. BBH signals will push these numbers to more than once a day with A+, 9 times a day with Voyager, and at least 27 times per day in NG networks. In fact such networks will detect BBHs to sub-1\% accuracies in the luminosity distance once a week with one CE and up to three times a day with three NG detectors.

\paragraph{Enabling multi-messenger astronomy} The synergy of GW and EM observations was beautifully demonstrated with the event GW170817. As such each network's potential to enable a follow-up in the EM spectrum---even without the detection of a loud EM transient such as a GRB---is paramount in determining the network's science capabilities. The main metric is thus the sky localization in relation to the FOVs of various EM telescopes, see Tab.~\ref{tab:em_fov}. We performed this study for BNS and BBH signals emitted from redshifts up to $z=0.5$.

Due to the relative quietness of BNS signals, the loudest, best-localized events of the cosmic BNS population already stem from the population at $z\le0.5$ and the conclusions, drawn for the cosmic BNS population above, also apply to the low-$z$ population. Hence, the A+, Voyager, and single NG detector network predominantly cater to the Rubin Observatory for follow-up surveys with a few to tens sub-$1\,{\rm deg^2}$ detections per year. In fact, these networks should further be able to localize rare, {\em golden} BNS signals to within $0.1\,{\rm deg^2}$ once per year, thus enabling EM surveys with EUCLID, Athena, Nancy Grace Roman, Chandra, Lynx, Swift-XRT, Keck and the 30 m-Telescope with such high-fidelity events. Networks with two or three NG detectors will deliver few to one hundred of such events to the aforementioned observatories. The triple-NG network should be able to localize rare, {\em golden} BNS signals to within $0.01\,{\rm deg^2}$, thus enabling EM follow-up with VLT and ELT.
In fact, the triple-NG network is the only network to observe {\em golden} BNS mergers that are well-localized both on the sky ($\Omega_{90}\leq0.01\,{\rm deg^2}$) and in distance ($\Delta D_L/D_L<0.01$), see Fig. \ref{fig:mma_BNS}.

BBH detections lack EM counterparts and therefore depend on the sky localization from the GW signal to identify the binary's host galaxy. Fortunately, the intrinsic loudness of massive systems and the signal contributions from higher modes for mass-asymmetric binaries improve the sky localization estimates across all six networks, allowing them to localize BBH signals to within $0.1\,{\rm deg^2}$ a few times per year (A+ and one CE) and every fourth (Voyager) or second (one ET) week. Thus, starting with the Voyager generation, frequent EM follow-up surveys are possible with the following telescopes: Rubin Observatory, EUCLID, Athena, Nancy Grace Roman, Chandra, Lynx, Swift-XRT, Keck and the 30 m-Telescope. In fact the double- and triple-NG networks will observe such well-localized events every other to more than once a day. They will further provide VLT and ELT with $\Omega_{90}\leq0.01\,{\rm deg^2}$-events a few times per year to more than once a week. The triple-NG network will consistently measure the three-dimensional localization of a few {\em golden} BBH events to better than $\Omega_{90}\leq0.001\,{\rm deg^2}$ and $\Delta D_L/D_L<0.01$, see Fig. \ref{fig:mma_BBH}, allowing for follow-up surveys by GMT and HST--WFC3 and enabling single-event, high-precision cosmology with dark sirens \cite{Borhanian:2020ypi}.

One particularly intriguing aspect of multi-messenger astronomy with GWs and EM radiation is the potential to trigger EW alerts for BNS coalescences ahead of the actual merger, thus enabling the EM observatories to record the events as early as possible and observe the merger in the EM spectrum as it happens. The important metric here, besides visibility, is the sky localization of the events several minutes before the merger to provide enough time for GW signal pipelines to issue an alert and telescopes to slew to the estimated sky location. In our study we found that A+ and Voyager detectors will not provide significant detection rates with the required sky localization at least 2 minutes before the merger. Similarly, a single CE network might only observe a couple such events per year early enough for the Rubin Observatory to follow-up, assuming 10-fold tiling. In contrast, the ET network should send weekly two- and biweekly five-minutes alerts. The double- and triple-NG networks on the other hand will push these numbers to 5--13 daily two-minutes, 2--6 daily five-minutes, and {\em approximately weekly ten-minutes} alerts to the Rubin Observatory. In fact, both networks would even provide EUCLID, Athena, Nancy Grace Roman, Chandra, Lynx, Swift-XRT, Keck and the 30 m-Telescope with two-minutes and 5-minutes EW alerts for a handful of golden events per year, see Tab.~\ref{tab:ew}.

\subsection{Limitations of the study} \label{sec:limitations}

The main caveat of this study is the use of the Fisher information formalism to provide measurement quality estimates. The formalism is well-known and tested, see \cite{Vallisneri:2007ev} for a review of its shortcomings and \cite{Wang:2022kia,Dupletsa:2024gfl} for two approaches to improve the reliability of the Fisher-based error estimates based on Derivative Approximation for Likelihoods and the incorporation of priors for the estimates, respectively. Yet, the formalism only provides estimates for Gaussian posteriors which is likely not the best assumption for the noise of these detectors; especially for signals at visibility threshold. Further, the reliance on numerical derivatives for LAL waveforms and the numerical inversion of the Fisher matrix are sources for numerical uncertainties affecting the quoted results. Further, we did not examine the quality of spin measurements since the addition of such parameters in the Fisher analysis lead to a high rate of ill-conditioned Fisher matrices for which the numerical inversion is not to be trusted.

Lastly, the chosen population distributions---chosen to be consistent with the LIGO and Virgo observations both in mass and redshift distributions---do not capture unexpected sources such as large merger populations of population III star remnants or primordial BHs at large redshifts beyond $z=10$. Besides, the chosen populations only include injections for which the component spins and the binary's orbital angular momentum are aligned and the chosen waveforms are non-precessing; thus the effects of spin-precession are not accounted for in the forecasts made in this study.

\subsection{Outlook and further studies}

Ultimately we can conclude that while the A+ and Voyager upgrades would do a tremendous job to increase the current detector facilities' lifespan and science capabilities. Yet, only two- and three-site NG networks will expand the detection reach significantly to observe binary black hole mergers from the edge of the observable Universe---a regime inaccessible to electromagnetic observations. These networks would be capable to observe almost every binary coalescences up to redshift $z=0.9$ and thus provide abundant detection rates enabling scientists to examine binary progenitor population and formation channels, map the large-scale structure of the universe, perform high-precision cosmology and tests of GR, etc. Further, the three-dimensional localization capabilities of such networks should enable a host of electromagnetic telescopes to not only follow-up the detections searching for counterparts and host galaxies, but actually even alert these observatories minutes before the actual mergers in the case of BNSs, allowing astronomers to record the mergers in the electromagnetic spectrum as it happens. Finally, the large rates would further provide redundancies for follow-up surveys to ensure that enough gravitational-wave events can be examined in the electromagnetic window when accounting for maintenance and already reserved observation time. Hence, the planned and proposed detector updates and new facilities will be an important addition for the fundamental physics, astrophysics and cosmology communities.

\begin{acknowledgements}
We thank the members of the Cosmic Explorer project team, in particular Stefan Ballmer, Duncan Brown, Matthew Evans, Anuradha Gupta, Evan Hall, Kevin Kuns, Philippe Landry, Geoffrey Lovelace, Jocelyn Read, David Shoemaker, Josh Smith, Rory Smith, Varun Srivastava, Salvatore Vitale, for many helpful conversations. We thank Ryan Magee in aiding us to verify and correct the sky area estimation of \textsc{gwbench}. We further thank Marica Branchesi, Samuele Ronchini, Michele Maggiore, Francesco Iacovelli, Michele Mancarella, and Stefano Foffa for confirming that the network efficiency and sky resolution of a network containing ET reported in this study is in good agreement with their own findings. S. B. acknowledges support from NSF Grant No. PHY-1836779, the Deutsche Forschungsgemeinschaft (DFG), Project MEMI No. BE6301/2-1, ERC Starting Grant No.~945155--GWmining, Cariplo Foundation Grant No.~2021-0555, MUR PRIN Grant No.~2022-Z9X4XS,  MUR Grant ``Progetto Dipartimenti di Eccellenza 2023-2027'' (BiCoQ), and the ICSC National Research Centre funded by NextGenerationEU.  B. S. S. was supported in part by NSF Grants No. PHY-1836779, No. PHY-2012083, No. AST-2006384, No. PHY-2207638, No. AST- 2307147, No. PHY-2308886, and No. PHYS-2309064.
\end{acknowledgements}

\appendix

\section{\textsc{gwbench} settings}\label{app:settings}

We used the waveform wrapper \texttt{wf\_model\_name = 'lal\_bns'} with the waveform model \texttt{wf\_other\_var\_dic = \{'approximant':'IMRPhenomD\_NRTidalv2'\}} for BNS injections and the wrapper \texttt{wf\_model\_name = 'lal\_bbh'} together with \texttt{wf\_other\_var\_dic = \{'approximant':'IMRPhenomXHM'\}} for BBH injections. The numerical derivatives were computed using step sizes of \texttt{step = 1e-9} and \texttt{step = 1e-6} for \texttt{IMRPhenomD\_NRTidalv2} and \texttt{IMRPhenomXHM}, respectively. The derivative method was set to \texttt{method = 'central'} or  \texttt{method = 'backward'} in case the chirpmass was close to 0.25 while the method and derivative orders were set to \texttt{order = 2} and \texttt{d\_order\_n = 1}, respectively. Further we set \texttt{use\_rot = 1}, \texttt{conv\_cos = None}, \texttt{conv\_log = ('Mc','DL','lam\_t')}. The injection sets were generated using the \texttt{injections} module of \texttt{gwbench}: \texttt{injections.injections\_CBC\_params\_redshift} with \texttt{redshifted = 1}. For BNSs we used the following seeds for the various redshift bins (zmin, zmax, seed): (0, 0.5, 7669), (0.5, 1, 3103), (1, 2, 4431), (2, 4, 5526), (4, 10, 7035), and (10, 50, 2785). For BBHs we used (0, 0.5, 5485), (0.5, 1, 1054), (1, 2, 46), (2, 4, 5553), (4, 10, 5998), and (10, 50, 4743).

\section{Data release}\label{app:data}
We will share the raw data for a number of networks---including the ones presented in this paper---together with the data used in a different, ongoing study once we publish the results of that study.

\begin{table*}
    \centering
    \caption{The best-fit parameters $a$, $b$, and $c$ of the sigmoid function, $f(z) = \left [(1+b)/(1+b\,e^{az}) \right ]^{c}$, fitted to the detection efficiency $\epsilon(z,\,\rho_*)$ curves of detector networks in Fig.~\ref{fig:detection_efficiency_rate} for two values of the SNR threshold $\rho_*=10$ and $\rho_*=100.$}
    \begin{tabular*}{\textwidth}{@{\extracolsep{\fill}}l|c c c|c c c}
    \hline\hline
     & \multicolumn{3}{c|}{$\rho_*=10$ \phantom{AA}} & \multicolumn{3}{c}{$\rho_*=100$ \phantom{AA}}  \\
    \hline
    Sigmoid parameter & $a$    & $b$    & $c$   & $a$    & $b$    & $c$      \\
    \hline
    \hline
    & \multicolumn{6}{c}{\emph{BNS}} \\
    \hline
    \hline
\aplus   & 64.62 & 0.005701 & 0.3495 & 743.5 & 0.003345 & 0.3067 \\
\voy     & 28.98 & 0.01056 & 0.3125 & 299.6 & 0.008898 & 0.3247 \\
\et      & 14.95 & 0.006631 & 0.1549 & 129.1 & 0.0088 & 0.2344 \\
\ce      & 5.471 & 0.1155 & 0.2049 & 40.74 & 0.07955 & 0.5162 \\
\EC      & 6.708 & 0.01416 & 0.1403 & 52.8 & 0.01481 & 0.3585 \\
\ECS     & 5.834 & 0.00928 & 0.09996 & 44.96 & 0.01035 & 0.3129 \\    \hline
    \hline
    & \multicolumn{6}{c}{\emph{BBH}} \\
    \hline
    \hline
\aplus   & 25.11 & 0.002056 & 0.08482 & 210.9 & 0.004143 & 0.1138 \\
\voy     & 14.74 & 0.001824 & 0.04888 & 95.27 & 0.006636 & 0.1123 \\
\et      & 14.2 & 5.08e-05 & 0.007811 & 43.1 & 0.004653 & 0.07757 \\
\ce      & 14.2 & 0.03823 & 0.003638 & 16.46 & 0.08902 & 0.1202 \\
\EC      & 2.792 & 0.006192 & 0.01262 & 19.55 & 0.00885 & 0.08601 \\
\ECS     & 0.1211 & 0.4108 & 0.2391 & 17.22 & 0.004554 & 0.06978 \\
    \hline
    \hline
    \end{tabular*}
    \label{tab:sigmoid_fit_params}
\end{table*}

\begin{table*}
    \centering
    \caption{The best-fit parameters $a$, $b$, and $c$ of the sigmoid function, $f(z) = \left [(1+b)/(1+b\,e^{az}) \right ]^{c}$, fitted to the detection efficiency $\epsilon(z,\,\rho_*)$ curves of detector networks in Fig.~\ref{fig:detection_efficiency_rate_ce_configs} for two values of the SNR threshold $\rho_*=10$ and $\rho_*=100.$}
    \begin{tabular*}{\textwidth}{@{\extracolsep{\fill}}l|c c c|c c c}
    \hline\hline
     & \multicolumn{3}{c|}{$\rho_*=10$ \phantom{AA}} & \multicolumn{3}{c}{$\rho_*=100$ \phantom{AA}}  \\
    \hline
    Sigmoid parameter & $a$    & $b$    & $c$   & $a$    & $b$    & $c$      \\
    \hline
    \hline
    & \multicolumn{6}{c}{\emph{BNS}} \\
    \hline
    \hline
\cetw    & 8.162 & 0.08435 & 0.2873 & 70.18 & 0.07338 & 0.4655 \\
\cefo    & 5.471 & 0.1155 & 0.2049 & 40.74 & 0.07955 & 0.5162 \\
\cetwfo  & 6.161 & 0.02666 & 0.153 & 48.81 & 0.02272 & 0.3887 \\
\cefofo  & 5.384 & 0.02244 & 0.127 & 43.41 & 0.01781 & 0.3592 \\
    \hline
    \hline
    & \multicolumn{6}{c}{\emph{BBH}} \\
    \hline
    \hline
\cetw    & 17.42 & 0.005799 & 0.007106 & 24.83 & 0.07936 & 0.1399 \\
\cefo    & 14.2 & 0.03823 & 0.003638 & 16.46 & 0.08902 & 0.1202 \\
\cetwfo  & 2.077 & 0.04043 & 0.01891 & 18.02 & 0.01749 & 0.09498 \\
\cefofo  & 0.06361 & 0.4927 & 0.719 & 15.64 & 0.0136 & 0.08679 \\
    \hline
    \hline
    \end{tabular*}
    \label{tab:sigmoid_fit_params_ce_configs}
\end{table*}

\section{Sigmoid best-fit parameters}\label{app:sigmoid}

Tabs.~\ref{tab:sigmoid_fit_params} and \ref{tab:sigmoid_fit_params_ce_configs} record the best-fit parameters $a$, $b$, and $c$ of the sigmoid curves fitted to the detector efficiencies $\epsilon(z)$ in Fig.~\ref{fig:detection_efficiency_rate} Sec.~\ref{sec:rate} and Fig.~\ref{fig:detection_efficiency_rate_ce_configs} in App.~\ref{app:ce_configs}, respectively.

\section{Visibility and measurement quality: histograms} \label{app:pdf}

\begin{figure*}
\includegraphics[width=0.99\textwidth]{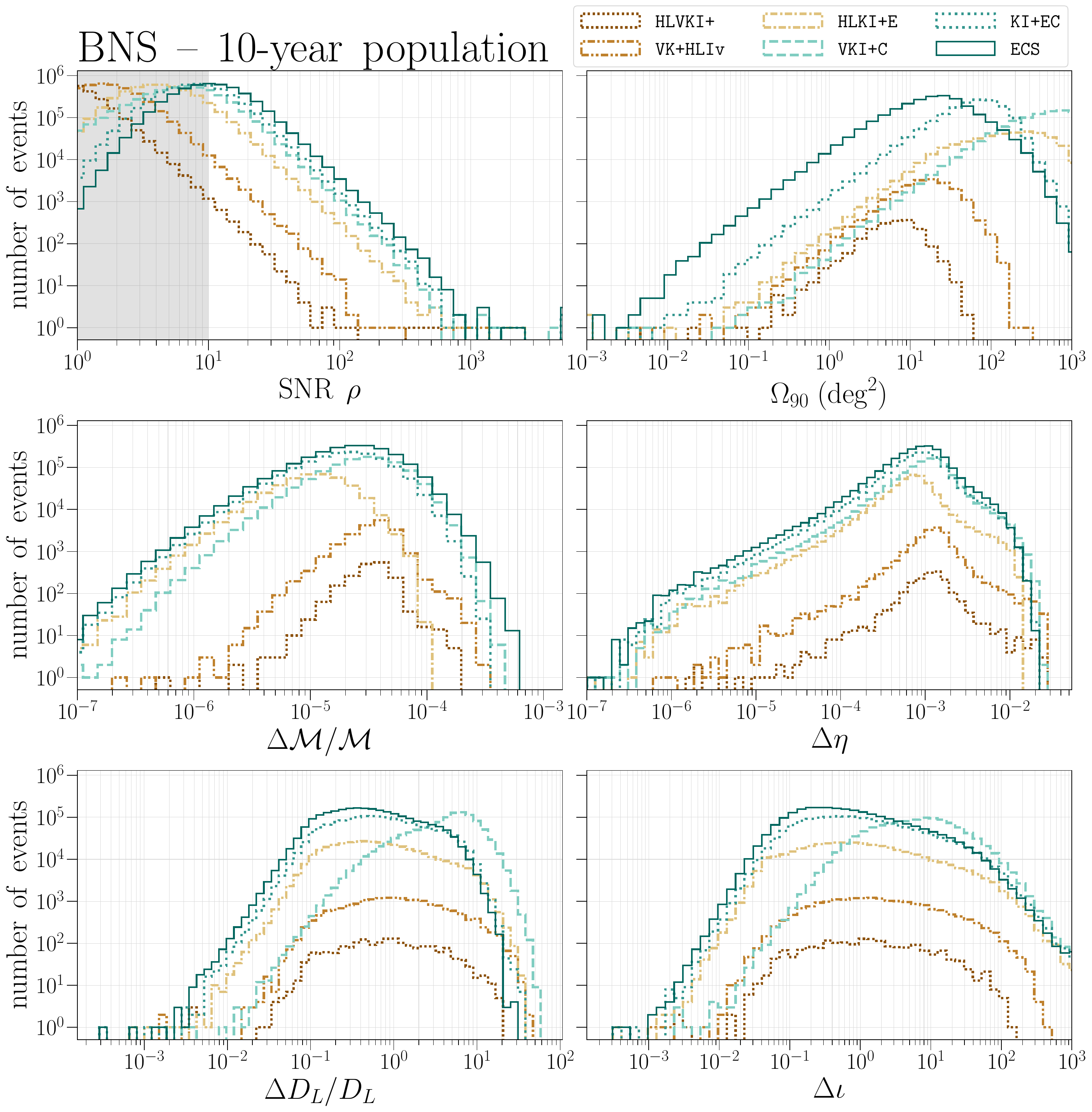}
\caption{Histograms of the \textbf{10 years} BNS injection sample used to generate Fig.~\ref{fig:BNS_visibility} for SNR $\rho$, 90\%-credible sky area $\Omega_{90}$, fractional errors on chripmass $\Delta \mathcal{M}/\mathcal{M}$ and luminosity distance $\Delta D_L/D_L$, and absolute errors on symmetric mass ratio $\Delta \eta$ and the inclination angle $\Delta\iota$ observed in the six studied A+, Voyager, and NG networks. The histograms were generated from $\sim 4.7\times10^{6}$ injections sampled according to Sec. \ref{sec:resampling}.  The non-SNR panels are obtained for events with SNR $\rho\geq10$, indicated by the non-shaded region in the SNR panel.}
\label{fig:BNS_visibility_pdf}
\end{figure*}

\begin{figure*}
\includegraphics[width=0.99\textwidth]{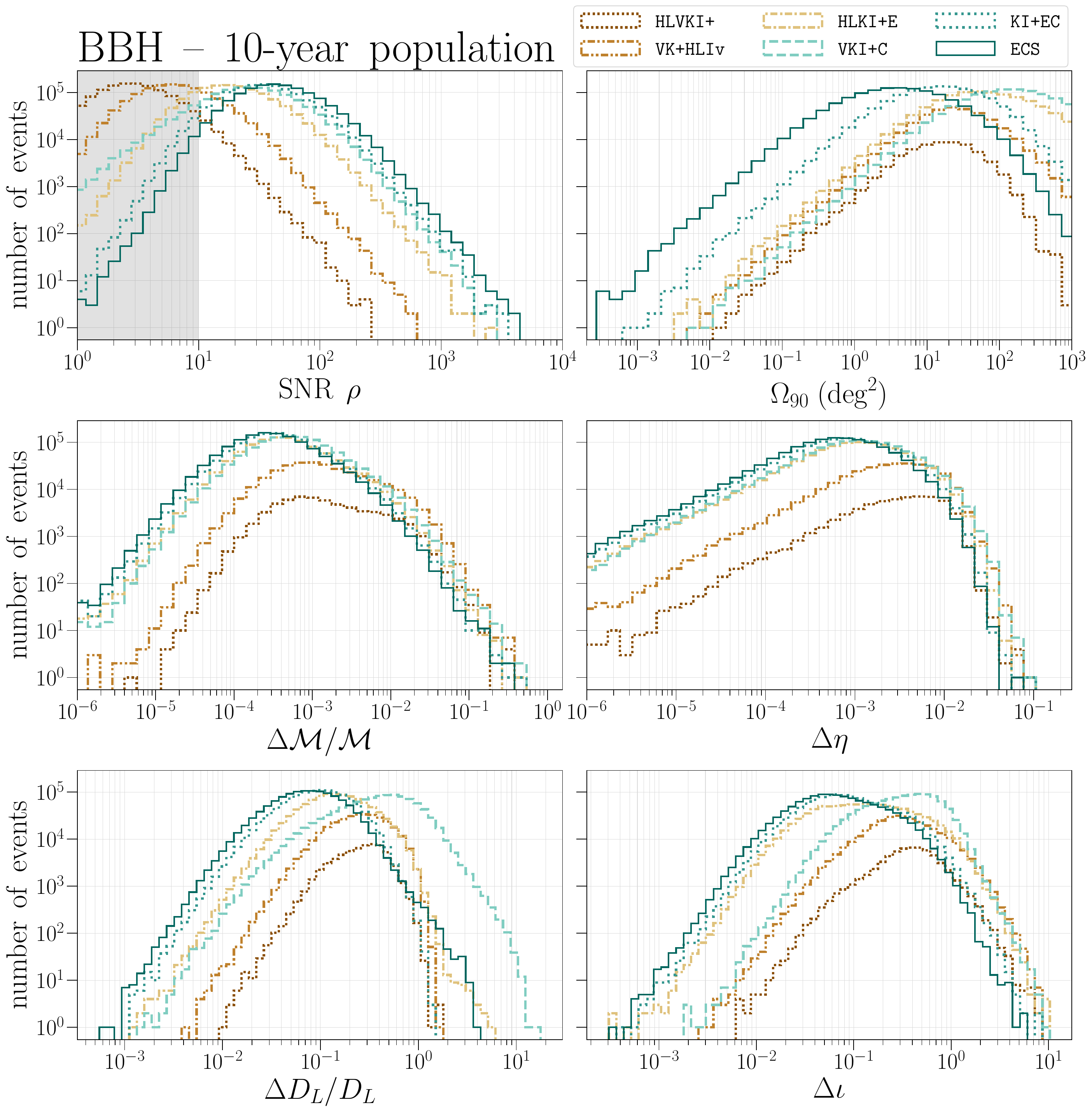}
\caption{Histograms of the \textbf{10 years} BBH injection sample used to generate Fig.~\ref{fig:BBH_visibility} for SNR $\rho$, 90\%-credible sky area $\Omega_{90}$, fractional errors on chripmass $\Delta \mathcal{M}/\mathcal{M}$ and luminosity distance $\Delta D_L/D_L$, and absolute errors on symmetric mass ratio $\Delta \eta$ and the inclination angle $\Delta\iota$ observed in the six studied A+, Voyager, and NG networks. The histograms were generated from $\sim 1.2\times10^{6}$ injections sampled according to Sec. \ref{sec:resampling}. The non-SNR panels are obtained for events with SNR $\rho\geq10$, indicated by the non-shaded region in the SNR panel.}
\label{fig:BBH_visibility_pdf}
\end{figure*}

In Sec.\,\ref{sec:visibility} we presented the cumulative density plots of the SNR, 90\%-credible sky area, and other parameters. Here we show the corresponding histograms for the respective parameters, presented on a log-log scale in Figs.~\ref{fig:BNS_visibility_pdf} and \ref{fig:BBH_visibility_pdf}. Since the total number of events accessible each year is in excess of a million for BNS and more than one-hundred thousand for BBH, we have shown the histograms over five orders of magnitude in density. This will help recognize events at the tail end of the distribution that cannot be easily inferred from the cumulative distribution. In order to minimize Monte Carlo errors we have used a number of events expected over a 10-year period. However, as discussed in Sec.\,\ref{sec:visibility}, numbers quoted in the text and various tables assume a 1-year observation period.

While events at the tail end of the distribution are rate, they would be very loud and their parameters will be measured with great precision. For example, only NG observatories have the potential to observe significant number of BNS (BBH) mergers with SNRs larger than 100 (1000) and localize sources to better than 10 arc min$^2$. We examine the science potential of these tail ends events in Sec.\,\ref{sec:rare_events}.

The density plots readily reveal the mode of the distributions, indicating where to expect most of the events to lie, and limitations of different networks, informing the best science return we can hope to extract. In fact, we see that the three generations (A+, Voyager, and NG) are qualitatively different with respect to every metric used in this study.

\section{Cosmic Explorer: Influence of Proposed Detector Configurations}\label{app:ce_configs}
\begin{figure*}
    \centering
    \includegraphics[width=0.5\textwidth]{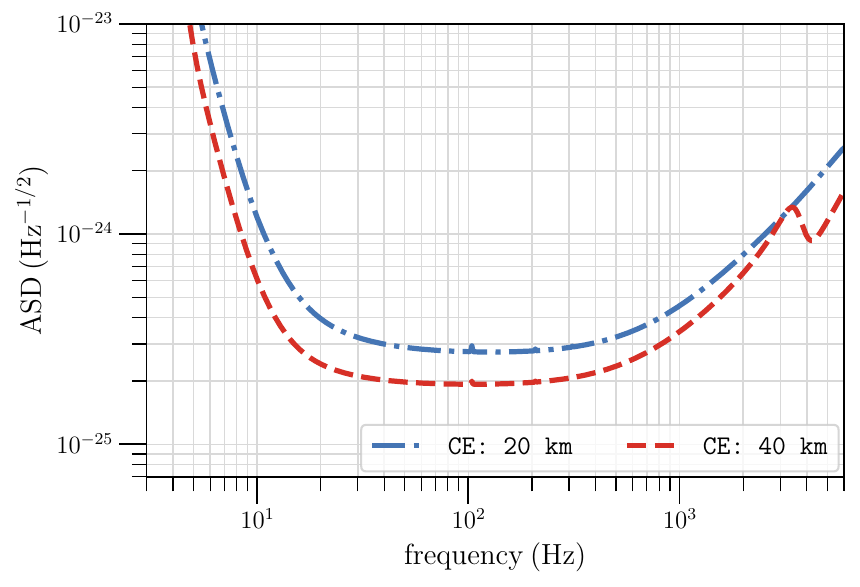}
    \caption{Sensitivity of the two examined Cosmic Explorer configurations with $20\,{\rm km}$ and $40\,{\rm km}$ arm length, respectively.
    The noise curves are taken from the \texttt{ce2\_20km\_cb} and \texttt{ce2\_40km\_cb} \texttt{.txt}-files inside \texttt{ce\_curves.zip} file at \href{https://dcc.cosmicexplorer.org/CE-T2000007}{https://dcc.cosmicexplorer.org/CE-T2000007}.}
    \label{fig:sensitivity_curves_ce_configs}
\end{figure*}

\begin{figure*}
\includegraphics[width=0.49\textwidth]{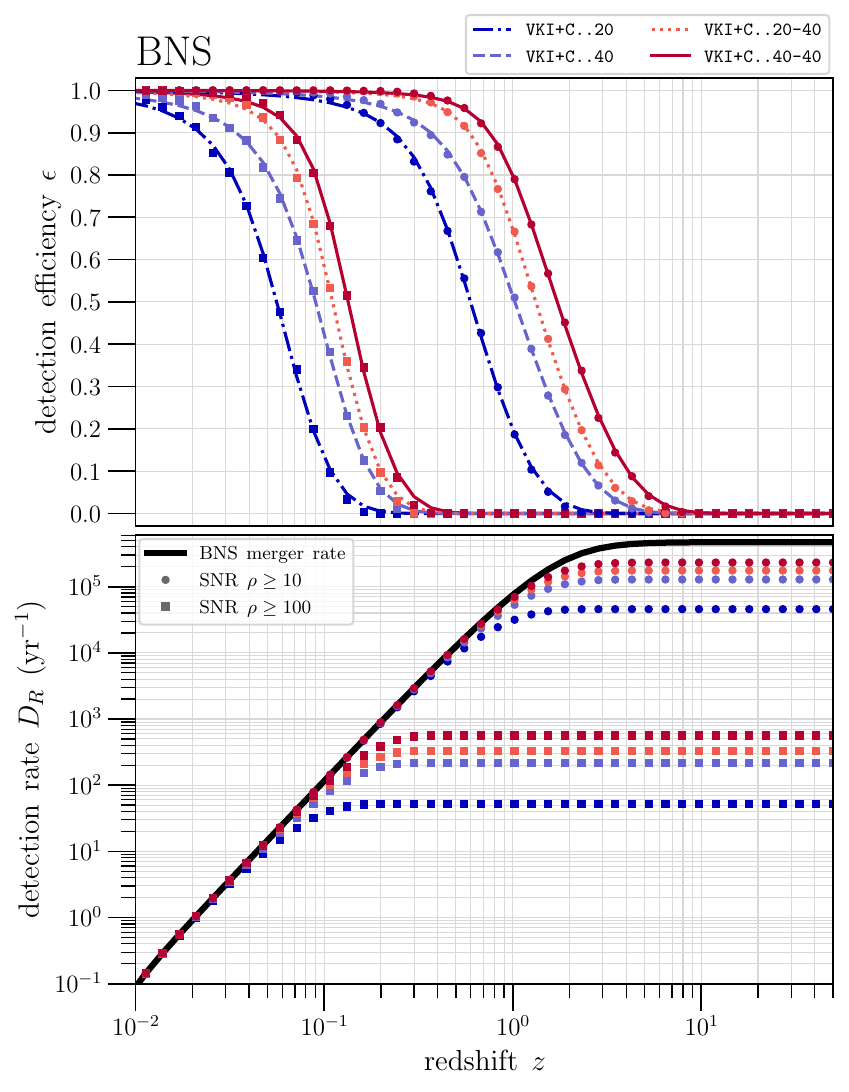}
\includegraphics[width=0.49\textwidth]{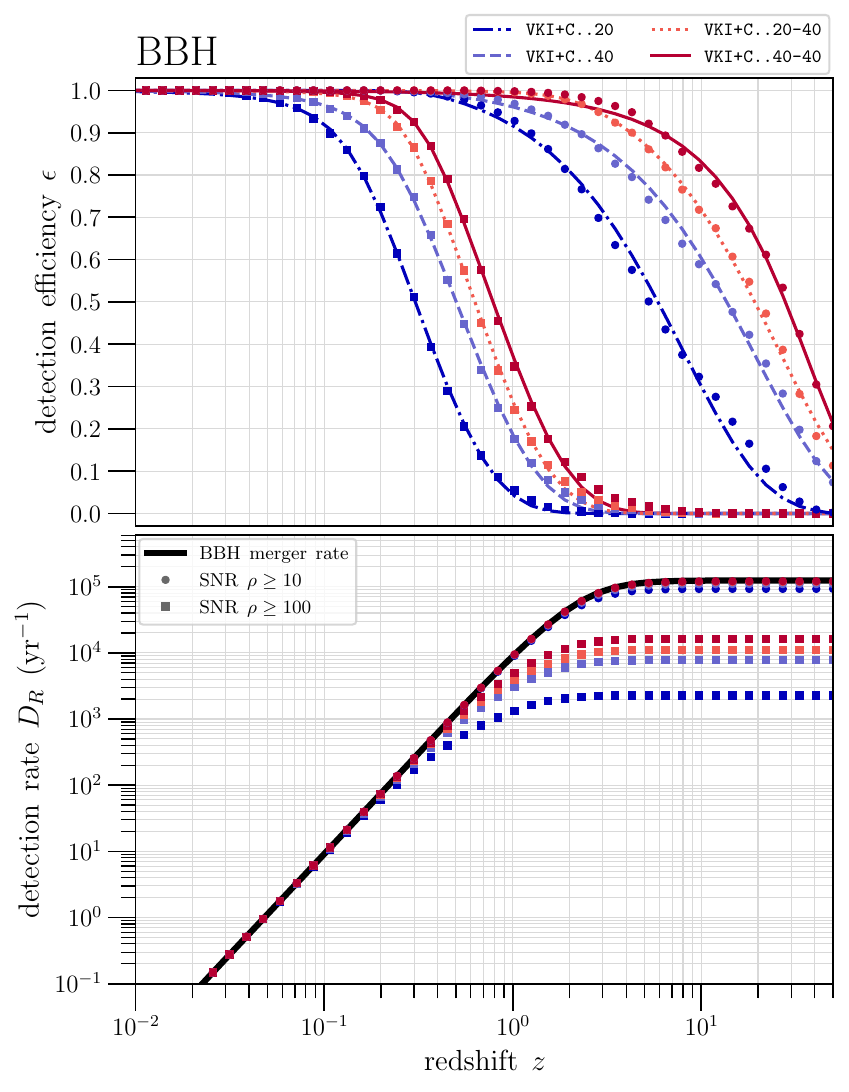}
\caption{Detection efficiencies $\epsilon$ and detection rates $D_R$, see Eqs.  \eqref{eq:efficiency} and \eqref{eq:det_rate}, respectively, of the four studied CE networks are plotted as functions of redshift $z$. The efficiency specifies the value at the indicated redshift $z$ while the detection rate corresponds to the integrated rate up to that redshift. The circles (squares) denote the values for events with SNR $\rho\geq10$ ($\rho\geq100$). The thick, black lines in the rate panels are the cosmic BNS and BBH merger rates, see Sec.~\ref{sec:rate}. The fit lines in the efficiency panels are sigmoid fits with $f_{\rm sigmoid}=\Big(\frac{1+b}{1+b\,\mathrm{e}^{ax}}\Big)^c$.}
\label{fig:detection_efficiency_rate_ce_configs}
\end{figure*}

\begin{figure*}
\includegraphics[width=\textwidth]{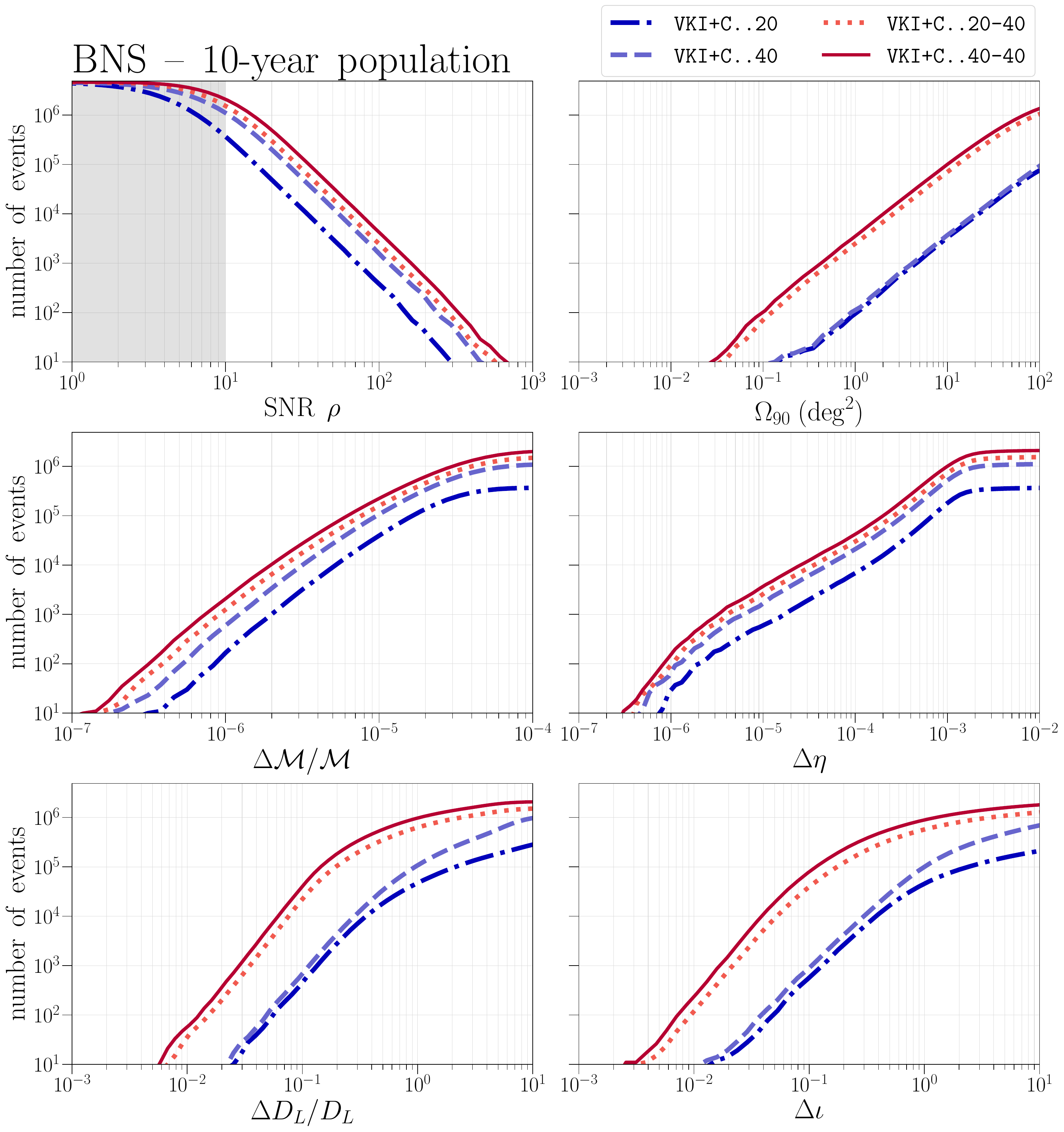}
\caption{Cumulative histograms for the SNR $\rho$, 90\%-credible sky area $\Omega_{90}$, fractional errors on chripmass $\Delta \mathcal{M}/\mathcal{M}$ and luminosity distance $\Delta D_L/D_L$, and absolute errors on symmetric mass ratio $\Delta \eta$ and inclination angle $\Delta\iota$ for BNS mergers observed in the four studied CE networks. The histograms were generated from $\sim 4.7\times10^{6}$ injections sampled according to Sec. \ref{sec:resampling}. The non-SNR panels are obtained for events with SNR $\rho\geq10$, indicated by the non-shaded region in the SNR panel. The SNR panel is flipped to highlight the behavior for large values.}
\label{fig:BNS_visibility_ce_configs}
\end{figure*}

\begin{figure*}
\includegraphics[width=\textwidth]{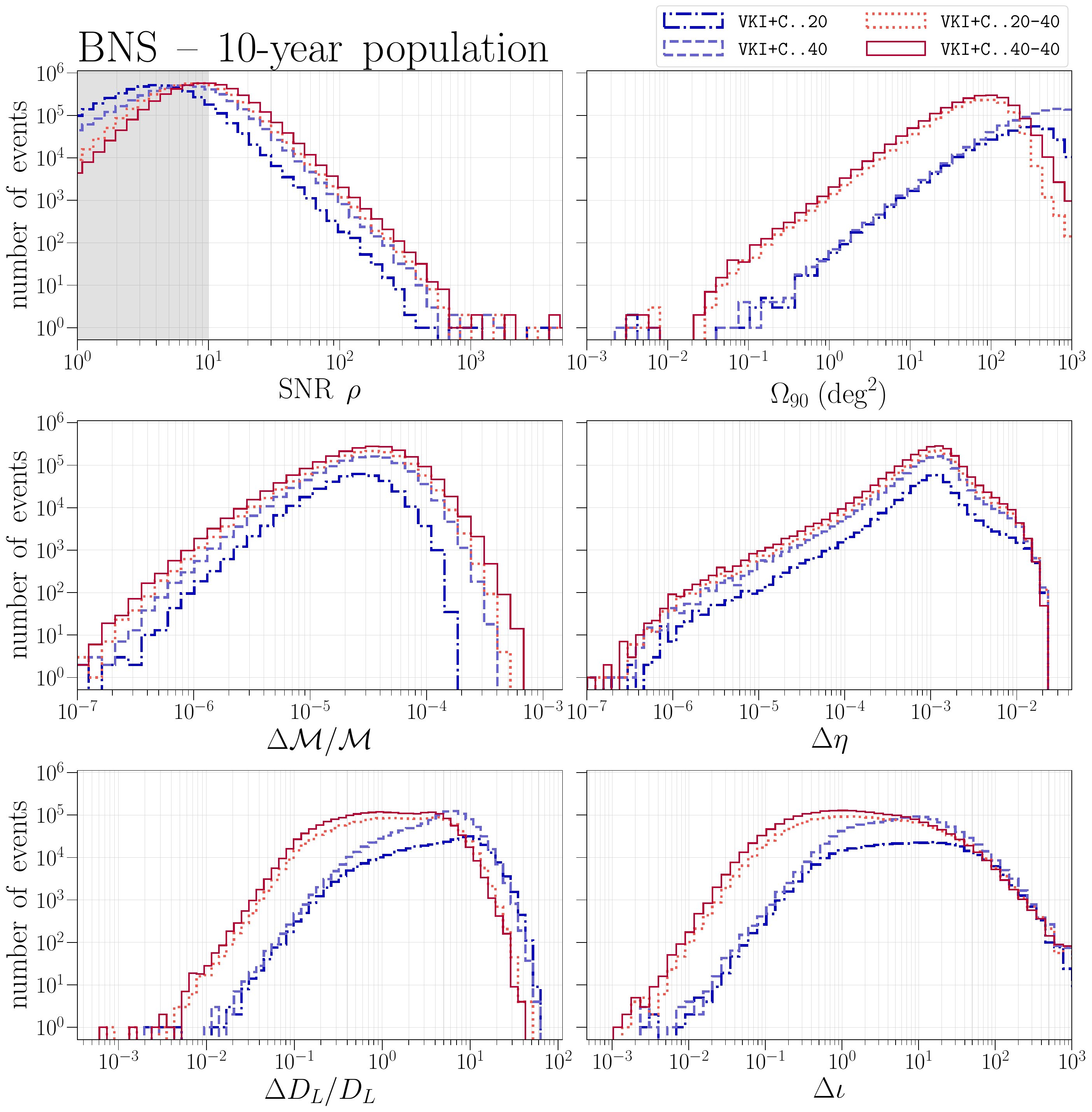}
\caption{Histograms of the \textbf{10 years} BNS injection sample used to generate Fig.~\ref{fig:BNS_visibility_ce_configs} for SNR $\rho$, 90\%-credible sky area $\Omega_{90}$, fractional errors on chripmass $\Delta \mathcal{M}/\mathcal{M}$ and luminosity distance $\Delta D_L/D_L$, and absolute errors on symmetric mass ratio $\Delta \eta$ and the inclination angle $\Delta\iota$ observed in the four studied CE networks. The histograms were generated from $\sim 4.7\times10^{6}$ injections sampled according to Sec. \ref{sec:resampling}. The non-SNR panels are obtained for events with SNR $\rho\geq10$, indicated by the non-shaded region in the SNR panel.}
\label{fig:BNS_visibility_pdf_ce_configs}
\end{figure*}

\begin{figure*}
\includegraphics[width=\textwidth]{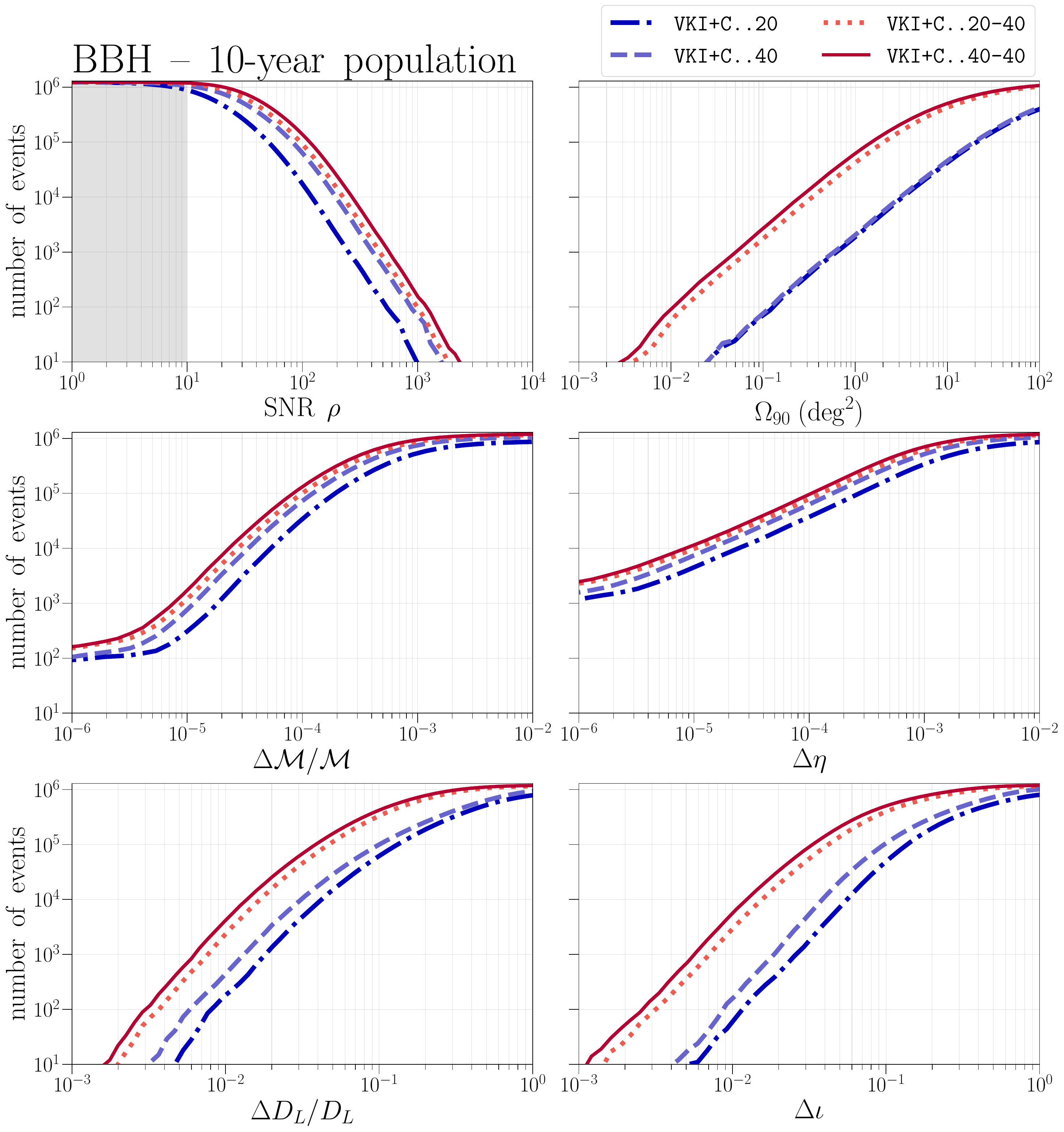}
\caption{Cumulative histograms for the SNR $\rho$, 90\%-credible sky area $\Omega_{90}$, fractional errors on chripmass $\Delta \mathcal{M}/\mathcal{M}$ and luminosity distance $\Delta D_L/D_L$, and absolute errors on symmetric mass ratio $\Delta \eta$ and inclination angle $\Delta\iota$ for BBH mergers observed in the four studied CE networks. The histograms were generated from $\sim 1.2\times10^{6}$ injections sampled according to Sec. \ref{sec:resampling}. The non-SNR panels are obtained for events with SNR $\rho\geq10$, indicated by the non-shaded region in the SNR panel. The SNR panel is flipped to highlight the behavior for large values.}
\label{fig:BBH_visibility_ce_configs}
\end{figure*}

\begin{figure*}
\includegraphics[width=\textwidth]{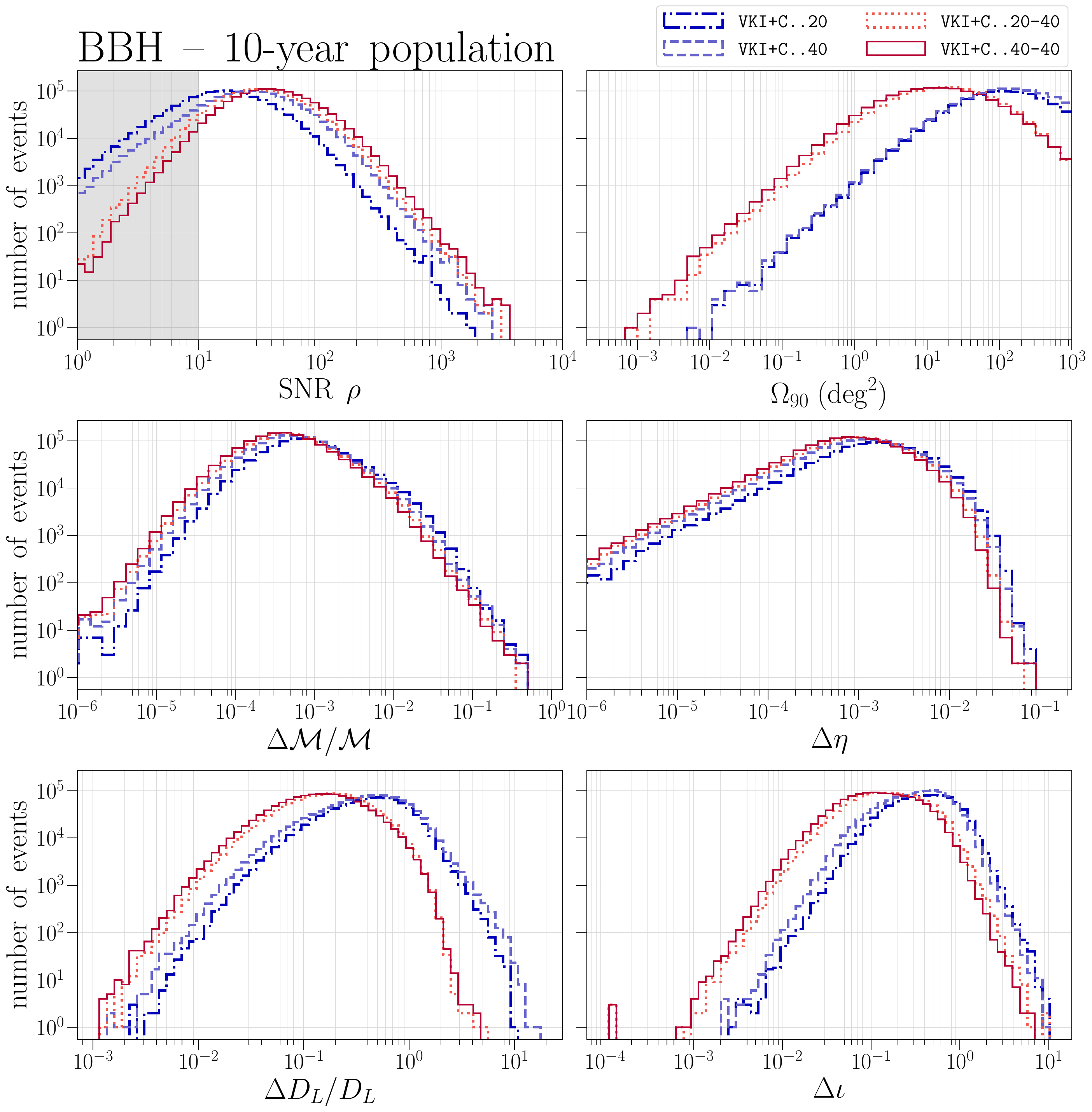}
\caption{Histograms of the \textbf{10 years} BBH injection sample used to generate Fig.~\ref{fig:BBH_visibility_ce_configs} for SNR $\rho$, 90\%-credible sky area $\Omega_{90}$, fractional errors on chripmass $\Delta \mathcal{M}/\mathcal{M}$ and luminosity distance $\Delta D_L/D_L$, and absolute errors on symmetric mass ratio $\Delta \eta$ and the inclination angle $\Delta\iota$ observed in the four studied CE networks. The histograms were generated from $\sim 1.2\times10^{6}$ injections sampled according to Sec. \ref{sec:resampling}. The non-SNR panels are obtained for events with SNR $\rho\geq10$, indicated by the non-shaded region in the SNR panel.}
\label{fig:BBH_visibility_pdf_ce_configs}
\end{figure*}

\begin{figure*}
\includegraphics[width=0.6\textwidth]{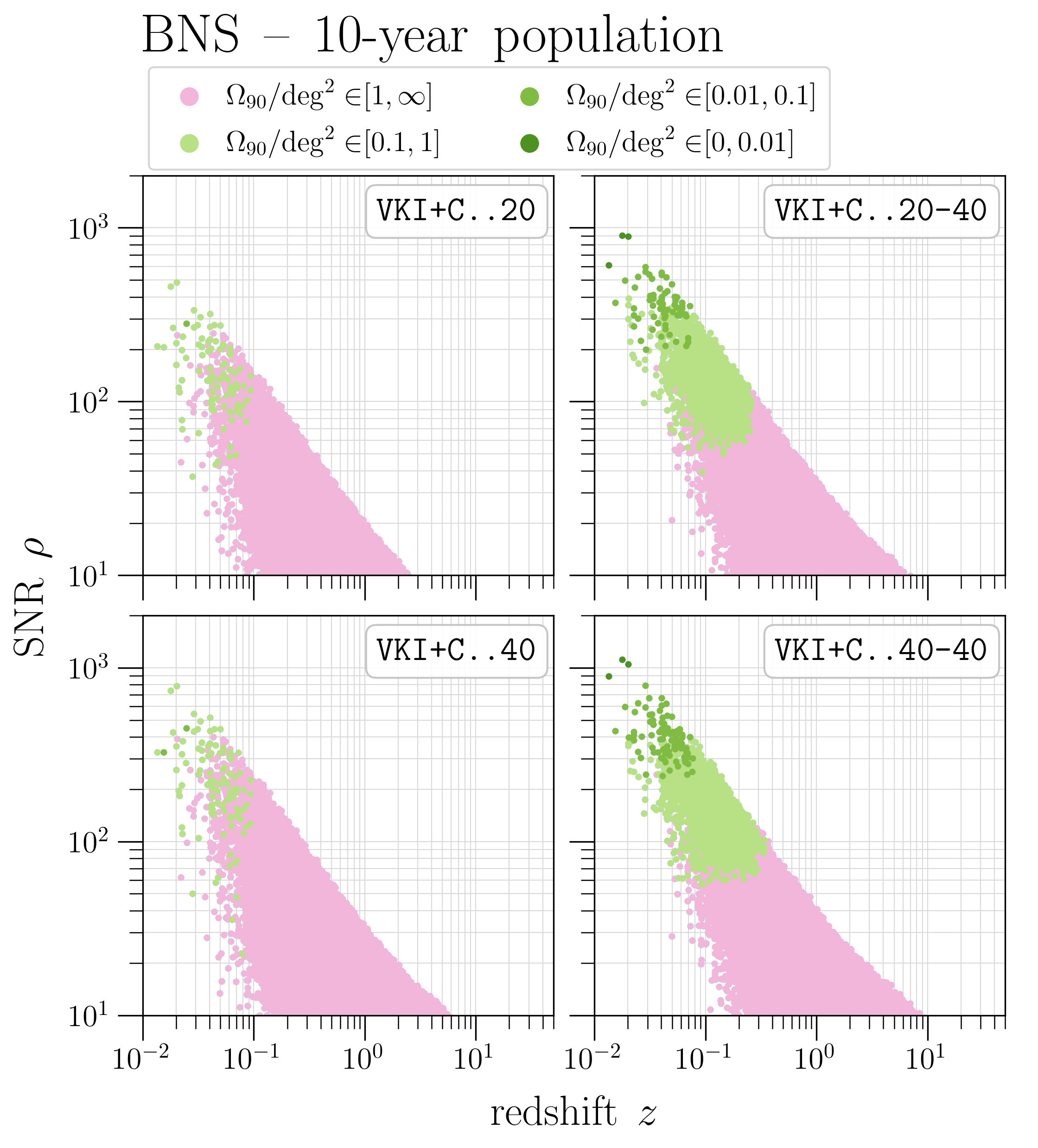}
\includegraphics[width=0.6\textwidth]{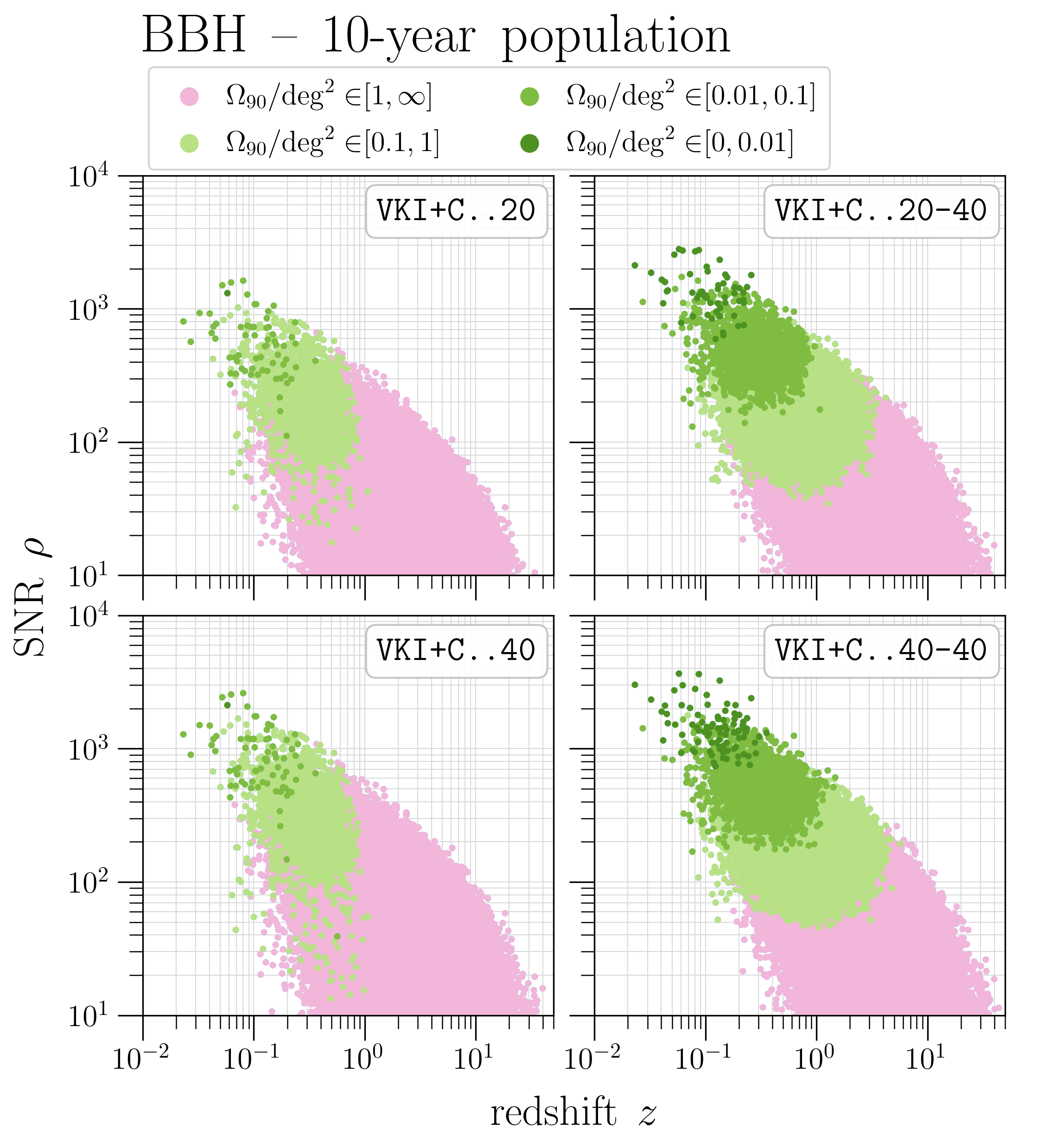}
\caption{The scatter plots illustrate the correlations between redshift $z$, SNR $\rho$, and 90\%-credible sky area $\Omega_{90}$ for BNS (\emph{top}) and BBH (\emph{bottom}) mergers in the four studied CE networks.}
\label{fig:visibility_scatter_ce_configs}
\end{figure*}

\begin{figure*}
\includegraphics[width=\textwidth]{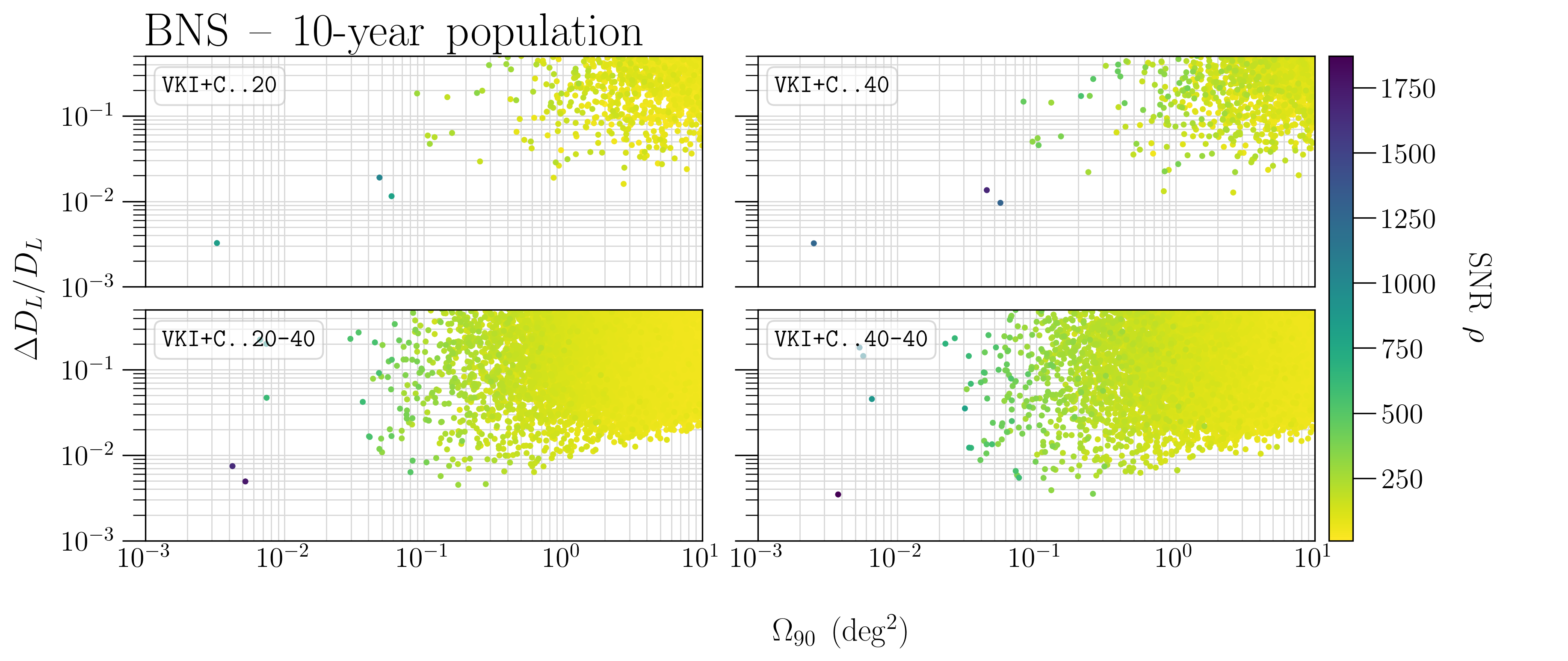}
\includegraphics[width=0.93\textwidth]{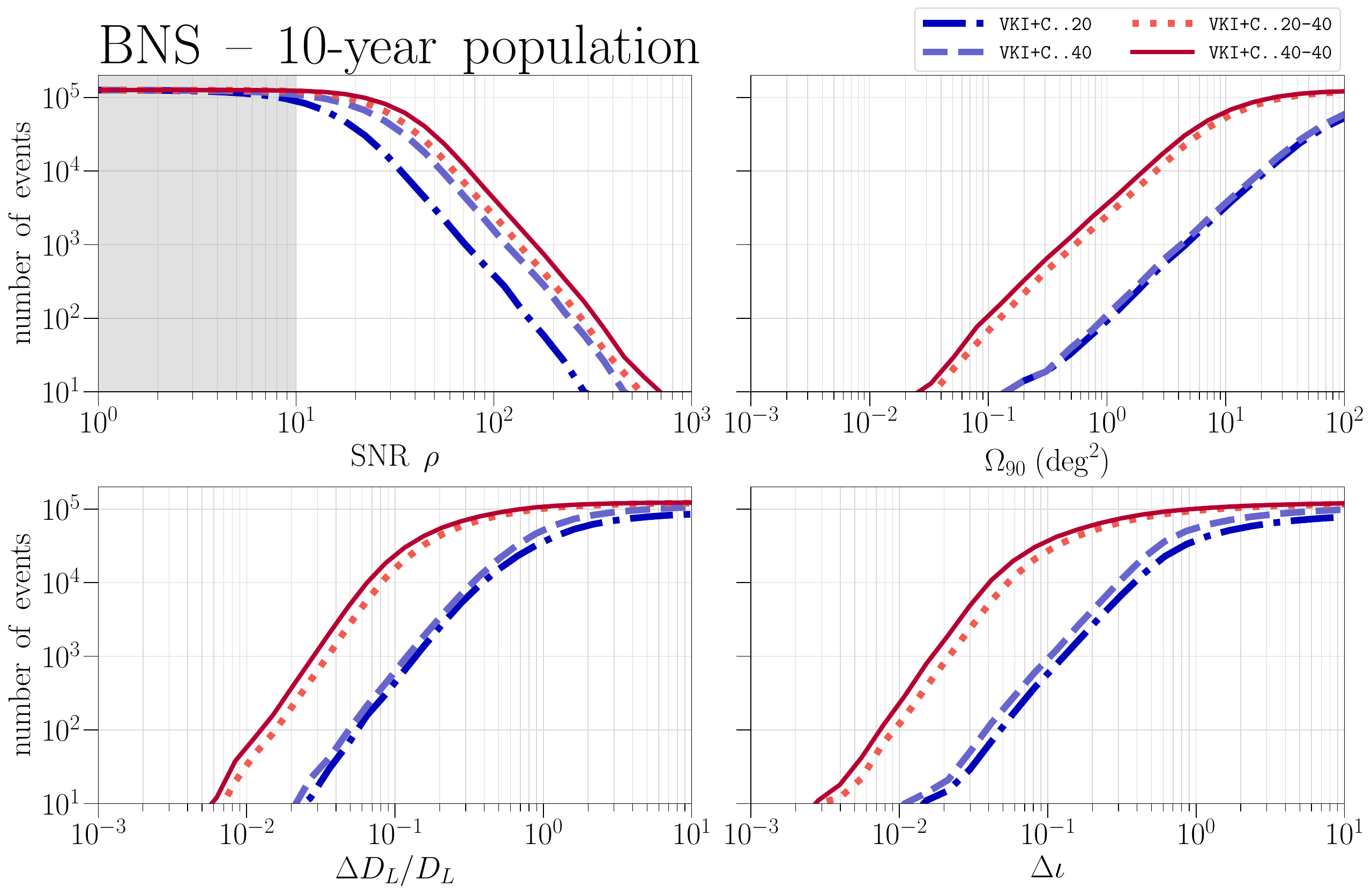}
\caption{\emph{Top}: The scatter plots illustrate the correlations between SNR $\rho$, 90\%-credible sky area $\Omega_{90}$, and fractional luminosity distance error $\Delta D_L/D_L$ for BNS mergers with SNR $\rho\geq10$ in the four studied CE networks for redshifts $z\leq0.5$. The color bar indicates the SNR of the events.\\
\emph{Bottom:} Cumulative histograms for the SNR $\rho$, 90\%-credible sky area $\Omega_{90}$, fractional luminosity distance errors $\Delta D_L/D_L$, and absolute errors on the inclination angle $\Delta\iota$ for BNS mergers observed in the four studied CE networks for redshifts $z\leq0.5$. The histograms were generated from $\sim 1.2\times10^{5}$ injections sampled according to Sec. \ref{sec:resampling}. The non-SNR panels are obtained for events with SNR $\rho\geq10$, indicated by the non-shaded region in the SNR panel. The SNR panel is flipped to highlight the behavior for large values.}
\label{fig:mma_BNS_ce_configs}
\end{figure*}

\begin{figure*}
\includegraphics[width=\textwidth]{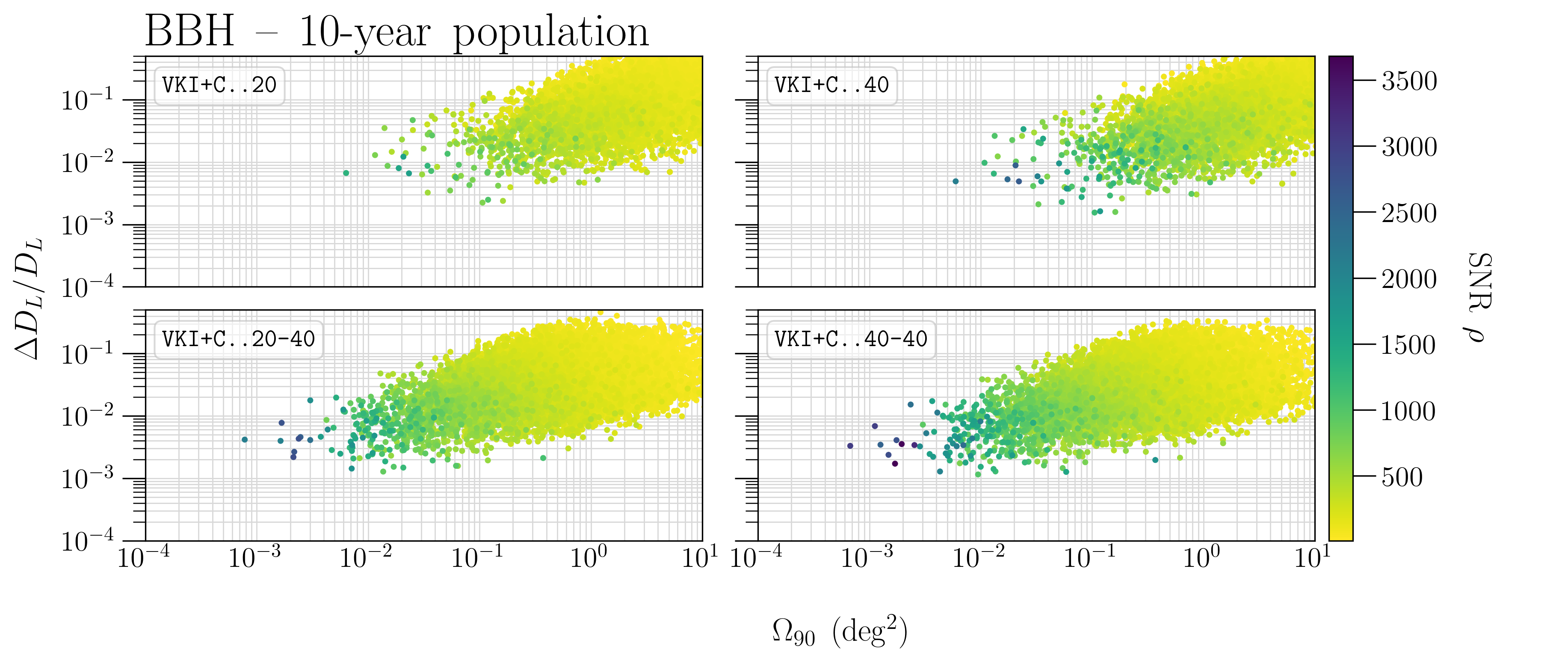}
\includegraphics[width=0.93\textwidth]{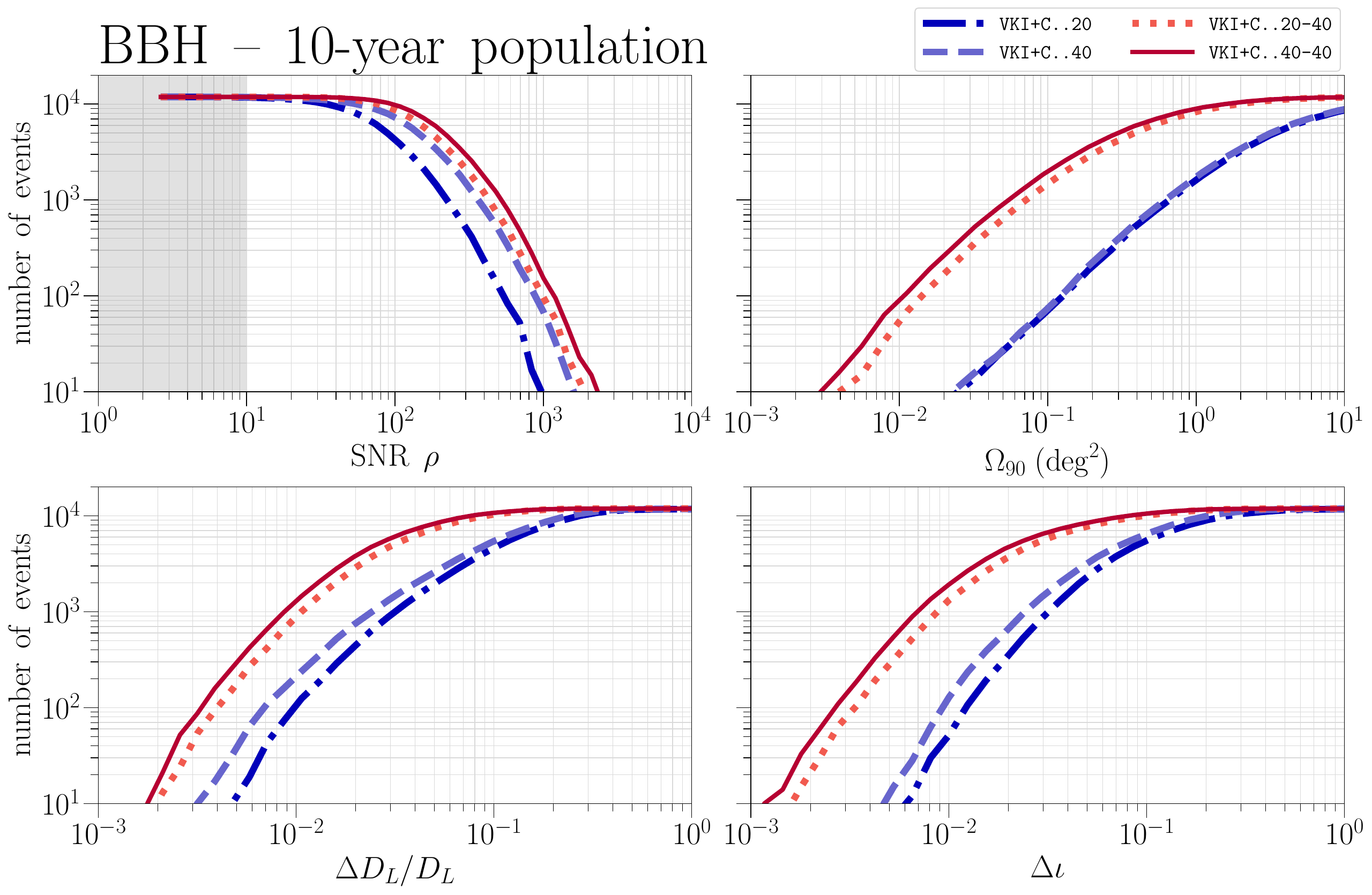}
\caption{\emph{Top}: The scatter plots illustrate the correlations between SNR $\rho$, 90\%-credible sky area $\Omega_{90}$, and fractional luminosity distance error $\Delta D_L/D_L$ for BBH mergers with SNR $\rho\geq10$ in the four studied CE networks for redshifts $z\leq0.5$. The color bar indicates the SNR of the events.\\
\emph{Bottom:} Cumulative histograms for the SNR $\rho$, 90\%-credible sky area $\Omega_{90}$, fractional luminosity distance errors $\Delta D_L/D_L$, and absolute errors on the inclination angle $\Delta\iota$ for BBH mergers observed in the four studied CE networks for redshifts $z\leq0.5$. The histograms were generated from $\sim 1.2\times10^{4}$ injections sampled according to Sec. \ref{sec:resampling}. The non-SNR panels are obtained for events with SNR $\rho\geq10$, indicated by the non-shaded region in the SNR panel. The SNR panel is flipped to highlight the behavior for large values.}
\label{fig:mma_BBH_ce_configs}
\end{figure*}

\begin{figure*}[b]
\includegraphics[width=0.99\textwidth]{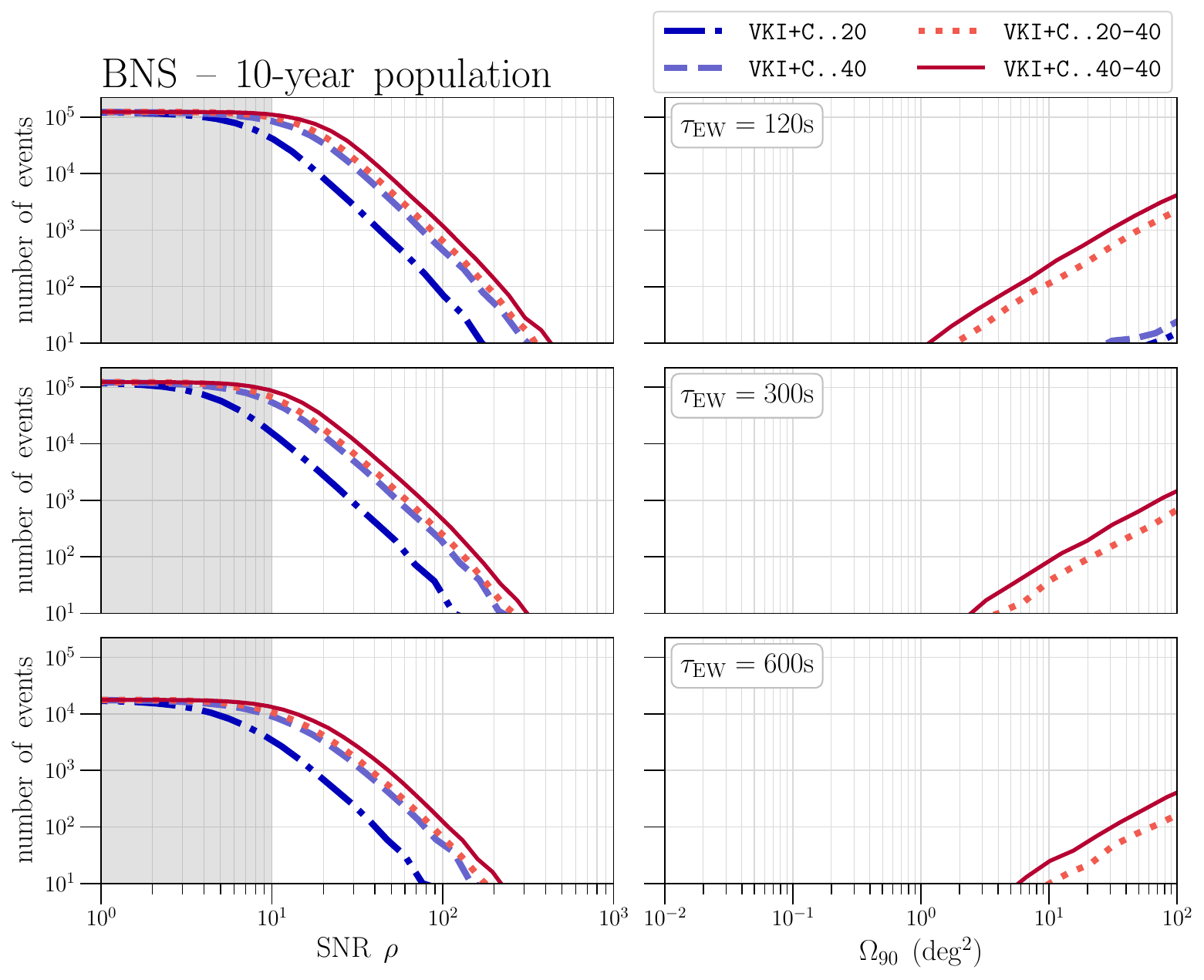}
\caption{Cumulative histograms for the SNR $\rho$ and 90\%-credible sky area $\Omega_{90}$ for BNS mergers observed in the four studied CE networks for redshifts $z\leq0.5$ 2, 5, and 10 minutes before merger. The histograms were generated from $\sim 1.2\times10^{5}$ injections sampled according to Sec. \ref{sec:resampling}. The $\Omega_{90}$ panel is obtained for events with SNR $\rho\geq10$, indicated by the non-shaded region in the SNR panel. The SNR panel is flipped to highlight the behavior for large values.
\\The cumulative histograms of SNR in the $\tau_{\rm EW}=600\,{\rm s}$ panel do not reach the value 1 because $\sim85\%$ of the events exhibited a cutoff-frequency $f_{\rm EW} < 11\,{\rm Hz}$. Since this is too close to the lower frequency cutoff of the used V+ sensitivity curve used in half our networks, we chose to abort these runs for all networks.}
\label{fig:ew_snr_sa90_ce_configs}
\end{figure*}

\begin{figure*}
    \centering
    \includegraphics[width=0.7\textwidth]{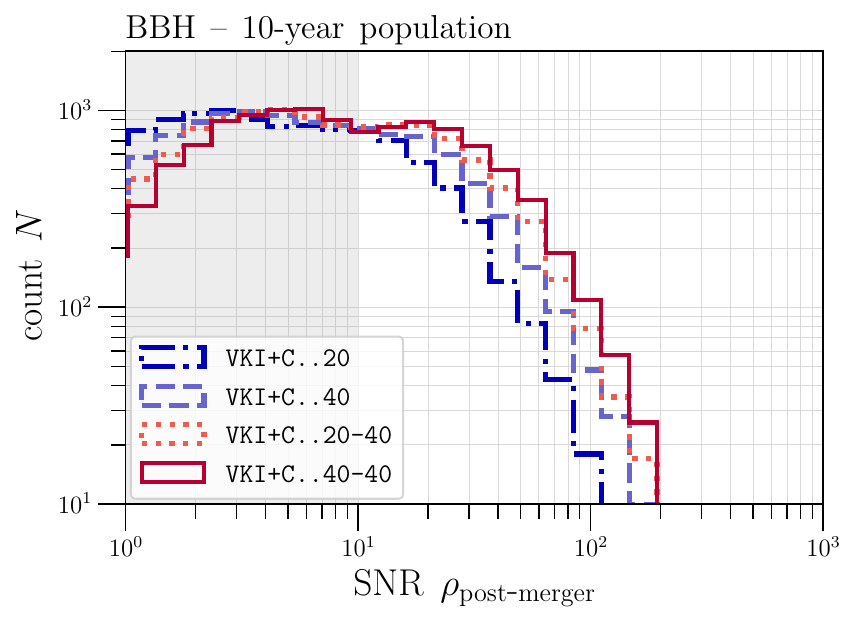}
    \caption{Histograms of the SNR $\rho_{{\rm post}\textnormal{-}{\rm merger}}$ in the post-merger signal of the BBH population in the four studied CE networks for redshifts $z\leq0.5$ and a observation time of \textbf{10 years}. The histograms were generated from $\sim 1.2\times10^{4}$ injections sampled according to Sec. \ref{sec:resampling}. }
    \label{fig:pm_ce_configs}
\end{figure*}

In developing the science case for a NG GW detector proposal, the Cosmic Explorer Project is investigating various configurations of CE detectors, e.g. with varying detector arm lengths and hence detectors sensitivity curves, see Fig. \ref{fig:sensitivity_curves_ce_configs}. Given the early state of the CE detector proposal, we included, for completeness, four  networks containing one or two CE detectors with either $20\,{\rm km}$ or $40\,{\rm km}$ arms to demonstrate the potential of NG networks without an ET detector: \cetw, \cefo, \cetwfo, and \cefofo. The \cefo\ network is the same as \ce\ and is included as a reference from the main body of this paper. Figs. \ref{fig:detection_efficiency_rate_ce_configs} to \ref{fig:pm_ce_configs} reiterate the plots presented throughout this paper for these four CE networks instead of the six studied A+, Voyager, and NG networks.

Comparing Tabs.~\ref{tab:reach_ce_configs} and \ref{tab:rates_snr_ce_configs} to \ref{tab:reach} and \ref{tab:rates_snr} indicates a single $20\,{\rm km}$ CE detector to be equivalent {\em in sensitivity} to one ET when either is embedded into a network of A+ detectors. Hence, the \cetw\ and \et\ (\cetwfo\ and \EC) networks yield comparable reaches, per-mille redshifts, and detection rates at the respective threshold SNR values.

As the comparison of the \ce\ and \et\ networks already predicted, {\em measurement quality} depicts a more nuanced picture which does not strictly follow the conclusions drawn based on sensitivity. Comparing Tabs.~\ref{tab:rates_sky_area_DLP_ce_configs} and \ref{tab:mma_rates_ce_configs} to \ref{tab:rates_sky_area_DL} and \ref{tab:mma_rates}, it is clear that the ET network outpaces both single CE networks, \cetw\ and \cefo, in detection rates of well-localized events in sky area and luminosity distance for BNSs and BBHs. In fact, the single ET network slightly outperforms both double-CE networks, \cetwfo\ and \cefofo, in sky localization for BNSs and matches the BBHs numbers of \cetwfo. The two $40\,{\rm km}$ CEs in \cefofo\ will detect more well-localized BBHs per year than \et, yet fall significantly behind the \EC\ network with one ET and one $40\,{\rm km}$ CE. The situation flips again in favor of the double-CE networks for luminosity distance estimation: both \cetwfo\ and \cefofo\ measure the luminosity distance to BNSs and BBHs consistently at a level between \EC\ and \ECS.

Since early warning detections heavily depend on a network's capability to detect GWs at an early stage in the binary evolution, the double-CE networks should observe EW BNS events more often than \et due to the higher sensitivity of CE, but much less frequently than either \EC\ or \ECS which combine the strengths of the ET and CE detectors (see Table~\ref{tab:ew_ce_configs} in comparison to Table~\ref{tab:ew}).

\begin{table*}[]
\caption{The reach $z_r$ and per-mille redshift $z_{pm}$ of the considered networks for BNS and BBH signals with SNRs $\rho\geq10$ or $\rho\geq100$. Here we define the reach (per-mille redshift) as the redshift at which a given network detects 50\% {\red (0.1\%)} of the injections with the specified SNR or louder. Please refer to the detection efficiency panels of Fig.~\ref{fig:detection_efficiency_rate_ce_configs} for a visual representation.}
\begin{tabular}{l|c|c|c|c|c|c|c|c}
\hline
\hline
           & \multicolumn{4}{c|}{BNS}                                 & \multicolumn{4}{c}{BBH}                                 \\
\hline
SNR $\rho\quad \;$ & \multicolumn{2}{c|}{$\geq10$} & \multicolumn{2}{c|}{$\geq100$} & \multicolumn{2}{c|}{$\geq10$} & \multicolumn{2}{c}{$\geq100$} \\
\hline
\hline
   & $\;\;\: z_r\;\;\:$        & $\;\;\: z_{pm}\;\;\: $       & $\;\;\: z_r\;\;\: $         & $\;\;\: z_{pm}\;\;\: $       & $\;\;\: z_r\;\;\: $        & $\;\;\: z_{pm}\;\;\: $       & $\;\;\: z_r\;\;\: $        & $\;\;\: z_{pm}\;\;\: $        \\
\hline
\hline
\cetw        &  0.6   & 3.3   & 0.06  & 0.25      &  5.9   & 41    & 0.3   & 2.1       \\ \hline
\cefo        &  1     & 6.6   & 0.09  & 0.39      &  14    & $>50$ & 0.5   & 3.6       \\ \hline
\cetwfo      &  1.3   & 7.9   & 0.11  & 0.44      &  19    & $>50$ & 0.63  & 4.3       \\ \hline
\cefofo      &  1.7   & 11    & 0.13  & 0.54      &  28    & $>50$ & 0.79  & 5.4       \\ \hline
\hline
\end{tabular}
\label{tab:reach_ce_configs}
\end{table*}

\begin{table*}
    \centering
        \caption{Cosmic merger rates (per year) of BNS and BBH mergers in the Universe and the number that would be observed by different detector networks each year with $\rho \geq 10, 30, 100,$ where $\rho$ is the signal-to-noise ratio. Due to uncertainty in the various quantities that go into the calculation these numbers are no more accurate than one or two significant figures.}
    \begin{tabular*}{\columnwidth}{@{\extracolsep{\fill}}l?r|r|r?r|r|r}
\hline
\hline
                    & \multicolumn{3}{c?}{BNS} & \multicolumn{3}{c}{BBH} \\
\hline
Cosmic rate         & \multicolumn{3}{c?}{$4.7 \times 10^5$} & \multicolumn{3}{c}{ $1.2 \times 10^5$} \\
\hline
\hline
SNR $\rho$          & \multicolumn{1}{c|}{$ \geq10$} &  \multicolumn{1}{c|}{$ \geq30$} & \multicolumn{1}{c?}{$\geq100$} &  \multicolumn{1}{c|}{$\geq10$} &  \multicolumn{1}{c|}{$ \geq30$} & \multicolumn{1}{c}{$\geq100$}\\
\hline
\hline
\cetw     & 37,000   & 1,400    & 38       & 89,000   & 27,000   & 1,700   \\
\cefo     & 110,000  & 6,000    & 160      & 110,000  & 53,000   & 6,400   \\
\cetwfo   & 150,000  & 9,200    & 250      & 120,000  & 66,000   & 9,200   \\
\cefofo   & 210,000  & 16,000   & 420      & 120,000  & 80,000   & 14,000  \\
\hline
\hline
    \end{tabular*}
    \label{tab:rates_snr_ce_configs}
\end{table*}
\begin{table*}[ht]
    \centering
    \begin{minipage}[b]{0.45\textwidth}
      \caption{Detection rates of BNS and BBH mergers from the full redshift range $z\in[0.02,50]$ to be observed by different detector networks each year with $\Omega_{90}/{\rm deg^2}\leq 1,\, 0.1,\, 0.01$ as well as $\Delta D_L/D_L \leq 0.1,\, 0.1$, where $\Omega_{90}$ is the 90\%-credible sky area and $D_L$ the luminosity distance. These detection rates are calculated for events with SNR $\rho\geq10$. Due to uncertainty in the various quantities that go into the calculation these numbers are no more accurate than one or two significant figures.}
    \begin{tabular*}{\columnwidth}{@{\extracolsep{\fill}}l?r|r|r?r|r}
\hline
\hline
Metric \phantom{AAA} & \multicolumn{3}{c?}{$\Omega_{90}$  $({\rm deg^2})$ \phantom{A}} & \multicolumn{2}{c}{$\Delta D_L/D_L$}\phantom{A} \\
\hline
Quality &  $\leq1$\phantom{A} &  $\leq 0.1$\phantom{A} &  $\leq 0.01$\phantom{A} &  $\leq 0.1$\phantom{A} &  $\leq0.01$\phantom{A}\\
\hline
\hline
\multicolumn{6}{c}{\emph{BNS}} \\
\hline
\hline
\cetw     & 10       & 1        & 0        & 46       & 0       \\
\cefo     & 11       & 1        & 0        & 67       & 0       \\
\cetwfo   & 240      & 7        & 1        & 2,100    & 4       \\
\cefofo   & 360      & 10       & 1        & 4,000    & 6       \\
\hline
\hline
\multicolumn{6}{c}{\emph{BBH}} \\
\hline
\hline
\cetw     & 190      & 7        & 0        & 6,200    & 18      \\
\cefo     & 210      & 8        & 0        & 10,000   & 45      \\
\cetwfo   & 4,400    & 180      & 6        & 34,000   & 240     \\
\cefofo   & 6,300    & 270      & 9        & 42,000   & 420     \\
\hline
\hline
    \end{tabular*}
    \label{tab:rates_sky_area_DLP_ce_configs}
    \end{minipage}
    \hspace{0.04\textwidth}
    \begin{minipage}[b]{0.45\textwidth}
     \caption{Detection rates of BNS and BBH mergers up to redshift $z=0.5$ to be observed by different detector networks each year with $\Omega_{90}/{\rm deg^2}\leq 1,\, 0.1,\, 0.01$ as well as $\Delta D_L/D_L \leq 0.1,\, 0.1$, where $\Omega_{90}$ is the 90\%-credible sky area and $D_L$ the luminosity distance. These detection rates are calculated for events with SNR $\rho\geq10$. Due to uncertainty in the various quantities that go into the calculation these numbers are no more accurate than one or two significant figures. The bare merger rates for BNSs and BBHs up to redshift $z=0.5$ are $\sim12,000\,{\rm yr^{-1}}$ and $\sim1,200\,{\rm yr^{-1}}$, respectively.}
    \begin{tabular*}{\columnwidth}{@{\extracolsep{\fill}}l?r|r|r?r|r}
\hline
\hline
Metric        & \multicolumn{3}{c?}{$\Omega_{90}$  $({\rm deg^2})$ \phantom{A}} & \multicolumn{2}{c}{$\Delta D_L/D_L$}\phantom{A} \\
\hline
Quality &  $\leq1$\phantom{A} &  $\leq 0.1$\phantom{A} &  $\leq 0.01$\phantom{A} &  $\leq 0.1$\phantom{A} &  $\leq0.01$\phantom{A}\\
\hline
\hline
\multicolumn{6}{c}{\emph{BNS}} \\
\hline
\hline
\cetw     & 9        & 1        & 0        & 43       & 0       \\
\cefo     & 10       & 1        & 0        & 56       & 0       \\
\cetwfo   & 230      & 7        & 1        & 1,400    & 3       \\
\cefofo   & 340      & 10       & 1        & 2,300    & 6       \\
\hline
\hline
\multicolumn{6}{c}{\emph{BBH}} \\
\hline
\hline
\cetw     & 160      & 7        & 0        & 470      & 11      \\
\cefo\phantom{AA}     & 180      & 8        & 0        & 550      & 21      \\
\cetwfo   & 840      & 150      & 5        & 1,000    & 88      \\
\cefofo   & 910      & 200      & 9        & 1,100    & 130     \\
\hline
\hline
    \end{tabular*}
    \label{tab:mma_rates_ce_configs}
    \end{minipage}
\end{table*}
\begin{table*}[b]
    \centering
    \caption{Number of BNS mergers events for redshifts $z\leq0.5$ observed in the four studied CE networks that can be localized to within a small region on the sky ($\Omega_{90}/{\rm deg^2} \leq 100, 20, 10, 1$) 2, 5, and 10 minutes before merger. For a given early warning time $\tau_{\rm EW}$ we determine the corresponding starting frequency $f_{\rm EW}$ using Eq. \eqref{eq:coalescence frequency}.}
    \begin{tabular*}{0.75\textwidth}{l @{\extracolsep{\fill}}|  cccc|cccc|cccc}
    \hline \hline
EW time & \multicolumn{4}{c|}{$\tau_{\rm EW}=120$ s} &
  \multicolumn{4}{c|}{$\tau_{\rm EW}=300$ s} &
  \multicolumn{4}{c}{$\tau_{\rm EW}=600$ s} \\
\hline
$\Omega_{90}$ (deg$^2$)\phantom{ABC} &
$\le 100$ &
$\le 20$ &
$\le 10$ &
$\le 1$ \phantom{A} &
$\le 100$ &
$\le 20$ &
$\le 10$ &
$\le 1$ \phantom{A} &
$\le 100$ &
$\le 20$ &
$\le 10$ &
$\le 1$ \phantom{A} \\
    \hline
    \hline
\cetw     & 1     & 0     & 0     & 0    \phantom{ABC} & 0     & 0     & 0     & 0    \phantom{ABC} & 0     & 0     & 0     & 0    \\
\cefo     & 2     & 0     & 0     & 0    \phantom{ABC} & 0     & 0     & 0     & 0    \phantom{ABC} & 0     & 0     & 0     & 0    \\
\cetwfo   & 170   & 21    & 9     & 0    \phantom{ABC} & 35    & 5     & 2     & 0    \phantom{ABC} & 1     & 0     & 0     & 0    \\
\cefofo   & 360   & 53    & 20    & 1    \phantom{ABC} & 97    & 13    & 6     & 0    \phantom{ABC} & 4     & 1     & 0     & 0    \\
\hline \hline
    \end{tabular*}
    \label{tab:ew_ce_configs}
\end{table*}

\clearpage

\bibliographystyle{JHEP}
\bibliography{refs}

\end{document}